\documentclass[12pt]{iopart}
\usepackage{iopams}
\usepackage[square,numbers,sort&compress]{natbib}
\usepackage{graphicx,amssymb,bm}
\usepackage{csquotes}
\usepackage{xcolor}
\usepackage[normalem]{ulem}
\usepackage{upgreek}
\newcommand{\refeq}[1]{(\ref{#1})}

\newcommand{\beqn}{\begin{eqnarray}}
\newcommand{\eeqn}{\end{eqnarray}}
\newcommand{\xp}{\times}
\newcommand{\fract}[2]{#1 / #2 }
\newcommand{\mb}[1]{\bm{#1}}
\newcommand{\eye}{\mb{I}}
\newcommand{\expo}[1]{\exp{\left[#1\right]}}
\newcommand{\imag}{{\rm{i}}}
\newcommand{\order}[1]{\Or\left( #1 \right) }
\newcommand{\ltsp}{c}
\newcommand{\bu}{\mb{b}}
\newcommand{\bvec}{\mb{B}}
\newcommand{\bmag}{B}
\newcommand{\bmagmax}{B_{\rm max}}

\newcommand{\wfreq}{\omega}
\newcommand{\wfreqhat}{\hat{\omega}}
\newcommand{\wfreqr}{\omega_{{\rm{r}}}}
\newcommand{\wfreqrhat}{\hat{\omega}_{{\rm{r}}}}

\newcommand{\spe}{s}
\newcommand{\pbetaprim}{\beta^\prime}
\newcommand{\pbeta}{\beta}
\newcommand{\pbetaeff}{\beta^{\rm eff}}
\newcommand{\pbetae}{\beta_{\rm{e}}}
\newcommand{\fldl}{\alpha}
\newcommand{\fldlz}{\alpha_0}
\newcommand{\flxl}{\psi}
\newcommand{\flxlz}{\psi_0}

\newcommand{\kkflxl}{k_\flxl}
\newcommand{\kkfldl}{k_\fldl}
\newcommand{\kperp}{k_\perp}
\newcommand{\kperpvec}{\mb{k}_{\perp}}
\newcommand{\kpara}{k_\|}
\newcommand{\kparageo}{k_\|^{\rm \scriptstyle g}}
\newcommand{\kparamode}{k_\|^{\rm \scriptstyle w}}

\newcommand{\xx}{x}
\newcommand{\ttime}{t}
\newcommand{\nbl}{\nabla}
\newcommand{\pvel}{\mb{v}}
\newcommand{\pvelprim}{\mb{v}^\prime}
\newcommand{\gyrophase}{\gamma}
\newcommand{\energy}{\varepsilon}
\newcommand{\pitch}{\lambda}
\newcommand{\sign}{\sigma}
\newcommand{\vpar}{v_\|}
\newcommand{\wpar}{w_\|}
\newcommand{\vmag}{v}
\newcommand{\vperp}{v_\perp}
\newcommand{\wperp}{w_\perp}
\newcommand{\wvel}{\mb{w}}

\newcommand{\Ufunc}{\mb{U}}
\newcommand{\lpar}{\theta}
\newcommand{\lchi}{\chi}
\newcommand{\kpar}{\bu \cdot \nbl \lpar}
\newcommand{\gpar}{g_\|}
\newcommand{\kparprim}{\bu \cdot \nbl \lpar^\prime}
\newcommand{\lscal}{a}
\newcommand{\toragl}{\zeta}
\newcommand{\saffac}{q}
\newcommand{\nufunc}{\nu}
\newcommand{\saffacprim}{\saffac^\prime}
\newcommand{\bcur}{I}
\newcommand{\jpar}{J_\|}
\newcommand{\jint}{j_\|}

\newcommand{\jintpm}{j^{\pm}_\|}
\newcommand{\jintplus}{j^{+}_\|}
\newcommand{\jintminus}{j^{-}_\|}

\newcommand{\transav}[1]{\left\langle{#1}\right\rangle^{\rm{t}}}
\newcommand{\bouncav}[1]{\left\langle{#1}\right\rangle^{\rm{b}}}
\newcommand{\gav}[2]{\left\langle{#1}\right\rangle^{\gyrophase}_{#2}}
\newcommand{\intv}[1]{\int   #1 \; d^3 \mb{v} }
\newcommand{\intvprim}[1]{\int   #1 \; d^3 \mb{v}^\prime }
\newcommand{\ptl}{\phi}

\newcommand{\apar}{A_{\|}}

\newcommand{\aptl}{\Psi}
\newcommand{\Daptlhat}{\Delta\widehat{ \aptl}}
\newcommand{\bpar}{B_{\|}}

\newcommand{\hhs}{h_{\spe }}
\newcommand{\hhe}{h_{{\rm{e}} }}
\newcommand{\hhepassing}{h^{\rm \scriptstyle passing}_{{\rm{e}} }}

\newcommand{\fulldistf}{f}
\newcommand{\fulldistfprim}{f^\prime}
\newcommand{\dlf}{\delta \! f}
\newcommand{\dlfs}{\delta \! f_\spe}
\newcommand{\eqlba}{F_{0\spe}}

\newcommand{\eqlbe}{F_{0e}}
\newcommand{\eqlbeprim}{F_{0{\rm{e}}}^\prime}
\newcommand{\hhez}{\hhe{}^{(0)}}
\newcommand{\hheh}{\hhe{}^{(1/2)}}
\newcommand{\hheo}{\hhe{}^{(1)}}
\newcommand{\HHe}{H_{\rm{e}}}

\newcommand{\bes}[1]{ J_0(#1)}
\newcommand{\besfirst}[1]{ J_1(#1)}
\newcommand{\bessn}[1]{J_{#1\spe}}
\newcommand{\besen}[1]{J_{#1{\rm{e}}}}

\newcommand{\gyrdvecs}{\bm{\rho}_\spe}

\newcommand{\gyrds}{\rho_{\spe}}
\newcommand{\gyrdi}{\rho_{\rm{i}}}
\newcommand{\gyrde}{\rho_{\rm{e}}}
\newcommand{\dens}{n}
\newcommand{\denss}{n_\spe}
\newcommand{\denssprim}{n_{\spe^\prime}}
\newcommand{\densi}{n_{\rm{i}}}
\newcommand{\dense}{n_{\rm{e}}}
\newcommand{\ddensi}{\delta \! n_{\rm{i}}}
\newcommand{\ddense}{\delta \! n_{\rm{e}}}
\newcommand{\ddenss}{\delta \! n_{\spe}}
\newcommand{\dupars}{\delta \! u_{\|,\spe}}

\newcommand{\dpress}{\delta \! p_{\perp,\spe}}
\newcommand{\pres}{p}

\newcommand{\temps}{T_\spe}
\newcommand{\tempi}{T_{\rm{i}}}
\newcommand{\tempe}{T_{\rm{e}}}

\newcommand{\vtheri}{v_{\rm{i}}}
\newcommand{\vthere}{v_{\rm{e}}}
\newcommand{\vthers}{v_{\spe}}

\newcommand{\zedeff}{Z_{\rm eff}}
\newcommand{\zeds}{Z_\spe}
\newcommand{\zedsprim}{Z_{\spe^\prime}}
\newcommand{\zedi}{Z_{\rm{i}}}
\newcommand{\charge}{e}
\newcommand{\tdrv}[2]{\frac{d #1}{d #2}}
\newcommand{\tdrvt}[2]{\fract{d #1}{d #2}}
\newcommand{\drv}[2]{\frac{\partial #1}{\partial #2}}
\newcommand{\drvt}[2]{{\partial #1}/{\partial #2}}

\newcommand{\ma}{m_\spe}
\newcommand{\me}{m_{{\rm{e}}}}

\newcommand{\md}{m_{{\rm D}}}
\newcommand{\massrt}{\left(m_{\rm{e}}/m_{\rm{i}}\right)^{1/2}}
\newcommand{\massr}{\left(\frac{m_{\rm{e}}}{m_{\rm{i}}}\right)^{1/2}}
\newcommand{\massru}{\left(\frac{m_{\rm{i}}}{m_{\rm{e}}}\right)^{1/2}}
\newcommand{\massrut}{\left({m_{\rm{i}}}/{m_{\rm{e}}}\right)^{1/2}}
\newcommand{\massrtp}[1]{\left({m_{\rm{e}}}/{m_{\rm{i}}}\right)^{#1}}
\newcommand{\massrutp}[1]{\left({m_{\rm{i}}}/{m_{\rm{e}}}\right)^{#1}}
\newcommand{\massrup}[1]{\left(\frac{m_{\rm{i}}}{m_{\rm{e}}}\right)^{#1}}

\newcommand{\vms}{\mb{v}_{{\rm{M}},{\spe}}}
\newcommand{\vme}{\mb{v}_{\rm{M,e}}}

\newcommand{\cycfs}{\Omega_\spe}
\newcommand{\cycfe}{\Omega_{\rm{e}}}
\newcommand{\cycferef}{\Omega_{\rm{e}}^{\scriptsize\rm{ref}}}
\newcommand{\gyrderef}{\rho_{\rm{e}}^{\scriptsize\rm{ref}}}
\newcommand{\lparp}{\lpar^{+}_b}
\newcommand{\lparm}{\lpar^{-}_b}
\newcommand{\lparpm}{\lpar^{\pm}_b}
\newcommand{\lparprim}{\lpar^{\prime}}

\newcommand{\coplandaue}{C_{\rm{ee}}}
\newcommand{\coplorentze}{C_{\rm{ei}}}
\newcommand{\copbothe}{\mathcal{C}}

\newcommand{\cops}{C^{\rm{GK}}_{\spe}}

\newcommand{\copehat}{\widehat{C}_{\rm{e}}}

\newcommand{\coloumblog}{\ln \Lambda}
\newcommand{\cope}{C^{\rm{GK}}_{\rm{e}}}

\newcommand{\cfreqhat}{\hat{\nu}}
\newcommand{\cfreqii}{\nu_{\rm{ii}}}
\newcommand{\cfreqei}{\nu_{\rm{ei}}}
\newcommand{\cfreqee}{\nu_{\rm{ee}}}

\newcommand{\cfreqssprim}{\nu_{\spe\spe^\prime}}
\newcommand{\geofunc}{G} 
\newcommand{\geoparams}{\bm{g}} 
\newcommand{\gneo}{g_{\rm \scriptscriptstyle N}} 
\newcommand{\gcla}{g_{\rm \scriptscriptstyle C}} 

\newcommand{\dske}{\delta_{\rm{e}}}
\newcommand{\wmaghat}{\hat{\omega}_{\rm{M}}}

\newcommand{\delt}{\Delta \ttime}
\newcommand{\lzed}{z}
\newcommand{\lzedmax}{z_{{\rm max}}}
\newcommand{\polm}{m}
\newcommand{\ptlhat}[1]{\hat{\ptl}_{#1}}
\newcommand{\sfunce}{\mathcal{S}_{{\rm e}}}

\newcommand{\bref}{B_{\scriptsize{\rm{ref}}}}

\newcommand{\shat}{\hat{s}}
\newcommand{\radial}{r}
\newcommand{\radialx}{x}
\newcommand{\binormal}{y}
\newcommand{\kky}{k_y}
\newcommand{\kkx}{k_x}
\newcommand{\kkr}{k_r}
\newcommand {\thetaz} {\theta_0}
 \newcommand{\negrid}{n_\varepsilon}
\newcommand{\ntwopi}{n_{2\pi}}
\newcommand{\ntheta}{n_\theta}
\newcommand{\npitch}{n_\lambda}
\newcommand{\rhominor}{\rho}
\newcommand{\rminor}{r}
\newcommand{\rmajor}{R}
\newcommand{\rmajormax}{R_{\rm max}}
\newcommand{\rmajormin}{R_{\rm min}}
\newcommand{\rmajgeo}{R_{\rm geo}}
\newcommand{\rmajorz}{R_0}
\newcommand{\aspect}{\epsilon}
\newcommand{\dpsidx}{d \flxl / d x}
\newcommand{\daldy}{d \fldl / d \binormal}
\newcommand{\kkxnorm}{\bref\rminor/ \saffac}
\newcommand{\kkynorm}{\bref\tdrvt{\rminor}{\flxl}}

\newcommand{\growth}{\gamma}
\newcommand{\growthhat}{\hat{\gamma}}

\newcommand{\lti}{L_{\tempi}}
\newcommand{\lte}{L_{\tempe}}
\newcommand{\lts}{L_{\temps}}

\newcommand{\rstars}{\rho_{\ast\spe}}

\newcommand{\wstar}{\omega_{\ast}}
\newcommand{\lln}{L_n}

\newcommand{\bflrs}{b_{\spe}}
\newcommand{\bflre}{b_{{\rm{e}}}}
\newcommand{\wstars}{\omega_{\ast,\spe}}
\newcommand{\wstarsn}{\omega^{\dens}_{\ast,\spe}}
\newcommand{\wstare}{\omega_{\ast,{\rm{e}}}}
\newcommand{\wstarehat}{\hat{\omega}_{\ast,{\rm{e}}}}
\newcommand{\etas}{\eta_\spe}
\newcommand{\etae}{\eta_{\rm{e}}}
\newcommand{\gstwo}{\texttt{GS2}}

\begin{document}

\title[]{New linear stability parameter to describe low-$\pbeta$ electromagnetic microinstabilities
 driven by passing electrons in axisymmetric toroidal geometry}
\author{M. R. Hardman$^{1}$\footnote{Current address: Tokamak Energy Ltd, 173 Brook Drive, Milton Park, Abingdon, OX14 4SD}
, F. I. Parra$^{2}$,    
  B. S. Patel$^3$, C. M. Roach$^3$,
 J. Ruiz Ruiz$^{1}$, M. Barnes$^{1}$, D. Dickinson$^{4}$,
 W. Dorland$^{5,1}$, J. F. Parisi$^{2}$, 
 D. St-Onge$^{1}$, and H. Wilson$^{4}$ 
\address{$^{1}$ Rudolf Peierls Centre for Theoretical Physics, University of Oxford, Oxford, OX1 3PU, UK} 
\address{$^2$Princeton Plasma Physics Laboratory, Princeton University, Princeton, NJ 08540, USA}
\address{ $^3$ Culham Centre for Fusion Energy, UKAEA, Abingdon OX14 3DB, UK} 
\address{$^4$ York Plasma Institute, Department of Physics, University of York, Heslington, York, YO10 5DD, UK}
\address{$^5$ Department of Physics, University of Maryland, College Park, Maryland 20742, USA}
}

\ead{\texttt{michael.hardman@physics.ox.ac.uk}} 
\begin{abstract}
    In magnetic confinement fusion devices, 
    the ratio of the plasma pressure to the magnetic field energy, $\pbeta$, 
    can become sufficiently large that 
    electromagnetic microinstabilities become unstable,
    driving turbulence that distorts or reconnects the equilibrium magnetic field.
    In this paper, a theory is proposed for electromagnetic, electron-driven linear instabilities
    that have 
    current layers localised to mode-rational surfaces and binormal wavelengths
    comparable to the ion gyroradius. 
    The model retains axisymmetric toroidal geometry
    with arbitrary shaping, and consists of orbit-averaged equations for the
    mode-rational surface layer, with a ballooning space kinetic matching condition 
    for passing electrons. The matching condition connects the current layer
    to the large scale electromagnetic fluctuations, and is derived
    in the limit that $\pbeta$ is comparable to the square root of the electron-to-ion-mass ratio. 
    Electromagnetic fluctuations only enter through the matching condition, allowing 
    for the identification of an effective $\pbeta$ that
    includes the effects of equilibrium flux surface shaping.
    The scaling predictions made by the asymptotic theory are tested with comparisons to results
    from linear simulations of micro-tearing and electrostatic
    microinstabilities in MAST discharge \#6252, showing excellent agreement.   
    In particular, it is demonstrated that the effective $\pbeta$ can explain
    the dependence of the local micro-tearing mode (MTM) growth rate on the ballooning parameter $\thetaz$ --
    possibly providing a route to optimise local flux surfaces for reduced MTM-driven transport. 

\end{abstract}
\maketitle
    \section{Introduction}\label{section:introduction}

 Turbulent transport often limits the performance of modern
 magnetic confinement fusion devices.
 Turbulence is driven by microinstabilities that are unstable
 to gradients in the mean plasma temperature and density profiles.
 It is difficult to suppress turbulent transport entirely:
 good confinement is achieved in regimes 
 where microinstability growth rates and 
 turbulent diffusivities are sufficiently
 small to allow for large density and temperature gradients.
 One such favourable regime is suggested by ideal magnetohydrodynamic (MHD)
 stability analysis of axisymmetric toroidal equilibria
 \cite{Strauss_1980_second_stability,LORTZ197849}: for a given
 magnetic equilibrium, there is both a minimum, and a maximum
 pressure gradient for instability. The existence of MHD-stable
 equilibria with large pressure gradients
 (equilibria with \enquote{second stability}) suggests a
 high-$\pbeta$ route to a reactor, where $\pbeta = 8\pi \pres/\bmag^2$ is the ratio of
  the total plasma pressure $\pres$ to the magnetic field energy $\bmag^2/8\pi$, with $\bmag$ the magnetic field strength. 

 High-$\pbeta$ equilibria suffer from unstable microinstabilities with 
 an electromagnetic character, such as the
 kinetic ballooning mode (KBM) \cite{Tang_1980KBM,Hastie_1981,Aleynikova2017KBM}
 or the microtearing mode (MTM)
 \cite{MTMPhysRevLett.44.994,GladdMTMPoF1981,Drake1983,ConnorCowleyHastieMTM_1990,Applegate_2007,Zocco_2015,Hamed2019_MTM,Patel_GKStability2021}.
 MTMs are driven by the electron temperature
 gradient (ETG) through both collisional
 \cite{GladdMTMPoF1981,Applegate_2007,Moradi_2013,Patel_GKStability2021}
 and collisionless \cite{Dickinson_MTMPed_2013,Predebon2013,Geng_2020,Patel_GKStability2021} mechanisms,
 despite a stable equilibrium current profile.
 In constrast, the macroscopic tearing mode \cite{FKRtearing1963} and drift-tearing modes
 \cite{DrakeLeePoF1977,Drake1983,Cowley_1986PoFtearing,ZoccoPoP2011,ConnorPPCF2012} 
 rely on finite resistivity and unstable current sheets
 in the magnetic equilibrium for instability.
 Electrostatic microinstabilities,
 which persist in the $\pbeta\rightarrow 0$ limit, may also appear
 \cite{Roach_2005,STroach2009PPCF}, 
 or be completely suppressed \cite{Patel_GKStability2021}.
 In particular, the suppression of the ion temperature gradient
 (ITG) instability and associated transport 
 is desirable; many studies have investigated the effects of finite $\pbeta$ on the ITG mode, see, e.g.,
 \cite{Bourdelle_2005,Citrin_2015,Zocco_2015_finitebetaITG,Ishizawa_2019}.
 Although there are a zoo of different instability mechanisms, microinstabilities
 share broad characteristics,
 and are well described by linearised, local gyrokinetics \cite{cattoPP78,Catto_1981},
 provided that $\rstars = \gyrds/\lscal \ll 1$.
 Here $\lscal $ is a typical equilibrium length scale, 
 $\gyrds = \vthers/\cycfs$ is the thermal gyroradius of particle species $\spe$, with
 $\vthers = \sqrt{2\temps/\ma}$ the thermal speed, $\cycfs = \zeds \charge \bmag/\ltsp \ma$
  the cyclotron frequency, $\temps$ the species temperature, $\ma$ the species mass,
 $\charge$ the proton charge, $\zeds$ the signed species charge number, and $\ltsp$ the speed of light.
 The wave number of the microinstability parallel to the magnetic field line $\kpara$ 
 is typically set by the connection length $\saffac \rmajor \sim \lscal$ between the inboard and outboard
 of the tokamak, i.e., $\kpara \saffac \rmajor\sim 1$. Here, $\saffac$ is the safety factor, and $\rmajor$
 is the major radius. The perpendicular wave number $\kperp$ satisfies $\kperp\gyrds\sim 1$,
 The frequency of the mode satisfies $\wfreq \sim \vthers/\lscal$. 
 The most dangerous microinstabilities are those with the longest wavelengths
 perpendicular to the magnetic field line -- larger turbulent eddies convect heat and particles faster. 
 
 In this paper we concern ourselves with the type of long-wavelength
 electromagnetic microinstability that is driven by passing electrons and 
 localised around mode-rational surfaces. We consider linear modes with binormal wavelengths 
 that are long compared to the thermal electron gyroradius; i.e., the binormal wavenumber $\kky$
 satisfies $\kky\gyrde \ll 1$.
 Reconnecting instabilities of this
 type are also referred to as MTMs. It is worth noting that
 there are electrostatic instabilities with many of the same characteristics
 \cite{HallatschekgiantelPRL2005,hardman_extended_tails}.
 Electromagnetic models of this type of linear instability have been developed
 in simple sheared-slab magnetic geometries \cite{MTMPhysRevLett.44.994,GladdMTMPoF1981,Drake1983,Zocco_2015},
 and in toroidal geometries with small inverse aspect ratio and circular flux surfaces \cite{ConnorCowleyHastieMTM_1990,Hamed2019_MTM}.
 However, a model using
 realistic axisymmetric toroidal geometry is required to make quantitative and
 qualitative comparisons to results from numerical simulations of experimental
 discharges. This is particularly relevant 
 in spherical tokamaks, where significant flux surface shaping is employed. 
 Here, we derive a model valid in full axisymmetric
 toroidal geometry with arbitrary cross-section. The form of the model reveals 
 the existance of an effective $\pbeta$ parameter that  
 takes into account the shaping of the local flux surface --  
 potentially opening a route to optimise local geometrical
 parameters for improved MTM stability. 

 Thanks to the presence of non-zero magnetic shear $\shat$, 
 long-wavelength, electron-driven instabilities have two
 distinct regions. First, there is a radial layer localised around the
 mode-rational surface with a width $\delta$. In this paper,
 we primarily consider the \enquote{collisionless} ordering where $\delta \sim \gyrde$.
 Second, there is an outer region: a region with large scales far
 from the rational surface. The scale of the outer region is tied to the choice of ordering for the 
 binormal wavenumber $\kky$, which sets a minimum to the value of $\kperp$.
 Because we are considering electron-driven
 microinstabilities at $\kky\gyrde \ll 1$ (rather than, e.g., MHD tearing instabilities),
 we may take $\kky\gyrdi\sim 1$ without loss 
 of generality. Hence, the radial scale of the outer region $\Delta \sim \gyrdi \gg \delta$. 
  In ballooning space \cite{ConnorProRSoc1979ballooning}, 
   modes with fine radial structure near the mode-rational
   surface appear with extended tails in the ballooning angle $\lpar$. 
   To see this, note the estimate for the radial wavenumber $\kkr \sim \kky \shat (\thetaz -\lpar)$, valid for $|\thetaz -\lpar| \gg 1$, with
   $\thetaz$ the ballooning parameter that is closely related to the poloidal position on the flux surface
   where the ballooning mode has $\kkr = 0$. Taking $-\pi < \thetaz \leq \pi$,
   we see that scales of $\kkr\gyrde \sim 1$ correspond 
   to $\lpar \sim (\kky\shat\gyrde)^{-1} \gg 1$, whereas scales of $\kkr\gyrdi \sim 1$ correspond to $\lpar \sim 1$.
   We can also view the separation in $\lpar$ between the width of the ballooning mode and the $2\pi$ scale of the toroidal geometry 
   as a separation between two $\kpara$: $\kparageo\sim1/\saffac\rmajor$ 
   associated with the connection length; and $\kparamode \sim \kky\shat\gyrde/\saffac\rmajor$
   associated with the width of the envelope of the mode in ballooning space.

 We choose an ordering for the various frequencies in the problem
 that maximises the physics retained in the model when we take $\kky\gyrde\ll 1$. We choose
\beqn  \kparageo\vthere  \gg \kparamode\vthere \sim \wstar \sim \cfreqee \sim \cfreqei \sim \wfreq, \label{eq:collisionless-ordering}\eeqn
 where $\wstar$ is the typical frequency associated with the drives of
 instability, and $\cfreqee$ and $\cfreqei$ are the electron-electron
 and the electron-ion collision frequencies, respectively.
 We refer to the ordering \refeq{eq:collisionless-ordering}
 as \enquote {collisionless} because 
 electrons make many transits of the device before undergoing a collision.
 Because collisions occur at the same rate at which energy is extracted
 from the equilibrium through $\wstar$, the ordering nonetheless
 admits collisional resonances. 
 The semicollisional limit of the MTM instability
 may be obtained by taking 
 $ \cfreqee / \wstar \gg 1$
 as a subsidiary ordering \cite{hardman_extended_tails}.
 
 The remainder of the paper is structured as follows.
 In section \ref{sec:lineargyrokinetic-electromagnetic}, 
 we give a very brief review of electromagnetic $\dlf$ gyrokinetics.
 Those familiar with gyrokinetic theory should skip
 directly to section \ref{section:reduced-model-equations}, where 
 we examine the physics of electron-driven instabilities
 in the limit of \beqn \pbeta \sim \massr \sim \kky\gyrde \ll 1.  \label{eq:beta-ordering}\eeqn
 This calculation allows us to obtain inner region equations and a kinetic matching condition
 for the inner region that is formulated in the ballooning space representation. 
 We show that, in this limit, $\pbeta$ only enters the model via the matching condition. 
 This allows for the definition of an effective $\pbeta$ parameter that depends on the local magnetic geometry,
 the binormal wavenumber $\kky$, and the ballooning parameter $\thetaz$. 
 In section \ref{section:numerical-evidence},
 we demonstrate that the scalings predicted by the
 orbit-averaged model are satisfied by MTMs
 simulated using the $\delta\! f$ gyrokinetic code \gstwo~\cite{KOTSCHENREUTHER1995CPC}
 in a well-studied Mega-Amp Spherical Tokamak (MAST) discharge
 \cite{Applegate_2007}. This suggests that the model equations
  are likely to describe the modes that arise in the experimental 
 tokamak plasma regimes that are of interest to the community.
 These results motivate a future numerical implementation of 
 the orbit-averaged model, which would require the development
 of a new orbit-averaged gyrokinetic code 
 with realistic geometry and collision operator.
 In section \ref{section:effective-beta-test}, we demonstrate
 that the geometrical dependence of the effective $\pbeta$ 
 parameter accurately predicts the 
 growth rate of the MAST MTM as a function $\thetaz$.
 Finally, in section \ref{section:discussion}, we discuss the possible impacts of our results.
 
 Included with the paper are appendices containing results that are referred 
 to in the main text. In \ref{section:kr-gyrde-sim-1-low-beta},
 we demonstrate how to obtain the orbit-averaged equations.
 In \ref{section:drift-kinetic-collisions}, 
 we give the electron collision operator appearing in the model.
 In \ref{section:kr-dske-sim-1-low-beta},
 we show that the electron-inertial scale $\dske = \gyrde/\sqrt{\pbeta}$
 can be neglected in the leading-order matching between the inner and outer regions.
 Finally, in \ref{section:resolutions}, we provide a description of the numerical resolutions
 used in the simulations presented in this paper.  

    \section{Linearised electromagnetic gyrokinetics}\label{sec:lineargyrokinetic-electromagnetic}
  The $\dlf$ gyrokinetic equations \cite{cattoPP78,Catto_1981} are derived in the magnetised plasma limit
  $\rstars \ll 1$, and are valid for fluctuations satisfying 
  $\wfreq \sim \rstars \cycfs \ll \cycfs$, $\kpara \lscal \sim \kperp \gyrds \sim 1$.
  The distribution function of particle species $\spe$ is decomposed into an 
  equilibrium (Maxwellian) piece $\eqlba$ and a $\rstars$ small perturbation $\dlfs$.
  In the ballooning space representation of linear modes, $\dlfs$ can be written in the form
  \beqn \dlfs(\lpar,\energy,\pitch,\sign, \gyrophase) = \expo{-\imag \kperpvec\cdot\gyrdvecs}\hhs(\lpar,\energy,\pitch,\sign) - \frac{\zeds \charge\ptl(\lpar)}{\temps}\eqlba \label{eq:dlfs-definition} \eeqn 
  where the nonadiabatic response $\hhs$ is independent of the gyrophase $\gyrophase$
  that measures the phase of the cyclotron orbit of the particle,
  and $-\fract{\zeds \charge\ptl(\lpar)\eqlba}{\temps}$ is referred to as the adiabatic,
  or Boltzmann, response. Here, $\kperpvec$ is the perpendicular wave vector,
  $\gyrdvecs = \bu \xp \pvel /\cycfs$ is the vector gyroradius,
  and $\bu = \bvec/\bmag$ is the equilibrium magnetic field direction vector, with $\bvec$ the equilibrium magnetic field.
  The vector   
  $\pvel$ is the particle velocity, $\energy = \ma \vmag^2/2$
  is the particle kinetic energy, with $\vmag = |\pvel|$, $\pitch = \vperp^2/\vmag^2 \bmag$
  is the pitch angle, with $\vperp = |\pvel - \vpar \bu|$ the perpendicular speed and 
  $\vpar = \pvel\cdot\bu$ the parallel velocity, and $\sign = \vpar/|\vpar|$ is the sign of $\vpar$.

 The linear gyrokinetic equation for modes of complex frequency $\wfreq$ is
\beqn \fl \vpar \kpar \drv{\hhs}{\lpar}
 +( \imag \kperpvec \cdot \vms -  \imag \wfreq) \hhs - \cops[\hhs] \nonumber \eeqn \beqn
 = \frac{\zeds \charge\eqlba}{\temps}\left( \imag \wstars - \imag \wfreq \right)
 \left[\bessn{0}\left(\ptl{}-\frac{\vpar}{\ltsp}\apar\right)
 + \frac{\bessn{1}}{\bflrs}\frac{\vperp^2 }{\ltsp \cycfs}\bpar\right]
 , \label{eq:lineargyrokinetic}\eeqn 
  where we have written the equation in terms of the nonadiabatic response $\hhs$ in the coordinates $(\lpar,\energy,\pitch,\sign)$,  
  and where $\vms =  (\bu/\cycfs) \xp \left( \vpar^2 \bu \cdot \nbl \bu + \fract{\vperp^2 \nbl\bmag}{2 \bmag}\right)$
  is the magnetic drift, and
  $\cops[\cdot]$ is the linearised gyrokinetic
  collision operator of the species $\spe$.
  The finite Larmor radius effects enter through $\bflrs = \kperp \vperp / \cycfs$,
  and the Bessel functions of the 0\textsuperscript{th} and 1\textsuperscript{st} kinds
  $\bessn{0}= \bes{\bflrs}$, and $\bessn{1}= \besfirst{\bflrs}$, respectively.
   The fluctuating electrostatic potential $\ptl$  is determined
   by quasineutrality
  \beqn \frac{\ddensi}{\densi} - \frac{\ddense}{\dense} = 
 \left(\frac{\zedi\tempe}{\tempi} + 1 \right)  \frac{\charge\ptl}{\tempe} \label{eq:qn}, \eeqn
 here written in terms of the \textit{nonadiabatic}, fluctuating density
 $\ddenss = \intv{ \bessn{0} \hhs{}}$ for a two-species plasma satisfying
 equilibrium quasineutrality $\zedi \densi = \dense$.
 The magnetic fluctuation $\bpar$ is determined by perpendicular pressure balance, i.e., 
 \beqn \bpar
  = -\frac{4\pi}{\bmag} \sum_{\spe} \dpress \label{eq:bpar}, \eeqn
  where  
 $ \dpress =\denss \temps \intv{\fract{ 2 \bessn{1}\vperp^2 \hhs}{\bflrs\vthers^2 \denss}}$ 
 is the \textit{nonadiabatic} fluctuating perpendicular pressure. 
 The parallel magnetic vector potential $\apar$ is determined by parallel Amp\`{e}re's law 
  \beqn \kperp^2 \apar
= \frac{4\pi}{\ltsp}\jpar \label{eq:apar}, \eeqn
  where the parallel current 
  $ \jpar = \sum_{\spe} \zeds \charge\denss \dupars$,  with the fluctuating parallel-flow velocity 
  $\dupars = \intv{\fract{\bessn{0}\vpar\hhs{}}{\denss}}$. 
  Gradients of equilibrium profiles appear through the drive frequency
 $\wstars = \wstarsn  \left(1 + \etas\left(\fract{\energy}{\temps}- \fract{3}{2}\right)\right)$,
  where the diamagnetic frequency 
  $ \wstarsn =  -(\fract{\ltsp \kkfldl \temps}{ \zeds \charge \denss})\drvt{\denss}{\flxl}$, 
  with $ \etas = \tdrvt{\ln \temps}{\ln \denss}$, $\flxl$ the poloidal flux, 
  $\fldl = \toragl - \saffac(\flxl) \lpar -\nufunc(\flxl,\lpar)$
  the dimensionless binormal coordinate, $\toragl$ the toroidal angle, $\nufunc(\flxl,\lpar)$
  a periodic function of $\lpar$ determined by flux-surface shaping,
  and $\kkfldl$ the dimensionless binormal wave number.
  The magnetic field can be written in terms of $\flxl$ and $\fldl$ using the Clebsch representation,
  i.e., $\bvec = \nbl\fldl\xp\nbl\flxl$.
  In this paper, we focus on axisymmetric magnetic fields that have the form
   $ \bvec = \bcur \nbl \toragl + \nbl \toragl \xp \nbl \flxl$, 
   where $\bcur = \bcur(\flxl)$ is the toroidal current function.
   The safety factor is the average magnetic field line pitch, i.e.,
   $ \saffac(\flxl) = \int^{2 \pi}_0 (\fract{\bvec \cdot \nbl \toragl}{\bvec \cdot \nbl \lpar}) \; d \lpar/2\pi $.
   An explicit expression for $\nufunc$ 
   may be obtained by equating the Clebsch and axisymmetric representations of $\bvec$.
    In terms of  $\flxl$ and $\fldl$, we can write the perpendicular wave vector as
    $\kperpvec = \kkfldl \nbl \fldl + \kkflxl \nbl \flxl$. 
    Note that the contravariant radial wave number has a component that is periodic
    with $\lpar$, and a component that grows secularly with $\lpar$:
    \beqn \kkr = \kperpvec \cdot \nbl \radial = (\thetaz - \lpar)\kkfldl \tdrv{\saffac}{\radial} |\nbl \radial|^2
  - \kkfldl(\saffac \nbl \lpar  + \nbl\nufunc )\cdot \nbl \radial, \eeqn
    where $\thetaz = \kkflxl/\saffacprim \kkfldl$, with $\saffacprim = \tdrvt{\saffac}{\flxl}$, and 
    $\radial = \radial(\flxl)$ is a flux label that has dimensions of length.
    We explicitly define local radial and binormal coordinates with units of length,
      $\radialx = (\flxl - \flxlz)(\dpsidx)^{-1}$ and
      $\binormal =  (\fldl - \fldlz)(\daldy)^{-1} $, respectively, where $(\flxlz,\fldlz)$
      are the coordinates of the field line of interest, and 
      $\dpsidx = \kkxnorm$ and $\daldy=\kkynorm$, with $\bref$ a reference $\bmag$.
    Then, the local binormal wave number $\kky =\kkfldl (\daldy)$, 
    and the local field-aligned radial wavenumber $\kkx =  \kkflxl (\dpsidx) $.

\section{An orbit-averaged gyrokinetic model valid in the limit of $\kky\gyrde \ll 1$}
\label{section:reduced-model-equations}

\begin{figure}
\begin{minipage}{0.49\textwidth}
\begin{center}
\includegraphics[clip, trim=0cm 0cm 0cm 0cm, width=1.0\textwidth]{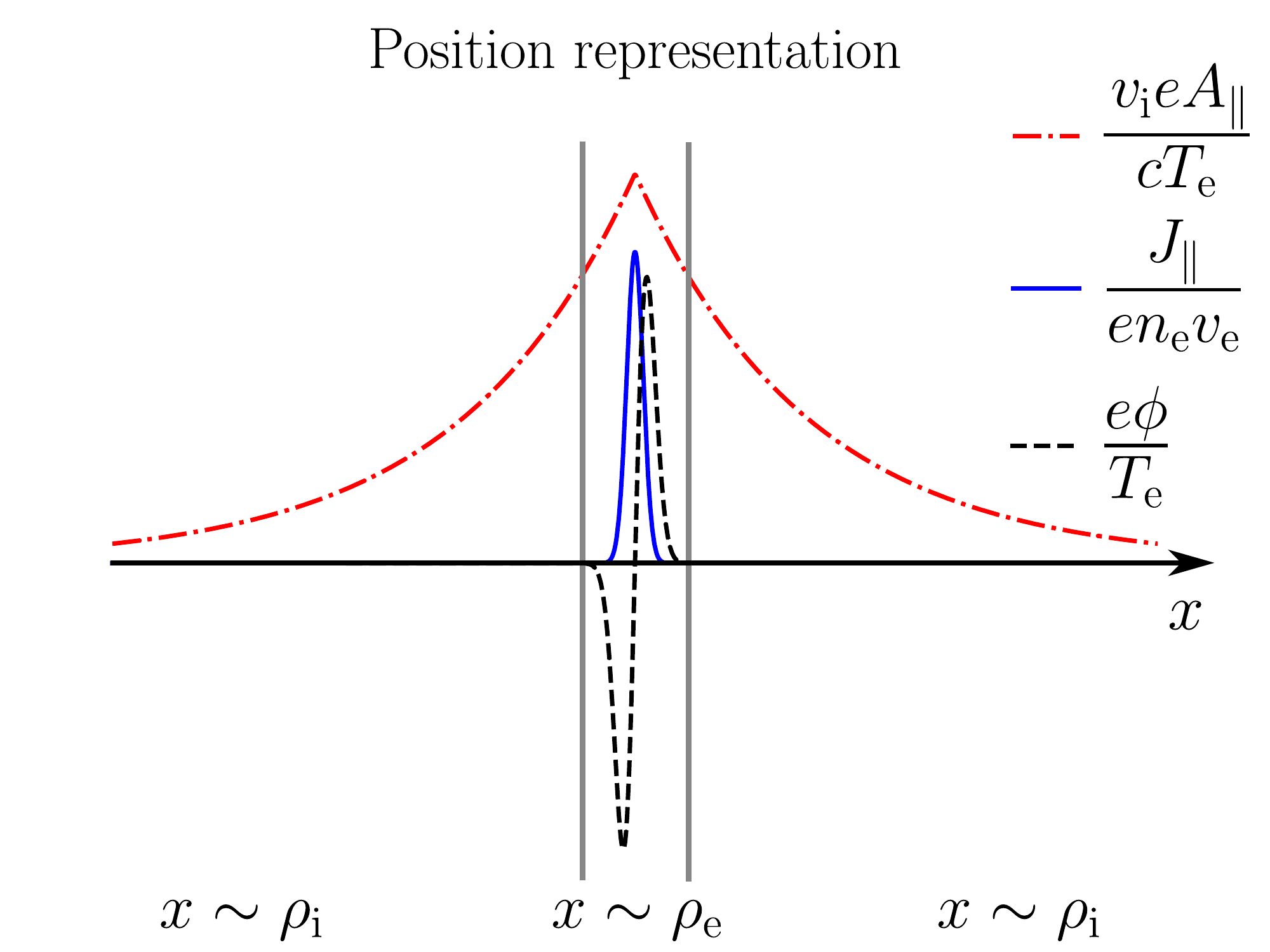}
\end{center}
\end{minipage}
\begin{minipage}{0.49\textwidth}
\begin{center}
\includegraphics[clip, trim=0cm 0cm 0cm 0cm, width=1.0\textwidth]{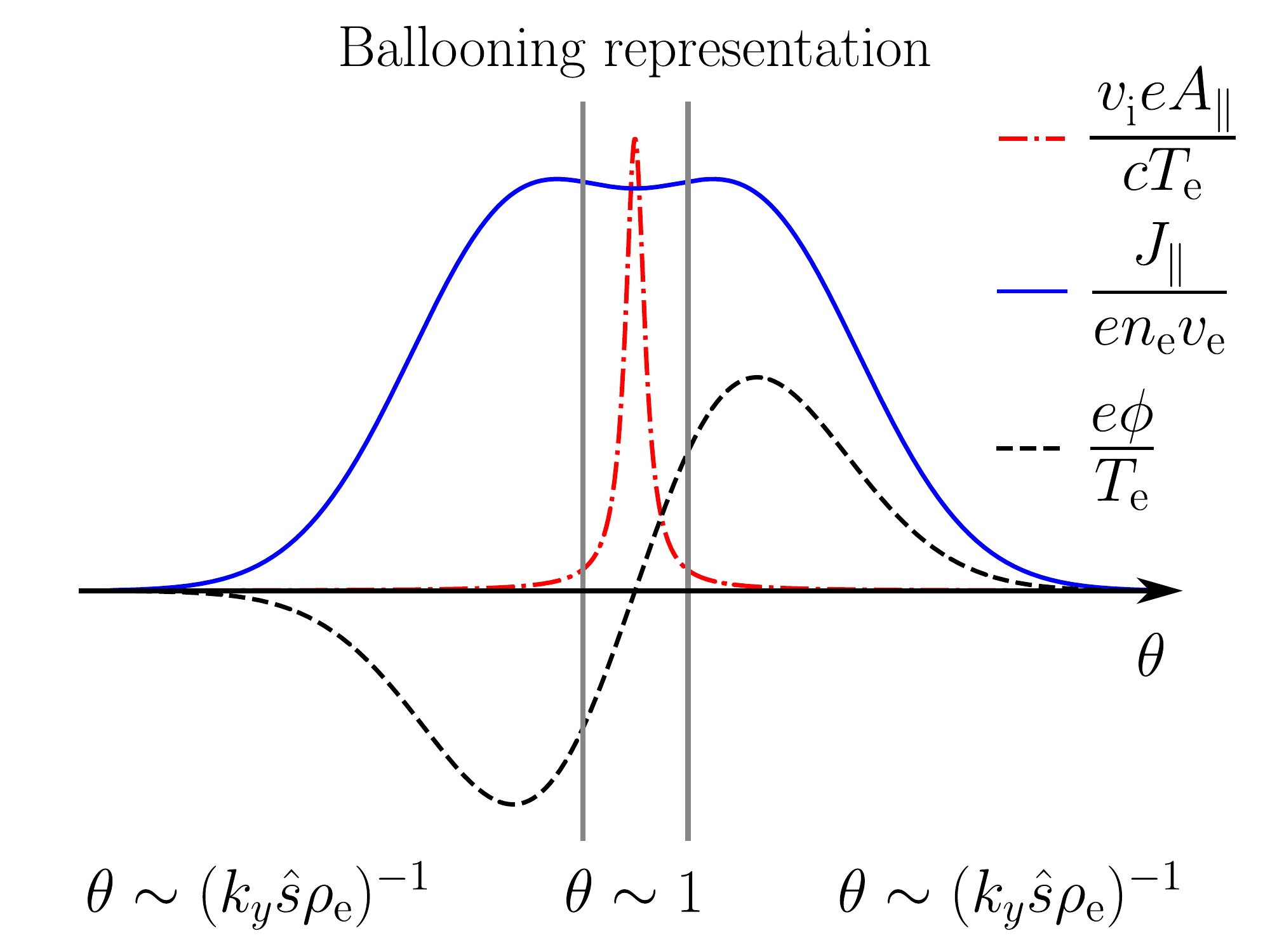}
\end{center}
\end{minipage}
\caption{ An illustration of a $\pbeta \ll 1$ micro-tearing instability.
 We sketch the eigenmodes in two different representations
 of the radial coordinate. 
 In the position representation, the large-scale macro-tearing stable $\apar$
 is driven by an electron current layer at small radial scales close
 to the rational surface ($\xx \sim \gyrde \ll \gyrdi$). In the ballooning representation,
 $\apar$ is well localised to ballooning angles of $\lpar \sim 1$,
 whereas the current $\jpar$ that drives $\apar$ is developed at large
 ballooning angles $\lpar \sim (\kky\shat \gyrde)^{-1} \gg 1$. The amplitude of the electrostatic potental $\ptl$
 is largest in the current layer, with fine radial structures in $\xx$
 and extended tails in $\lpar$.}
 \label{fig:position-ballooning-representations}
\end{figure}

To obtain the reduced model equations whilst retaining general
 axisymmetric toroidal geometry,
 we carry out the microinstability calculation in ballooning space:
 the ballooning space representation allows for simple representation of the operators
 in the gyrokinetic equations.
 A simple ballooning-space picture emerges 
 for long-wavelength, reconnecting modes localised near mode-rational surfaces
 in the limits \refeq{eq:collisionless-ordering} and \refeq{eq:beta-ordering},
 in terms of the electrostatic potential $\ptl$,
 the current $\jpar$ and the parallel vector potential $\apar$.
 A schematic of a reconnecting mode
 is given in figure \ref{fig:position-ballooning-representations}:
 in the more familiar position representation 
 there is an electron current layer near the rational surface and an
 outer region with a long-wavelength (macro-tearing stable) $\apar$.
 In the ballooning representation 
 the electron current layer
 appears as an extended \enquote{tail} at large ballooning angles
 $\lpar \sim (\kky\shat\gyrde)^{-1} \gg 1$.
 Reconnection takes place through an $\apar$ that is localised near $\lpar \sim 1$
 in ballooning angle -- $\apar$ is generated by a current
 that is developed through the electron physics at large $\lpar$.
 
 \subsection{The inner region -- the electron current layer}\label{section:inner-region} 
  We first consider the equations for the electron current layer. The derivation proceeds
  along the same lines as in the electrostatic calculation presented in \cite{hardman_extended_tails}.
  We present the results of the calculation in particularly convenient notation
  for understanding the structure of the model. For completeness, technical details are contained
  in \ref{section:kr-gyrde-sim-1-low-beta}. 
  
  At large ballooning angles $\lpar \sim (\kky\shat\gyrde)^{-1} \gg 1$ the gyrokinetic
  system of equations undergoes several simplifications. 
  First, the nonadiabatic response of ions becomes small, and can be neglected to leading-order.
  This is due to the large ion gyro orbit compared to the scales of interest, i.e., $\kkr \gyrdi \gg 1$.
  Second, two separated scales emerge in the structure of the eigenmodes:
  the width of the mode in the ballooning space $\lpar \sim (\kky\shat\gyrde)^{-1}$
  becomes much larger than the scale of the periodic geometric variation due to the toroidal
  geometry $\lpar \sim 2 \pi \sim 1$. This fact permits a 
  multiscale analysis where we introduce a normalised extended ballooning angle 
  $\lzed = \kky\shat|\gyrderef|\lpar \sim 1$ to describe the envelope of the mode, 
  whilst retaining $\lpar$ to describe poloidal geometric variation.
  Here, $\gyrderef = \vthere/\cycferef$ and $\cycferef = - \charge \bref/\me \ltsp$.
  Third, the leading order balance in the electron gyrokinetic equation becomes
  one between parallel streaming and radial magnetic drifts. This allows us to determine
  that the electron nonadiabatic response $\hhe$ 
  takes the leading-order form \beqn \hhe(\lzed,\lpar) = 
  \HHe (\lzed) \expo{-\imag \gneo \lzed }, \label{eq:hhe}\eeqn
where $\HHe$ is a smooth function of $\lzed$,
 \beqn \gneo (\lpar) = \frac{ \saffac\bcur\wpar }{ \rminor\bmag}, \eeqn and $\wpar = \vpar/\vthere$. 
  The phase $\expo{-\imag  \gneo \lzed }$
  contains all the $\lpar$ variation in $\hhe$, and 
  is due to the finite width of 
  electron banana orbits in the torus.
  By going to first order in the asymptotic expansion in $\massrt\sim\kky|\gyrderef|$, we can find orbit-averaged
  equations that describe the envelope of the eigenmode $\HHe$, and thus
  determine $\hhe$ and the fields to leading-order. 
  
  The inner region equations for $\HHe$ in the low-$\pbeta$
  ordering \refeq{eq:beta-ordering} are as follows. For passing electrons, we have that 
  \beqn \fl \transav{\wpar \gpar} \shat \drv{\HHe}{\lzed} 
    + \imag \left(\transav{\wmaghat} - \wfreqhat\right) \; \HHe
    -\cfreqhat\transav{\copehat\left[\HHe\right]}   
     \nonumber\eeqn \beqn  
     = - \imag \left(\wstarehat - \wfreqhat\right)\eqlbe
     \transav{\expo{\imag \gneo \lzed}\bes{\bflre} \frac{\charge\ptl}{\tempe}}
       \label{eq:electron_firstorder_passing_normalised},\eeqn 
    where we have written the equation in a dimensionless fashion. Note that 
    the potential $\ptl = \ptl(\lpar,\lzed)$. Here, 
    $\gpar = \saffac \rmajorz \kpar $ with $\rmajorz $ a reference major radius,
  \beqn  \transav{\cdot}=
\frac{ \int^{\pi}_{-\pi}  d \lpar \;
\fract{(\cdot)}{\wpar \gpar}  }{\int^{\pi}_{-\pi} d \lpar  \fract{}{\wpar \gpar} }
  , \label{eq:transitav}\eeqn    
 is the transit average, and
    \beqn \wmaghat = -\saffac\rmajorz \tdrv{\flxl}{\rminor}\frac{\bu}{\bmag}\xp\left(
    \wpar^2 \bu\cdot\nbl\bu + \wperp^2 \frac{\nbl \bmag}{2\bmag}\right)\cdot\left(\nbl\fldl+\lpar\nbl\saffac\right)
      - \wpar\gpar \gneo \shat    \eeqn
      is a normalised modified magnetic drift frequency, with $\wperp = \vperp/\vthere$.
      The term $\propto \gneo$ in $\wmaghat$ arises from the phase in the definition \refeq{eq:hhe} of $\HHe$. 
    The normalised frequency is $\wfreqhat = \saffac\rmajorz\wfreq/\kky|\gyrderef|\vthere$, and 
$\wstarehat = -(\saffac \rmajorz/2\lln)(1 + \etae (\energy/\tempe - 3/2))$  
is the drive of instability, with $\lln = -\tdrvt{\rminor}{\ln \dense}$. 
 The parameter       
       $\cfreqhat = \fract{\saffac\rmajorz \cfreqee}{\kky|\gyrderef|\vthere}$
      is a normalised collision frequency, and 
        \beqn \copehat[\HHe] = \frac{1}{\cfreqee}
  \gav{\expo{\imag \left(\gneo + \gcla\right)\lzed}
  \copbothe\Big[\expo{-\imag \left(\gneo + \gcla\right)\lzed}  \HHe\Big]}{}
  \label{eq:collision_operator}\eeqn
  is the collision operator, 
  with $\copbothe\left[\cdot\right]$ the drift-kinetic 
  electron collision operator defined in \ref{section:drift-kinetic-collisions} by
  equation \refeq{eq:DK-collisions},
  $\gav{\cdot}{}$ the gyrophase average at fixed $\energy$ and $\pitch$,
  $\gcla = \saffac (\nbl\flxl \cdot \bu \xp \wvel) \lzed/\rminor \bmag$,
    and $\wvel = \pvel/\vthere$.
    The function $\bflre$ 
    in equation \refeq{eq:electron_firstorder_passing_normalised} 
    takes the leading-order value
    \beqn \bflre = - \saffac |\lzed \nbl\flxl| \wperp / \rminor \bmag. \label{eq:bflre}\eeqn
    
    For trapped electrons, we obtain the bounce-averaged equation
    \beqn \fl   
     \imag \left(\bouncav{\wmaghat} - \wfreqhat\right) \; \HHe  -\cfreqhat\bouncav{\copehat\left[\HHe\right]}   
     = - \imag \left(\wstarehat - \wfreqhat\right)\eqlbe
     \bouncav{\expo{\imag \gneo\lzed}\bes{\bflre} \frac{\charge\ptl}{\tempe}}
       \label{eq:electron_firstorder_trapped_normalised},\eeqn
        where the bounce average is defined by
    \beqn \bouncav{\cdot}= \frac{ \sum_\sign \int^{\lparp}_{\lparm}   d \lpar
    \; \fract{(\cdot)}{|\wpar| \gpar} }{2 \int^{\lparp}_{\lparm} d \lpar  \fract{}{|\wpar| \gpar} }  , \label{eq:bounceav}\eeqn
     and $\lparpm$ are the upper and lower bounce points in $\lpar$,
     respectively, satisfying $\wpar(\lparpm)=0$.
    Note the absence of the parallel streaming term in equation
    \refeq{eq:electron_firstorder_trapped_normalised}:
    trapped particles are unable to pass between magnetic wells,
    forcing them to remain local in $\lzed$.

    Finally, to determine the electrostatic potential $\ptl$ that is necessary to solve equations 
    \refeq{eq:electron_firstorder_passing_normalised} and 
    \refeq{eq:electron_firstorder_trapped_normalised}, we have the quasineutrality relation
    at $\kkr\gyrde \sim 1$:
       \beqn \left(\frac{\zedi \tempe}{\tempi} + 1 \right)\frac{\charge \ptl}{\tempe}
  = - \frac{1}{\dense}\intv{  \expo{-\imag \gneo \lzed} \bes{\bflre}\HHe}  \label{eq:qninner},\eeqn
    where we have used that the
    contribution from the ion nonadiabatic response to quasineutrality is small \cite{hardman_extended_tails}.
    Note that even though $\HHe = \HHe(\lzed)$,
    $\ptl = \ptl(\lpar,\lzed)$ due to the prescence of trapped particles, and due to the 
    $\lpar$ dependence of the functions $\expo{-\imag \gneo \lzed}$ and $\bes{\bflre}$, and the Jacobian of the velocity integral $d^3 \pvel = (2\pi \bmag \energy/\me^2|\vpar|)d \energy d\pitch$.
    The orbit averages in equations  \refeq{eq:electron_firstorder_passing_normalised} and
    \refeq{eq:electron_firstorder_trapped_normalised} 
    can be calculated explicitly
    if the potential is expanded in poloidal harmonics, i.e., if we write
    $ \ptl(\lpar,\lzed) = \sum_{\polm} \ptlhat{\polm}(\lzed)\expo{\imag \polm \lpar}$.
    
   The parallel vector potential $\apar$ 
   and the perturbation to the magnetic field strength $\bpar$ do not appear
  to leading-order in equations 
    \refeq{eq:electron_firstorder_passing_normalised} and 
    \refeq{eq:electron_firstorder_trapped_normalised}
    due to the smallness of $\pbeta$ assumed by the ordering \refeq{eq:beta-ordering}.
    That the contribution from $\bpar$ is small may be verified
    directly by inspecting equation \refeq{eq:bpar}.
    To see that the contribution from $\apar$ may also be neglected in the 
    $\kkr\gyrde \sim 1 $ region, 
    we consider a normalised form of Amp\`{e}re's law,
    \beqn \frac{\left(\kperp\gyrderef\right)^2}{2\pbetae} \frac{\vthere}{\ltsp}\frac{\charge \apar}{\tempe}
     = \frac{\jpar}{\charge \dense \vthere},\label{eq:ampere-nondim}\eeqn
    with $\pbetae = 8\pi \dense \tempe / \bref^2 \sim \pbeta$. 
    The current $\jpar$ is determined through the definition
    \beqn \jpar = - \charge \intv{\vpar \bes{\bflre} \expo{-\imag \gneo \lzed} \HHe},
    \label{eq:current-inner-region}\eeqn
    where we have used that the nonadiabatic ion response is small. 
    Since the even and odd in $\vpar$ parts of $\HHe$ are mixed by
    equation \refeq{eq:electron_firstorder_passing_normalised},
    the current in the $\kkr\gyrde\sim 1$ region must satisfy the ordering
    \beqn \frac{\jpar}{\charge \dense \vthere} \sim \frac{\charge \ptl}{\tempe},
    \label{eq:current-ordering}\eeqn
    and hence, with the ordering \refeq{eq:beta-ordering},
    $\apar$ satisfies
    \beqn \frac{\vthere}{\ltsp}\frac{\charge\apar}{\tempe}
    \sim \pbeta \frac{\charge \ptl}{\tempe} \sim \massr \frac{\charge \ptl}{\tempe} \ll \frac{\charge \ptl}{\tempe}.\eeqn
    For $\pbeta \ll 1$, the $\kkr\gyrde \sim 1$ inner region is electrostatic to leading order.
  
 \subsection{The outer region: determining the matching condition}\label{section:outer-region}
 
    To solve the gyrokinetic layer equations 
    \refeq{eq:electron_firstorder_passing_normalised}, 
    \refeq{eq:electron_firstorder_trapped_normalised}, and \refeq{eq:qninner},
    we must impose a matching condition at $\lzed = 0^\pm$ on the incoming distribution of passing particles.
    This is clear from the cartoon
    given in figure \ref{fig:position-ballooning-representations}
    -- in ballooning space we must connect $\lzed < 0$ to $\lzed > 0$ 
    with an appropriate condition. 
    We now determine the appropriate matching condition in the presence of electromagnetic
    fluctuations, for the limit $\pbeta \sim \massrt \ll 1$, by considering
    the physics of the outer region.
    As the discussion here and in section \ref{section:subsidiary-limits}
    will demonstrate, the $\pbeta$ ordering \refeq{eq:beta-ordering} is special: $\pbeta\sim \massrt$ 
    is the largest $\pbeta$ for which the electrons are able to carry a $\vthere$-sized
    large-scale current; it is also the smallest $\pbeta$ where $\apar$ 
    is large enough to perturb the electrons from the equilibrium magnetic field line, 
    and hence, introduce electromagnetic physics into the mode.

    We begin the calculation by assuming that the passing part of $\hhe$, $\hhepassing$, satisfies
    \beqn \hhepassing \sim \frac{\charge\ptl}{\tempe}\eqlbe \label{eq:large-tail-ordering}\eeqn
    in the outer region of the mode, ruling out the appearance of
    ITG modes and TEMs \cite{hardman_extended_tails}, for which $\hhepassing = \order{\massrt \fract{\charge\ptl\eqlbe}{\tempe}}$.
    The natural radial wavenumber scale of the outer region is $\kkr\gyrdi\sim1$.
    This is a consequence of the ordering $\kky\gyrdi \sim 1$.
    Using the orderings \refeq{eq:collisionless-ordering}, \refeq{eq:beta-ordering},
    and \refeq{eq:current-ordering}, we find that the leading-order
    electron gyrokinetic equation in this region takes the form
    \beqn \vpar \kpar \drv{\hhe}{\lpar}=  \imag \left(\wstare - \wfreq\right)\frac{\charge\eqlbe}{\tempe}
  \frac{\vpar}{\ltsp}\apar 
  , \label{eq:outer-electron-0} \eeqn 
  where we have ordered 
   \beqn \frac{\vthere}{\ltsp}\frac{\charge\apar}{\tempe}
    \sim \massru\frac{\charge \ptl}{\tempe}. \label{eq:apar-ordering-outer}\eeqn
   We justify the ordering \refeq{eq:apar-ordering-outer}
   by noting that in the outer region the leading-order current satisfies
   \beqn \bvec\cdot\nbl\lpar\drv{}{\lpar}\left(\frac{\jpar}{\bmag}\right) = 0.
   \label{eq:current-continuity}\eeqn
   Equation \refeq{eq:current-continuity} is obtained by first integrating equation
   \refeq{eq:outer-electron-0} over all velocities, and second, by noting that
   the ion contribution to $\jpar$ is dominated by
   the electron contribution by a factor of $\vthere/\vtheri \sim \massrut$.
   Equation \refeq{eq:current-continuity} states that $\jpar/\bmag$ is constant across
   the outer region: hence, $\jpar$ satisfies equation \refeq{eq:current-inner-region} and ordering \refeq{eq:current-ordering} 
    everywhere. 
   Using this result in equation \refeq{eq:ampere-nondim} gives 
   the ordering \refeq{eq:apar-ordering-outer}.
    It is worth noting that an outer solution with constant
    $\jpar/\bmag$ has $\apar \propto \bmag/\kperp^2$.
   
   The matching condition for the passing particle distribution function is obtained by integrating
   equation \refeq{eq:outer-electron-0} in $\lpar$. We find that the total
   jump in $\hhe$ across the outer region is given by 
   \beqn \Delta \hhe =  \imag \frac{\left(\wstare - \wfreq\right)}{\ltsp}
   \frac{\charge\eqlbe}{\tempe}\Delta \aptl,\label{eq:hhe-jump}\eeqn
   where 
   \beqn \Delta \aptl = 
   \int^{\infty}_{-\infty} \frac{\apar(\lparprim)}{\kparprim} d \lparprim. \label{eq:DeltaPsi}\eeqn
  The constant $\Delta\aptl$ measures the net deviation of the perturbed field
  line from the equilibrium flux surface. If $|\Delta \aptl| > 0$, then the mode reconnects
  field lines. We can see this by considering the flux surface integral 
  $ \int^\pi_{-\pi} \fract{(\cdot)\; d \lpar }{ \bvec\cdot\nbl\lpar} $
  of the perturbed magnetic field $\delta\!\bvec = -\bu \xp \nbl \apar$ in the position representation.
  We find that the surface-integrated radial magnetic field of a mode at the rational surface is
  \beqn \left.\int^\pi_{-\pi} \frac{\delta\!\bvec \cdot \nbl \radial  }{ \bvec\cdot\nbl\lpar}\; d \lpar\right|_{\xx =0}  = -  \frac{\kky \Delta\aptl}{\bref} ,\label{eq:radial-B}\eeqn
  where $\xx$ is the radial position and $\xx =0$ is the location of the rational surface, and
  where we have converted the poloidal integral of a fourier sum evaluated at $\xx =0$ into an integral over the ballooning coordinate.
  
  We can write the matching condition in a convenient form by combining equations
    \refeq{eq:hhe}, and \refeq{eq:current-continuity}-\refeq{eq:DeltaPsi}:
    we use equation \refeq{eq:hhe} to show that $\Delta \hhe = \left[\HHe(z)\right]^{\lzed=0^+}_{\lzed=0^-}$, and
    we use that $\jpar/\bmag$ is a constant across $\lpar \sim 1$ 
    to evaluate $\Delta\aptl$ explicitly.
    In terms of dimensionless variables,  we find that the jump that should be 
    imposed on the passing part of the electron distribution function $\HHe$ at $\lzed = 0$ is
 \beqn \left[\frac{\HHe(\lzed)}{\eqlbe}\right]^{\lzed = 0^+}_{\lzed = 0^-}
 =  \imag\pbetaeff\left(\wstarehat -\wfreqhat\right)  \jint(0). 
 \label{eq:leading-order-matching-lowbeta}\eeqn
 Here, we have defined a dimensionless current-like quantity 
 \beqn \jint(\lzed) = \jintplus(\lzed) - \jintminus(\lzed), \eeqn
 with \beqn \jintpm(\lzed) = - \int^\infty_0  \int^{1/\bmagmax}_0  
 \HHe(\lzed,\energy,\pitch,\sign= \pm1)\frac{2\pi\energy \bref }{\dense\me^2 \vthere} 
 d \pitch d \energy, \label{eq:jintpm}\eeqn and 
 we have defined an effective plasma $\pbeta$:
\beqn \pbetaeff = \pbetae \frac{2\pi \geofunc(\thetaz)}{\shat \kky|\gyrderef|}. \label{eq:pbetaeff} \eeqn
 The parameter $\pbetaeff$ depends on $(\kky,\thetaz)$ and magnetic geometry, through the factors $\kky|\gyrderef|$ and
\beqn \geofunc(\thetaz) = \frac{\shat}{\pi} \int^\infty_{-\infty} 
\frac{\bmag}{\bref}\frac{\kky^2}{\kperp^2}\frac{d \lpar}{\gpar}.\label{eq:geofunc}\eeqn
 The normalisation of $\geofunc(\thetaz)$ is chosen so that in a magnetic geometry with circular flux surfaces and $\aspect = \rminor/\rmajor \rightarrow 0$, 
 $\geofunc(\thetaz) = 1 + \order{\aspect}$.
 In sections \ref{section:MTM-mass-scan} and \ref{section:ETG-mass-scan}, we use $\jintplus$
 to visualise $\HHe$ for forward-going particles. The matching condition \refeq{eq:leading-order-matching-lowbeta}
 leads us to expect a discontinuity in $\jintplus(\lzed)$ at $\lzed =0$. Nonetheless, $\jint$, which is $\propto \jpar/\bmag$ for $\lzed \ll 1$,
 is continuous across $\lzed =0$ due to the fact that the jump in $\HHe$ is even in $\vpar$.
 
 In the calculation presented above, it might seem that we have neglected to
 consider the impact of the electron inertial scale $\dske$
 that is intermediate to $\gyrde$ and $\gyrdi$: $\gyrde \ll \dske \ll \gyrdi$.
 The scale of $\kkr\dske\sim 1$ is often singled out as interesting because
 of the characterising property that (by equations \refeq{eq:current-ordering}
 and \refeq{eq:ampere-nondim}) $\vthere \apar/ \ltsp \sim \ptl$
 and so both fields enter simultaneously into the source of
 the electron gyrokinetic equation. In \ref{section:kr-dske-sim-1-low-beta},
 we prove that, in the orderings \refeq{eq:collisionless-ordering} and \refeq{eq:beta-ordering}
 considered here,
 the leading order distribution function envelope $\HHe$
 is constant across the $\kkr\dske\sim 1$ region.
 This means that fluctuations at the scale of $\kkr\dske \sim 1$ do not modify the matching condition \refeq{eq:leading-order-matching-lowbeta}.
 
 Note that the current appearing in equation \refeq{eq:leading-order-matching-lowbeta}
 is evaluated purely from $\HHe$. 
 If the microinstability is unstable to leading-order,
 then the microinstability is independent of the ion nonadiabatic response
 to leading-order. If the microinstability happens to be stable at leading order,
 then the nonadiabatic ion response
 must be included in the calculation to find a $\massrt$ small growth (or damping) rate. 
  
  Finally, we comment that the $\pbeta \ll 1$ ballooning-space calculation presented above is  
  related to the usual position-space $\pbeta \ll 1$ matching for tearing modes at large toroidal mode number. 
  In position space, described by the radial coordinate $\xx$ measuring position from the rational surface,
  taking $\pbeta \ll 1$ gives $\apar \propto \expo{-\kky|\xx|}$ in the outer region ($\xx \sim \gyrdi$)
  with $\apar(\xx)$ a constant in the inner region where $\xx \sim \gyrde$. 
  The matching is carried out by ensuring $\Delta^\prime = d \ln \apar(\xx)/d \xx$ is the same at the boundary of the inner and outer regions.
  In ballooning space, we have shown above that taking $\pbeta \ll 1$ gives $\jpar/\bmag$ a constant in the outer region
  ($\lpar \sim 1$), with $\apar(\lpar) \propto \bmag/\kperp^2$. 
  As a result, the inner region ($\lzed = \kky\shat|\gyrderef|\lpar \sim 1$)
  observes a constant $\Delta \aptl$ in the jump in electron distribution function $\hhe$ across the outer region, cf. equation \refeq{eq:hhe-jump}.
  In the ballooning space representation, the constant $\Delta \aptl$ forces fluctuations in the inner region just as 
  a constant $\apar(\xx=0)$ forces the inner region fluctuations in the position space representation. 
  Note that in a sheared slab the ballooning-space outer solution $1/\kperp^{2} = \kky^{-2}(1 + \shat^2 (\lpar-\thetaz)^2)^{-1}$
  is proportional to the fourier transform of the position-space outer solution
  $\expo{-\kky|\xx|}$ -- demonstrating that the \enquote{constant-$\apar$} and \enquote{constant-$\Delta\aptl$}  approximations are equivalent, cf. \cite{Hamed2019_MTM}.
  
 \subsection{Subsidiary limits}\label{section:subsidiary-limits}
 
 The electrostatic limit of the matching condition \refeq {eq:leading-order-matching-lowbeta}
 is simple. If we take $\pbetaeff$ to be small ($\pbetae \ll \kky|\gyrderef|$)
 then \refeq{eq:leading-order-matching-lowbeta}
 takes the leading-order form
 \beqn \left[\frac{\HHe(\lzed)}{\eqlbe}\right]^{\lzed = 0^+}_{\lzed = 0^-} = 0, \label{eq:leading-order-matching-electrostatic}\eeqn
 which is the matching condition for electrostatic,
 electron-driven instabilities localised to the
 mode-rational surface \cite{hardman_extended_tails}. 
 Note that the matching condition \refeq{eq:leading-order-matching-electrostatic} is independent of $\thetaz$, and hence,
 in the electrostatic limit 
 the leading-order mode frequency and growth rate are independent of $\thetaz$ \cite{hardman_extended_tails}.
 
 It is also interesting to consider the subsidiary limit  $ \pbetaeff \gg 1$
  ($ \pbetae\gg \kky|\gyrderef| $). 
 For this limit, it is instructive to write the matching condition
 in the form 
 \beqn \left[\frac{\HHe(\lzed)}{\eqlbe}\right]^{\lzed = 0^+}_{\lzed = 0^-}
 =  \imag\left(\wstarehat -\wfreqhat\right)  \Daptlhat, 
 \label{eq:leading-order-matching-lowbeta-subsidiary}\eeqn
 where  
 \beqn \Daptlhat = \frac{\vthere \kky |\gyrderef|}{\saffac\rmajorz \ltsp}
 \frac{\charge \Delta \aptl}{\tempe} = \pbetaeff \jint(0). \eeqn
 It is clear from equation \refeq{eq:leading-order-matching-lowbeta-subsidiary} that
 taking $\pbetaeff \gg 1$ cannot change the relationship between the sizes of $\HHe$ and $\Daptlhat$.
 This is also apparent in equation \refeq{eq:outer-electron-0}:
 electron parallel streaming must always balance
 the magnetic field line perturbations from $\apar$, once $\apar$ is sufficiently large,
 and so the ordering \refeq{eq:apar-ordering-outer}
 must remain satisfied. Instead of making $\Daptlhat$ large, taking $\pbetaeff \gg 1$ 
 results in a constraint on the current produced by the layer. 
 To preserve the ordering $\HHe/\eqlbe \sim \Daptlhat \sim  \charge \ptl/\tempe$ as $\pbetaeff\rightarrow \infty$,
 we must have that \beqn \jint(0) = \order{\frac{1}{\pbetaeff}\frac{\charge\ptl}{\tempe}}.\label{eq:current-shielding}\eeqn
 Solving the model equations in this limit requires us to 
 regard $\Daptlhat$ as a free parameter in \refeq{eq:leading-order-matching-lowbeta-subsidiary}.
  We fix $\Daptlhat$ relative to $\charge\ptl/\tempe$ 
  by finding the value of $\Daptlhat$ that makes equation \refeq{eq:current-shielding}
  satisfied, whilst simultaneously finding $\wfreqhat$ 
  such that equations  \refeq{eq:electron_firstorder_passing_normalised}, 
    \refeq{eq:electron_firstorder_trapped_normalised}, \refeq{eq:qninner},
    and \refeq{eq:leading-order-matching-lowbeta-subsidiary} are satisfied. 
 Note that the parameter $\pbetaeff$ disappears from the leading-order
 eigenvalue problem when $\pbetaeff \gg 1$. This means that the
 frequency $\wfreqhat$ and eigenmodes become independent
 of $\pbetaeff$ once $\pbetaeff$ is sufficiently large.

 The constraint \refeq{eq:current-shielding} appears in sheared-slab models as the requirement 
 that large-scale electron currents cannot leak out of the rational-surface layer.
 In the extreme case that $\pbetae$ becomes of order unity, then $\pbetaeff\sim \massrut$
 and equation \refeq{eq:current-shielding} suggests that
 the large-scale current produced by the electron layer $\jpar \sim \charge \dense \vtheri (\charge \ptl/\tempe)$.
 This current is comparable to the current carried by the ions in the outer region, breaking the assumption 
 that the ion current can be neglected in the calculation of $\jpar$. 
 Separate papers will consider this $\pbetae\sim 1$ limit, and resolve this issue.

   \subsection{Scaling predictions}\label{eq:scaling-predictions} 
   
 A significant prediction made by the theory is that the leading-order
 dispersion relation should depend on physics parameters in a precise manner.
 By inspecting the form of the model equations \refeq{eq:electron_firstorder_passing_normalised}, 
    \refeq{eq:electron_firstorder_trapped_normalised}, \refeq{eq:qninner},
    and \refeq{eq:leading-order-matching-lowbeta}, we observe that
 \beqn \wfreq = \frac{\vthere}{\saffac\rmajorz}\kky|\gyrderef| \;
 \wfreqhat\left(\pbetaeff,\cfreqhat,\lscal/\lln,\lscal/\lte,\tau,\zedi,\geoparams\right),\label{eq:dispersion-relation}\eeqn
 where we have indicated that $\wfreqhat$ is a function of physics parameters, 
 with $\tau = \tempi/\zedi\tempe$ and $\geoparams$ a vector of geometrical parameters that are needed to describe the local flux surface.
 Note that a dispersion relation of the form \refeq{eq:dispersion-relation} implies that 
 $\wfreq$ depends on $\kky$ only through the overall
 linear pre-factor, $\pbetaeff$, and $\cfreqhat$.
 The definition of $\pbetaeff$, equation \refeq{eq:pbetaeff},
 also provides an interesting explanation for why electron-driven
 electromagnetic modes tend to be observed at the
 longest wavelengths: $\pbetaeff$ is larger for smaller $\kky|\gyrderef|$. 
 We might expect that below a certain critical $\pbetaeff$ the electromagnetic
 microinstabilities are stable, and hence, there is likely to be
 a maximum $\kky$ for instability. In the limit $\pbetaeff \gg 1$,
 the model predicts that $\wfreqhat$ tends to a constant, and hence
 we expect that $\wfreq$ goes linearly with $\kky|\gyrderef|$ for the very smallest $\kky$.
 Whether or not there is a stronger
 cut-off at small $\kky|\gyrderef|$ ($\pbetaeff\gg1$) can only be determined by a 
 theory that handles a $\pbeta \sim 1$ ordering.
 
 The collisionality dependence of $\wfreqhat$ through the parameter 
 $\cfreqhat = \fract{\saffac\rmajorz \cfreqee}{\kky|\gyrderef|\vthere} $
 states that for a fixed collision frequency $\cfreqee$,
 modes of smaller $\kky|\gyrderef|$ are more collisional. 
 Taking $\cfreqhat \rightarrow \infty$ in the model 
 equations \refeq{eq:electron_firstorder_passing_normalised}, 
 \refeq{eq:electron_firstorder_trapped_normalised} and  \refeq{eq:qninner}
 yields the semicollisional limit \cite{hardman_extended_tails}, whereas taking
 $\cfreqhat \rightarrow 0$ yields the collisionless limit. Hence,
 depending on the value of $\saffac\rmajor\cfreqee/\vthere$, the $\kky$ spectrum 
 of a $\kky|\gyrderef|\ll1$ MTM potentially spans modes in the semicollisional limit
 $\cfreqhat \gg 1$ at very low $\kky|\gyrderef|$,
 to modes in the collisionless limit ($\cfreqhat\ll1$) at more moderate $\kky|\gyrderef|$.

    \section{Numerical evidence for the $\pbetae \sim \massrt$ theory}\label{section:numerical-evidence}

   Finding a general analytical solution of the model equations
      \refeq{eq:electron_firstorder_passing_normalised}, 
    \refeq{eq:electron_firstorder_trapped_normalised}, \refeq{eq:qninner}, and 
    \refeq{eq:leading-order-matching-lowbeta}
 is challenging. A numerical implementation is likely to be necessary
 to solve the model in magnetic geometries of interest.
 To test the analytical theory, we compare the intrinsic
 scalings predicted by the theory to numerical results using the  $\dlf$ gyrokinetic code \gstwo~\cite{KOTSCHENREUTHER1995CPC}. 
 We choose to examine microinstabilities in an experimentally relevant local equilibrium
 taken from close to mid-radius in a MAST H-mode plasma (MAST discharge \#6252).
 This discharge was examined previously using gyrokinetic simulations \cite{Applegate_2007}, demonstrating 
 the qualitative behaviour of MTMs in MAST by performing scans in $\cfreqee$, $\lscal/\lte$, $\pbeta$ and $\kky$. 
 We note that the scans in $\cfreqee$ (at a fixed $\kky$) showed that for $\cfreqee$ too large or small, the MTM is stabilised.
 Similar behaviour has been observed in other discharges \cite{MTMWGuttenfelder2011PRL,Moradi_2013}.

 We examine fastest-growing eigenmodes using
 \gstwo~as an initial value solver, including kinetic ions, kinetic electrons and all 
 three fields $\ptl$, $\apar$, and $\bpar$.
 For simplicity, we consider a two-species plasma
consisting of ions and electrons, with $\zedi = 1$ and $\tempi = \tempe$.
 We specify the magnetic geometry using the local Miller parameterisation \cite{Miller98},
 through the following \gstwo~ input parameters:
 the reference major radius $\rmajorz = (\rmajormax + \rmajormin)/2$
   with $\rmajormax$ and $\rmajormin$ the maximum and minimum major
   radial positions on the flux surface, respectively;
 the minor radius $\rminor$ of the flux surface of interest;
 the safety factor $\saffac $; the magnetic shear 
 $\shat = (\saffac/\radial)\tdrvt{\saffac}{\radial} $; 
 $\pbetae = 8\pi \dense \tempe / \bref^2$, with $\bref = \bcur/\rmajgeo$; 
 the normalised pressure gradient $\pbetaprim = (8\pi/\bref^2) \tdrvt{\pres}{\rhominor}$,
 with $\pres$ the total equilibrium pressure, $\rhominor = \rminor/\lscal$,
 and the normalising
 length $\lscal$ the half-diameter of the last closed flux surface;
 the Shafranov shift derivative $\tdrvt{\rmajorz}{\rminor} $; the elongation $\kappa$;
 the elongation derivative $\kappa^\prime = \tdrvt{\kappa}{\rhominor} $; the triangularity $\delta$; and
 the triangularity derivative $\delta^\prime = \tdrvt{\delta}{\rhominor}$.
 The equilibrium profiles are set by the normalised gradients 
 $\lscal/\lts = -  \tdrvt{\ln \temps}{\rhominor}$ and 
 $\lscal/\lln = - \tdrvt{\ln \denss}{\rhominor}$.
 We treat the collision frequencies $\cfreqee = \cfreqei/\zedi$ and $\cfreqii$
 as independent input parameters, defined according to the \gstwo~ convention \cite{barnesPoP09} 
 \beqn \cfreqssprim = 
 \frac{\sqrt{2}\pi \denssprim \zeds^2\zedsprim^2 \charge^4 \ln \Lambda }{\ma^{1/2} \temps^{3/2}}. 
 \label{eq:cfreqss}\eeqn 
 The Coloumb logarithm $\ln \Lambda$ in equation \refeq{eq:cfreqss} has the physical value 
 $\coloumblog \approx 17$ \cite{HazeltineMeiss}.
 The local equilibrium parameters for MAST discharge \#6252 are 
 provided in Table \ref{table:params}. The choice of $\rmajgeo$ corresponds to choosing $\bref = 0.458 {\rm{T}}$,
 the toroidal magnetic field on the magnetic axis in the MAST discharge \cite{Applegate_2007}.
 The numerical resolutions are described in \ref{section:resolutions}. 
 
\begin{table}[htb]
    \begin{center}
\begin{tabular}{cccc} 
\hline
 
      $\rmajorz/\lscal$ & 1.57 & $\rmajgeo/\lscal$ & 1.46  \\ 
 $\rminor/\lscal$ & 0.552 & $\pbetaprim$ & -0.248 \\ 
 $\saffac$ & 1.34   & $\pbetae$ & 0.0494  \\ 
  $\shat$ & 0.538    & $\lscal/\lte$ & 2.70  \\
 $d\rmajorz / d\rminor$ & -0.146 & $\lscal/\lln$ & -0.230  \\
$\kappa $ & 1.47 &  $\lscal/\lti$ & 2.70  \\
$\kappa^\prime $ & 0.0512  & $\lscal \cfreqee /\vtheri$ & 0.303 \\
$\delta $ & 0.162 &$\lscal \cfreqii /\vtheri$ & 0.00884  \\
$\delta^\prime $ & 0.333 &    $\zedeff$ & 1.0   \\
\hline
\end{tabular}
\end{center}
\caption{Local equilibrium and  Miller geometry 
 parameters used to study micro-stability in MAST discharge \#6252.
 Previous micro-stability studies of this discharge utilised a 
 numerical equilibrium, cf. \cite{Applegate_2007}. 
 The parameters are defined in the main text.}
\label{table:params}
    \end{table}
 
 \begin{figure}
\begin{minipage}{0.49\textwidth}
\begin{center}
\includegraphics[clip, trim=0cm 0cm 0cm 0cm, page=1, width=1.0\textwidth]{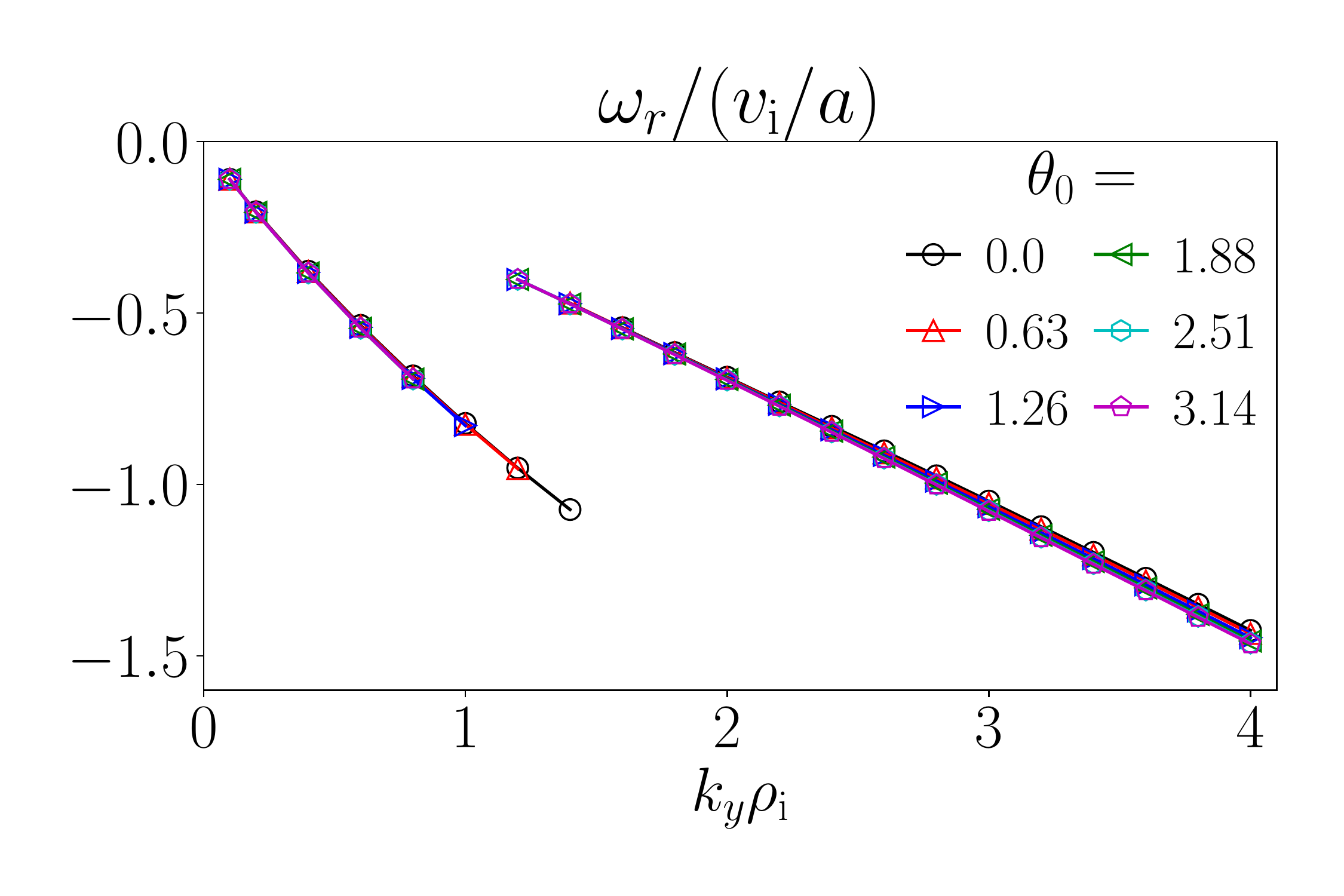}
\end{center}
\end{minipage}
\begin{minipage}{0.49\textwidth}
\begin{center}
\includegraphics[clip, trim=0cm 0cm 0cm 0cm,  page=1, width=1.0\textwidth]{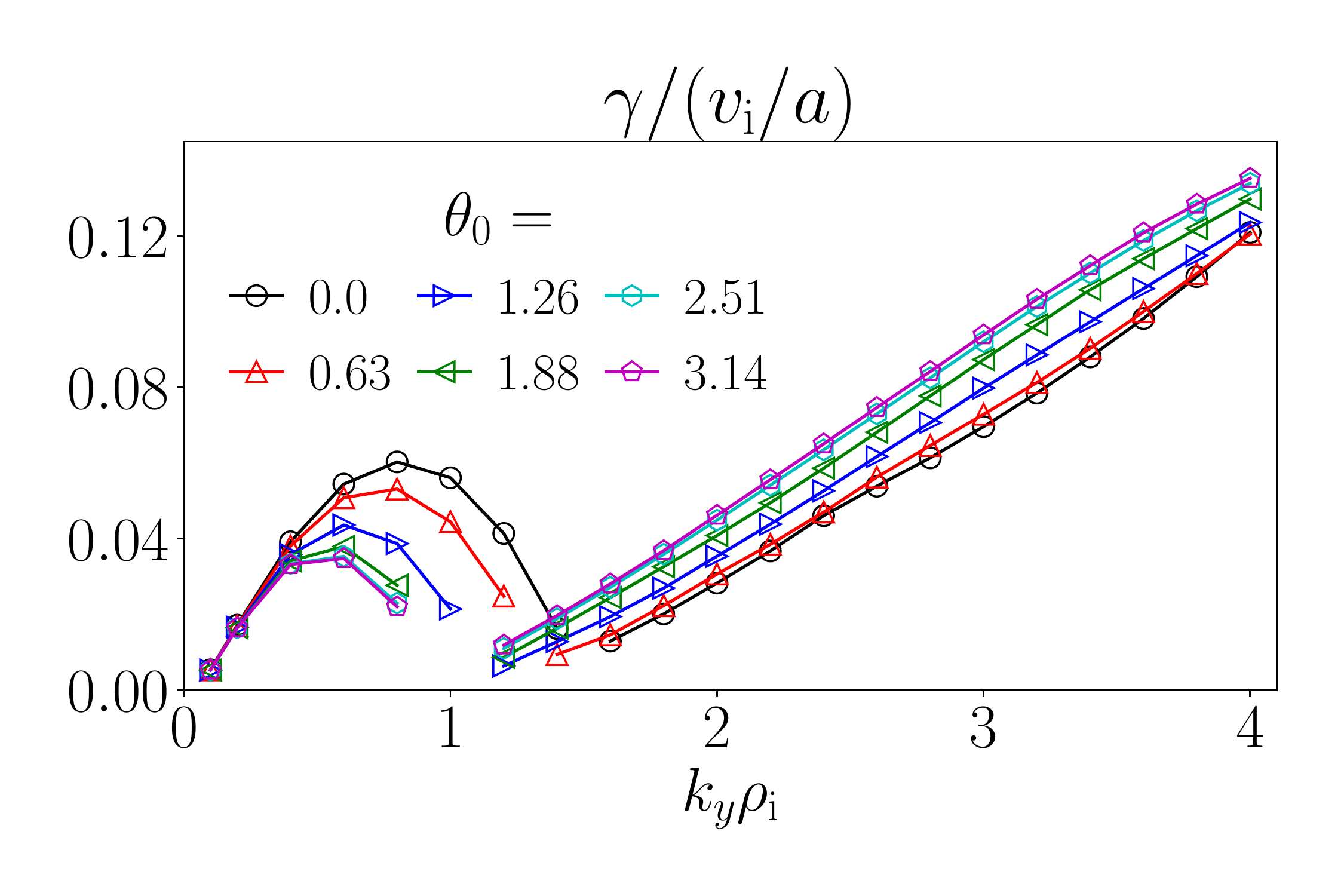}
\end{center}
\end{minipage}
\caption{ The real frequency $\wfreqr$ (left) and growth rate $\growth$ (right) 
 as a function of $(\kky\gyrdi,\thetaz)$ in the MAST discharge \#6252,
 for the physical mass ratio $\massrut = 61$. 
 The modes to the left of the discontinuity in $\wfreqr$ are the MTMs 
 identified in \cite{Applegate_2007}, whereas the modes to the right of the discontinuity
 are electrostatic, electron-driven modes with extended tails; see, e.g., \cite{HallatschekgiantelPRL2005,hardman_extended_tails}.}
 \label{fig:ky-spectrum-MAST-6252}
\end{figure}

\subsection{The wavenumber spectrum}\label{section:ky-spectrum}

\begin{figure}
\begin{minipage}{0.49\textwidth}
\begin{center}
\includegraphics[clip, trim=0cm 0cm 0cm 0cm, page=1, width=1.0\textwidth]{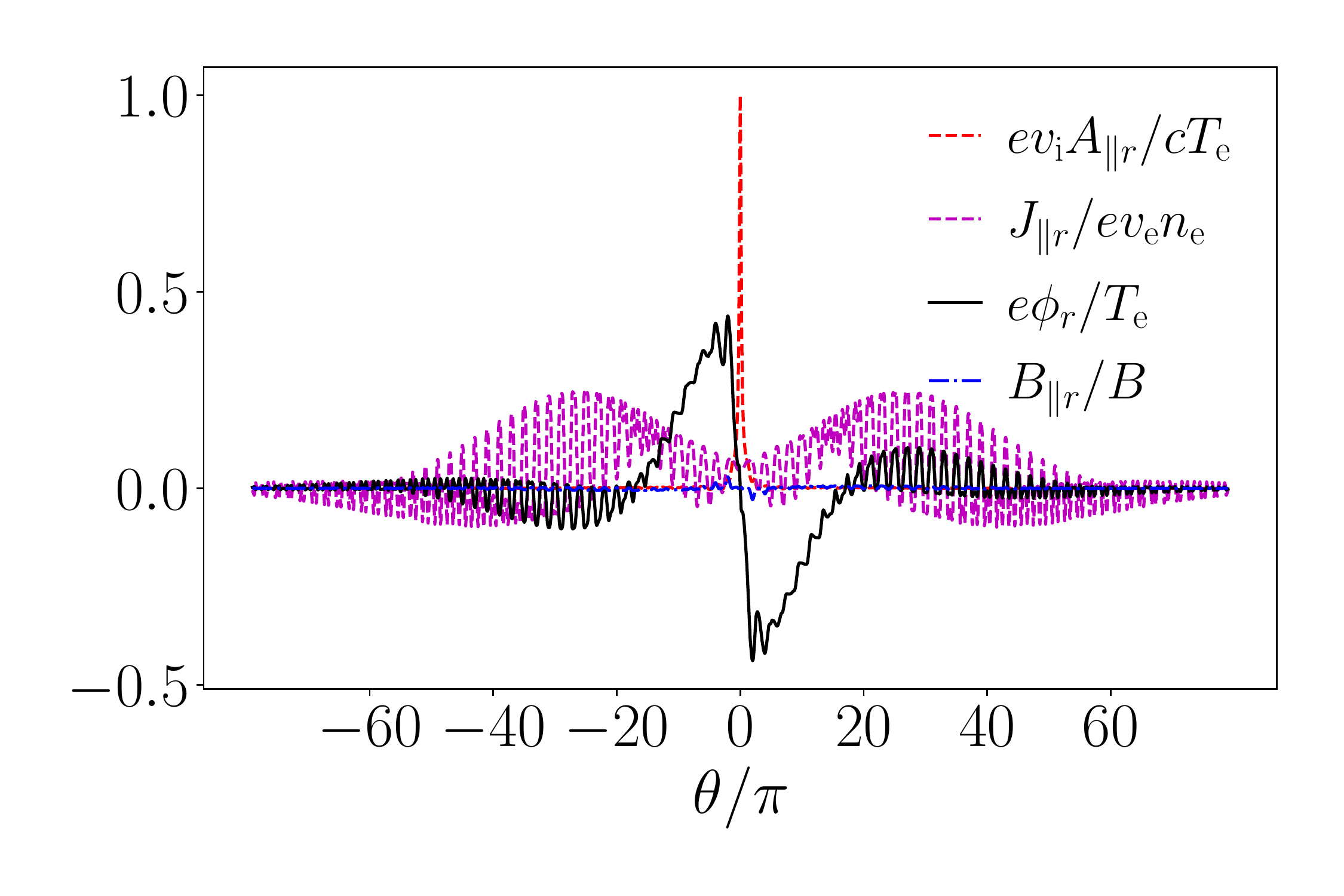}
\end{center}
\end{minipage}
\begin{minipage}{0.49\textwidth}
\begin{center}
\includegraphics[clip, trim=0cm 0cm 0cm 0cm,  page=1, width=1.0\textwidth]{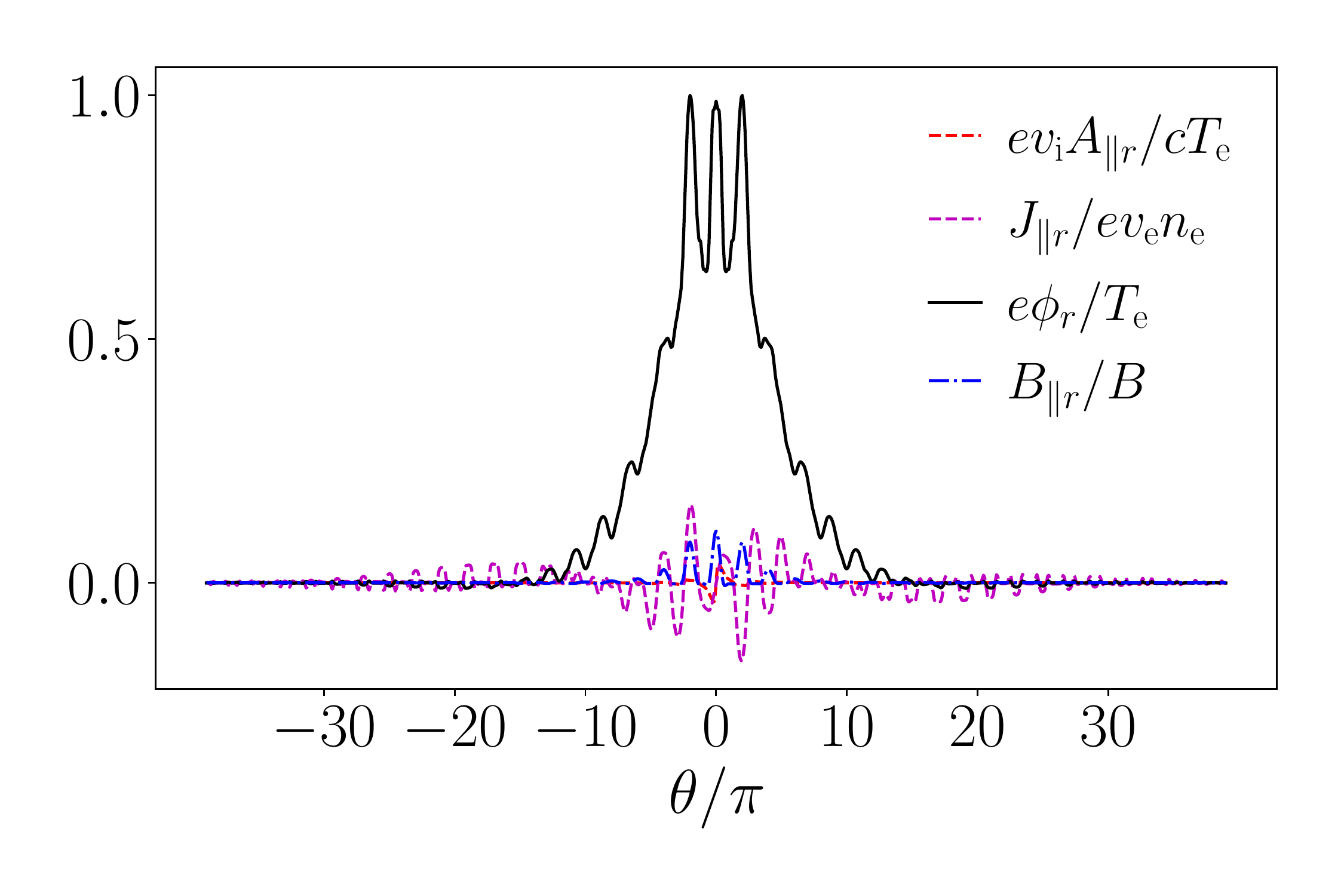}
\end{center}
\end{minipage}
\caption{ Eigenmodes from the wavenumber spectrum shown in figure \ref{fig:ky-spectrum-MAST-6252}.
 We consider the MTM at $(\kky\gyrdi,\thetaz) = (0.8,0.0)$ (left) and 
 the electrostatic mode at $(\kky\gyrdi,\thetaz) = (2.2,0.0)$ (right).
 For the MTM, we normalise the fields with the value of $\charge \vtheri\apar/\ltsp \tempe$ at the $\lpar$ location where $|\apar|$ is maximum. 
 For the electrostatic mode, we normalise the fields with the value of $\charge\ptl/\tempe$ 
 at the $\lpar$ location where $|\ptl|$ is maximum. We plot only the real parts of the eigenmodes, 
 noting that the imaginary parts have similar sizes. Note that 
 $\bpar$ is small for both modes, and that $\jpar$ and $\apar$ are only significant 
 in the MTM.}
 \label{fig:eigenmodes-MAST-6252}
\end{figure}

We first calculate the fastest-growing microinstabilities as a function of $(\kky\gyrdi,\thetaz)$,
 for the equilibrium parameters provided in table \ref{table:params},
 and the physical Deuterium-ion-to-electron mass ratio $\sqrt{\md/\me} = 61$.
 The  real frequency $\wfreqr$ and growth rate $\growth$  
 are plotted in figure \ref{fig:ky-spectrum-MAST-6252}. The discontinuity in $\wfreqr$ 
 indicates a transition between different mode branches: 
 to the left of the discontinuity, we find the MTMs that were 
 previously reported by Applegate et al. \cite{Applegate_2007};
 and to the right of the discontinuity, we find 
 electrostatic, electron-driven modes with extended tails
 in ballooning angle. The presence of the electrostatic modes
 was not reported in \cite{Applegate_2007},
 where only tearing-parity modes at $\thetaz  = 0 $ were considered.
 We show typical eigenmodes in figure \ref{fig:eigenmodes-MAST-6252}:
 we consider a MTM at $(\kky\gyrdi,\thetaz) = (0.8,0.0)$ and an electrostatic mode 
 at $(\kky\gyrdi,\thetaz) = (2.2,0.0)$. 
  Note that both instabilities have potentials $\ptl$ 
 with an extended structure in $\lpar$. The electrostatic mode has small contributions 
 from $\apar$ and $\bpar$, and carries little current.
 The MTM has features that match 
 the expectations given by the theory for electromagnetic modes 
 (cf. figure \ref{fig:position-ballooning-representations}).
 The potential $\ptl$ and $\jpar$ are extended, $\apar$ is significant only for $\lpar \sim 1$,
 and $\bpar$ can be neglected because it is small everywhere.
 The normalisations suggest that the orderings in section \ref{section:reduced-model-equations}
 are valid. We now provide a more precise test of the theory.

\subsection{Testing the $\massrt$ scalings: the MTM}\label{section:MTM-mass-scan}

\begin{figure}
\begin{minipage}{0.49\textwidth}
\begin{center}
\includegraphics[clip, trim=0cm 0cm 0cm 0cm, page=1, width=1.0\textwidth]{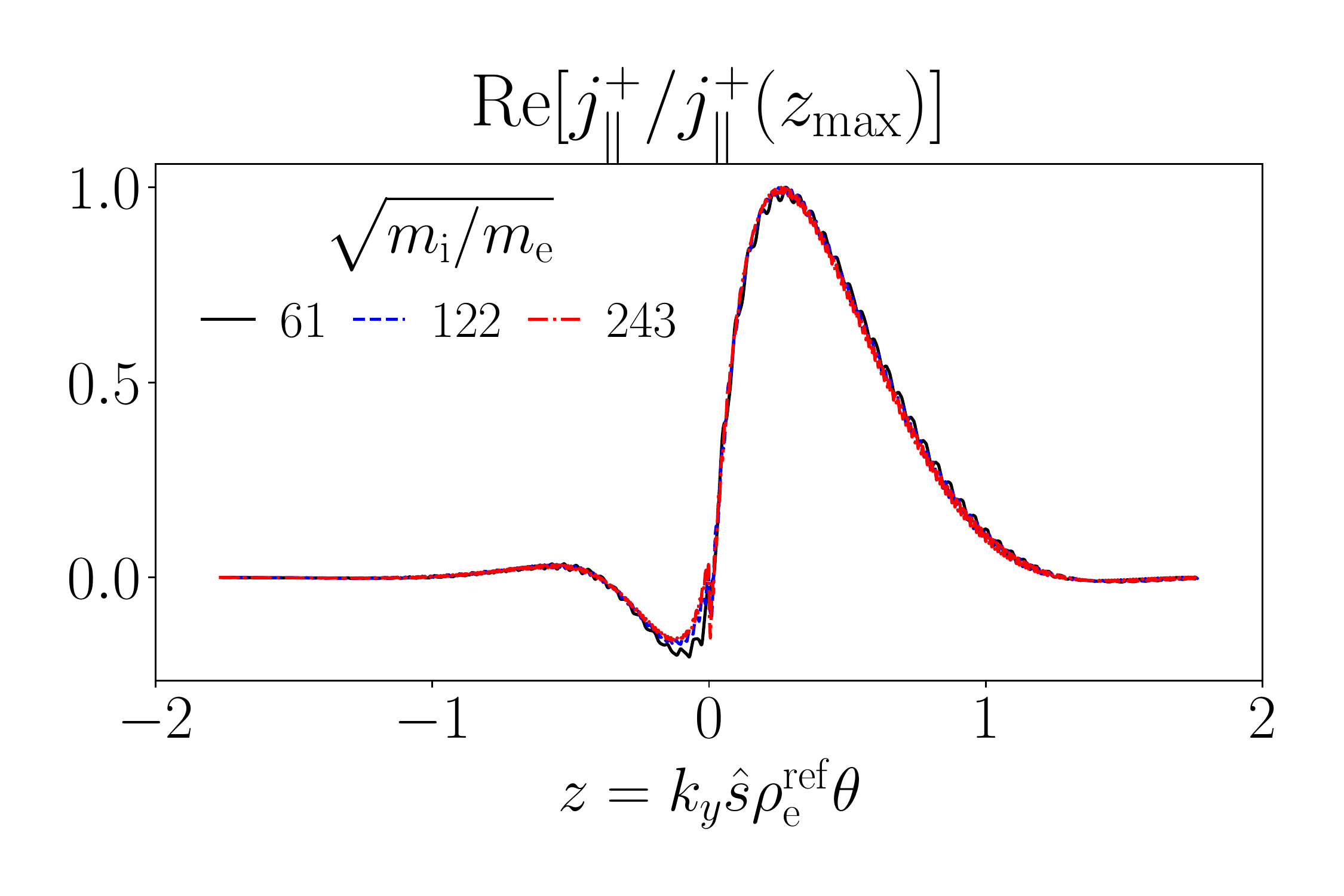}
\end{center}
\end{minipage}
\begin{minipage}{0.49\textwidth}
\begin{center}
\includegraphics[clip, trim=0cm 0cm 0cm 0cm,  page=1, width=1.0\textwidth]{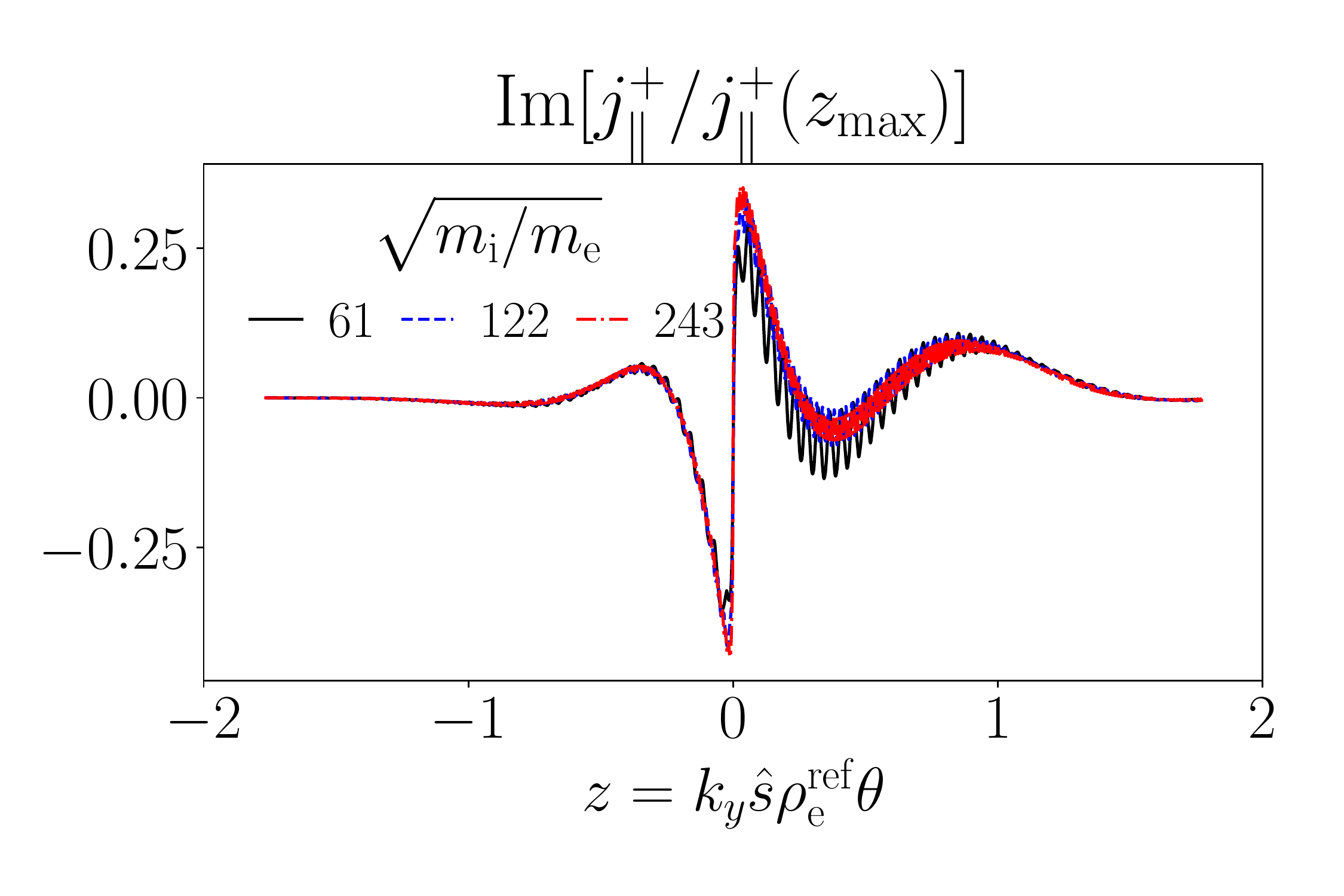}
\end{center}
\end{minipage}
\caption{The current-like quantity $\jintplus$, defined by equation \refeq{eq:jintpm}, 
for the MTM at $(\kky\gyrdi,\thetaz)= (0.8,0.0)$. 
The quantity $\jintplus$ is a measure of the
 forward-going electron distribution function $\HHe$ 
 that is defined by equation \refeq{eq:hhe}.
 Note the smoothness of $\jintplus$ for $|\lzed| \sim 1$,
 and the discontinuity at $\lzed \approx 0$. }
 \label{fig:eigenmodes-electromagnetic-jintplus}
\end{figure}

\begin{figure}
\begin{minipage}{0.49\textwidth}
\begin{center}
\includegraphics[clip, trim=0cm 0cm 0cm 0cm, page=1, width=1.0\textwidth]{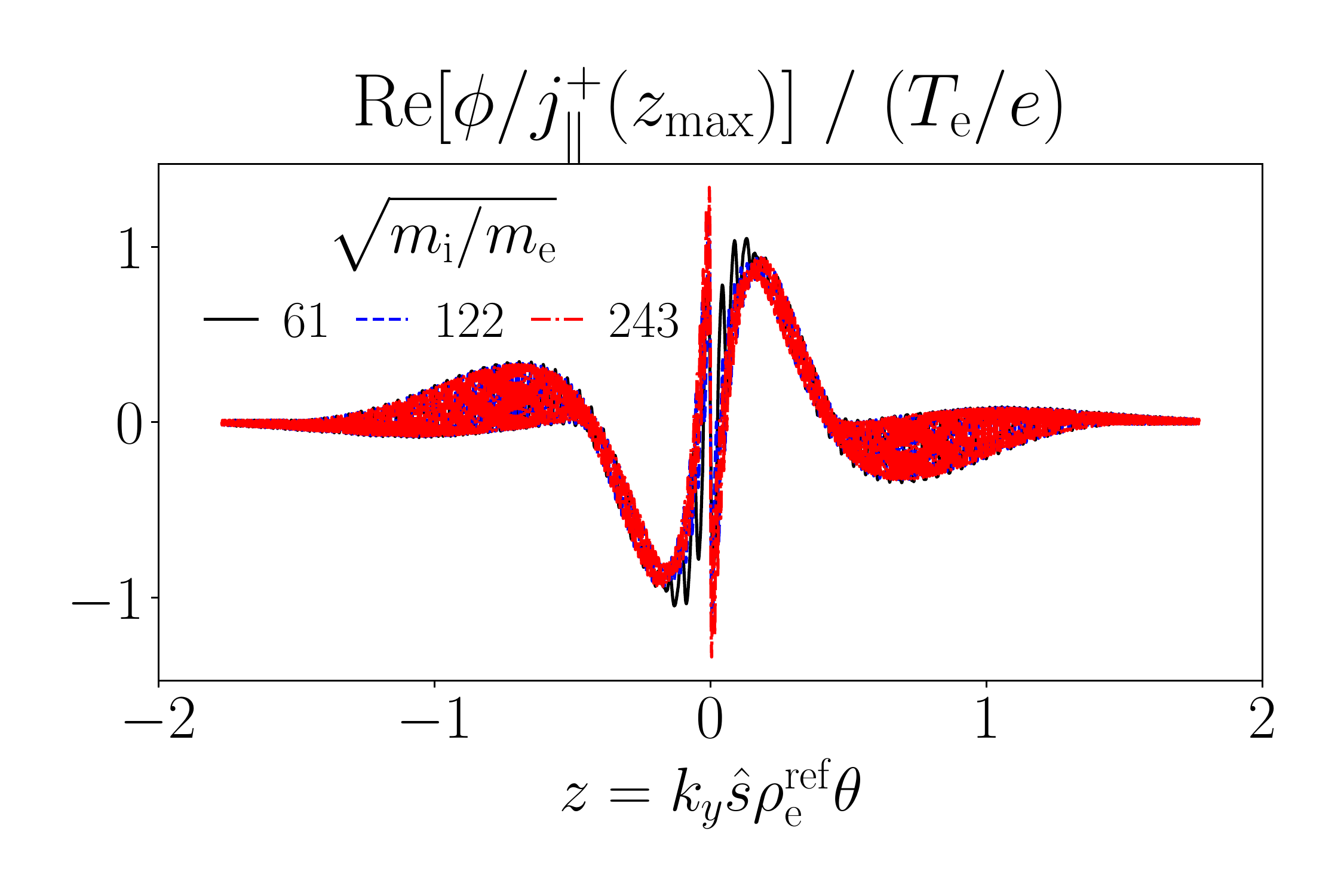}
\end{center}
\end{minipage}
\begin{minipage}{0.49\textwidth}
\begin{center}
\includegraphics[clip, trim=0cm 0cm 0cm 0cm,  page=1, width=1.0\textwidth]{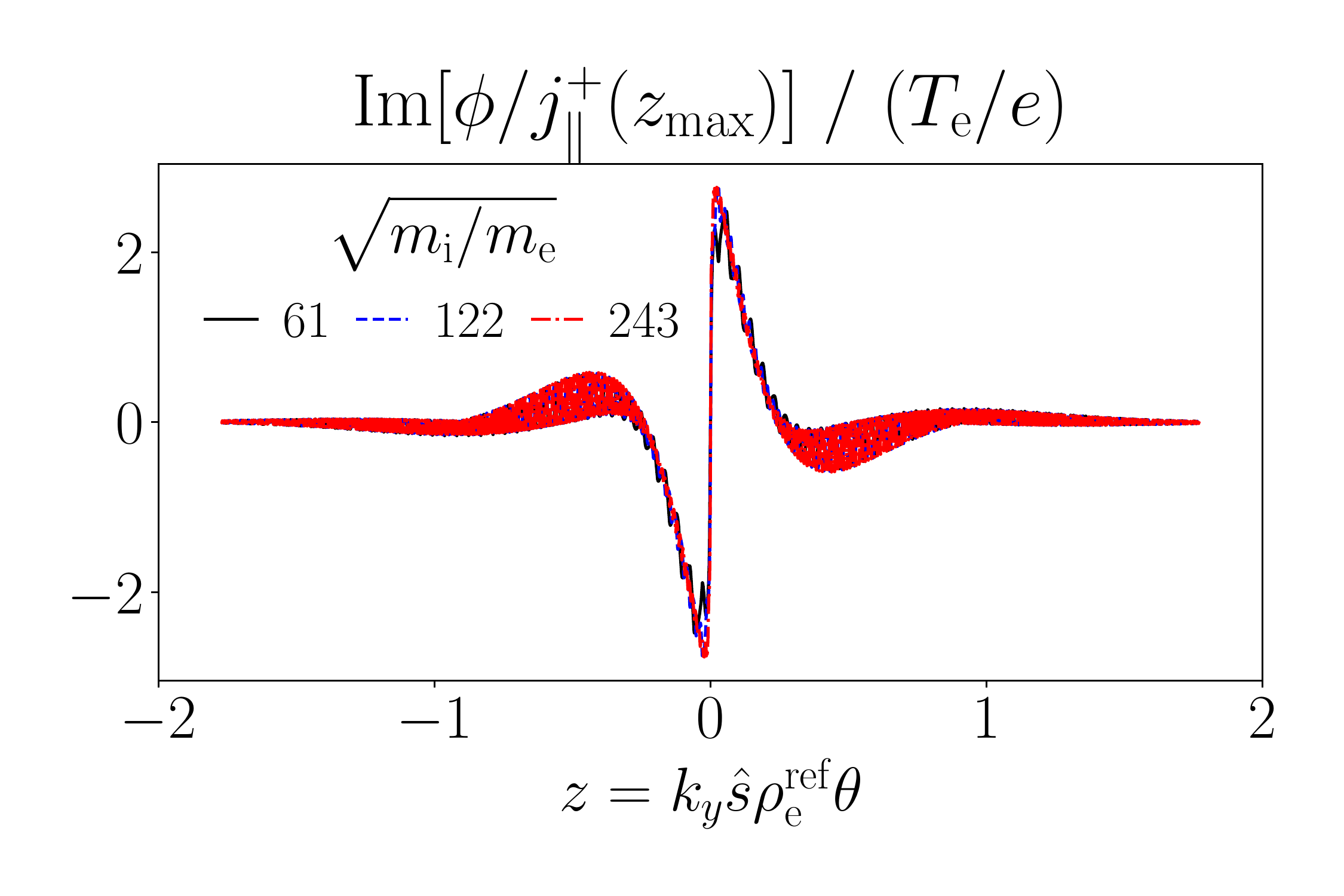}
\end{center}
\end{minipage}
\caption{ The real and imaginary parts of the electrostatic potential $\ptl$ for the
MTM at $(\kky\gyrdi,\thetaz)= (0.8,0.0)$. The potential is normalised by 
$\jintplus$ where $|\jintplus(\lzed)|$ is maximum. That the curves
 overlay for different $\massrut$ indicates that the ordering 
 \refeq{eq:large-tail-ordering} is satisfied, and that the width of the mode
 obeys $\lzed = \kky\shat\gyrderef\lpar \sim1$. }
 \label{fig:eigenmodes-electromagnetic-phi}
\end{figure}

We can demonstrate that both the MTMs and the electrostatic,
 electron-driven modes shown 
 in figure \ref{fig:eigenmodes-MAST-6252} satisfy the correct scalings 
 to be described by the analytical theory. To test the scalings,
 we perform linear simulations at fixed $\kky\gyrdi$ for varying $\massrut$. 
 Crucially, when we double $\massrut$ we halve $\pbetae$
 so that the ordering \refeq{eq:beta-ordering} remains satisfied. 
 This means that we scale $\pbetae \propto \massrt$ so that $\pbetae/\kky\gyrde\sim \pbetaeff$
 is held fixed at a given $\kky\gyrdi$.
 We hold $\lscal \cfreqee/\vtheri$ fixed in the scan, consistent with the ordering 
  \refeq{eq:collisionless-ordering}, and we hold fixed the other geometrical 
  parameters in table \ref{table:params}, including $\pbetaprim$. 
 We plot the eigenmodes as a function of the expansion parameter $\massrut$, and demonstrate 
 that the width of the eigenmode scales like $\kky\shat\gyrderef\lpar \sim 1$,
 and that the orderings for the fields are satisfied.

\begin{figure}
\begin{minipage}{0.49\textwidth}
\begin{center}
\includegraphics[clip, trim=0cm 0cm 0cm 0cm, page=1, width=1.0\textwidth]{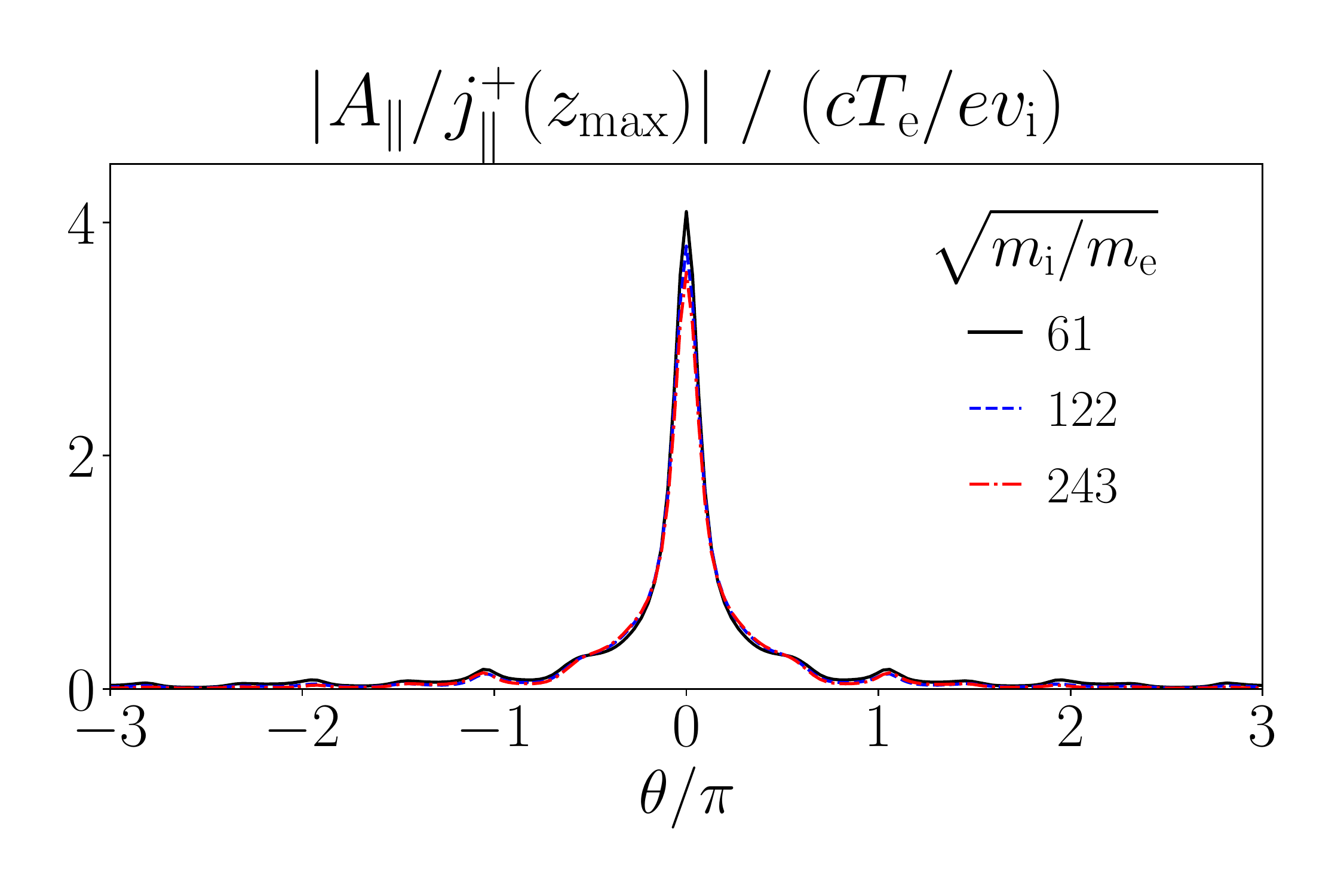}
\end{center}
\end{minipage}
\begin{minipage}{0.49\textwidth}
\begin{center}
\includegraphics[clip, trim=0cm 0cm 0cm 0cm,  page=1, width=1.0\textwidth]{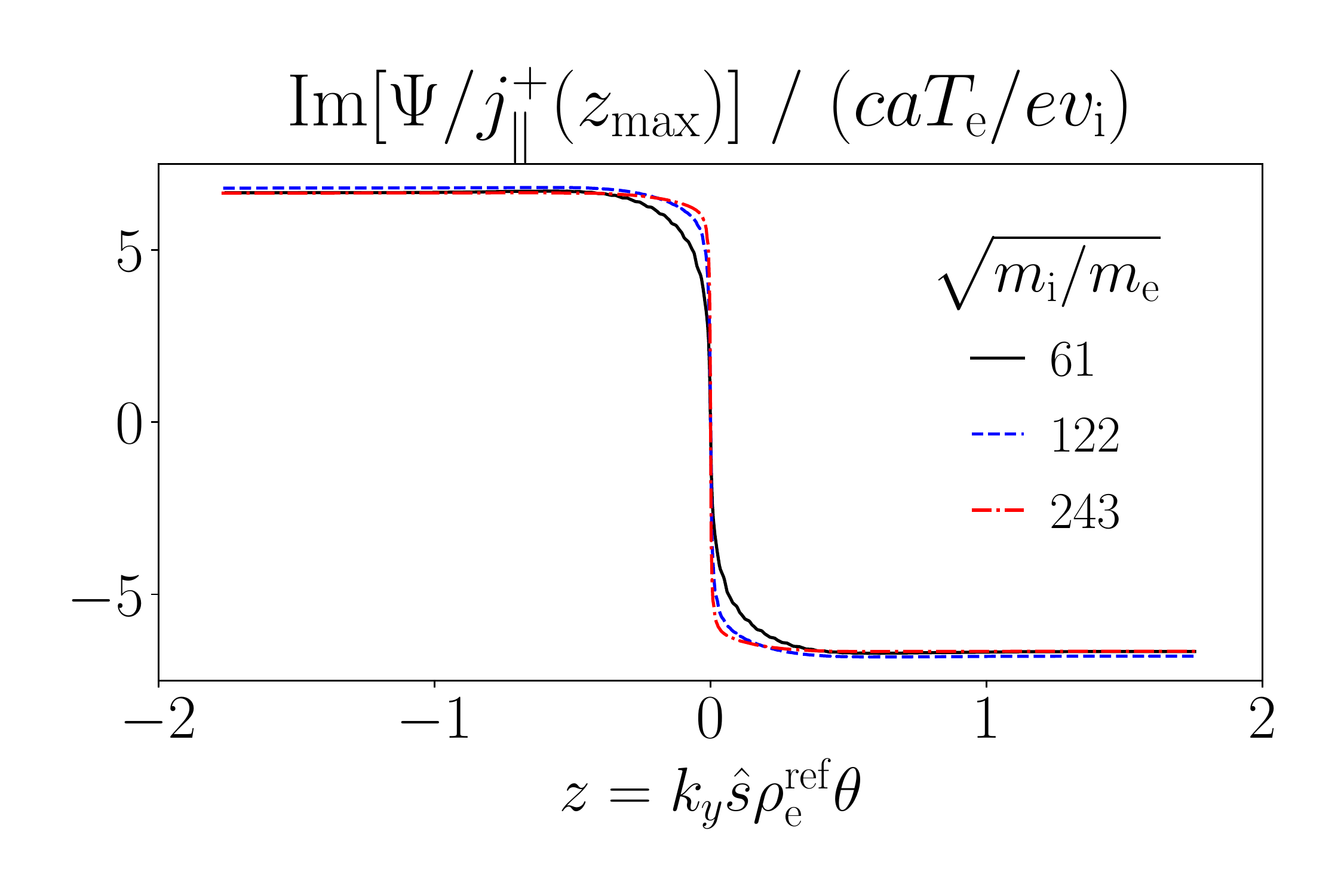}
\end{center}
\end{minipage}
\caption{  Plots of the parallel vector potential $\apar$
 and the auxiliary field $\aptl$
 (defined by equation \refeq{eq:aptl-definition}), for the
    MTM at $(\kky\gyrdi,\thetaz)= (0.8,0.0)$. 
     We  normalise the eigenmodes by 
$\jintplus$ where $|\jintplus(\lzed)|$ is maximum, 
cf. figure \ref{fig:eigenmodes-electromagnetic-jintplus}.
    Note that $\apar$
    is well localised to $\lpar \sim 1$ as $ \massrt \rightarrow 0$,
    and that the ordering \refeq{eq:apar-ordering-outer} is satisfied.
    The field $\aptl$ resembles a Heavyside function as
    $\massrt \rightarrow 0$, consistent with the asymptotic theory,
    where the jump $\Delta\aptl$ is entirely due to the $\apar$ at $\lpar \sim 1$, cf equation \refeq{eq:DeltaPsi}.
    The curves of $\aptl$ overlay for $\lzed\sim1$, consistent with
    the ordering \refeq{eq:apar-ordering-outer} and the matching condition \refeq{eq:hhe-jump}.}
 \label{fig:eigenmodes-electromagnetic-apar-Psi}
\end{figure}
 
 We first consider the MTM at $(\kky\gyrdi,\thetaz) = (0.8,0.0)$. In figure \ref{fig:eigenmodes-electromagnetic-jintplus},
 we visualise the forward-going part of the electron
 distribution function $\HHe$ using the real and imaginary parts of 
 the current-like quantity $\jintplus$ 
 (defined by equation \refeq{eq:jintpm}, where $\HHe$ is calculated from
 the \gstwo~$\hhe$ eigenmode via equation \refeq{eq:hhe}). We note that $\jintplus$ is a
 smooth function of $\lzed$ for $|\lzed| \sim 1$, 
 verifying a key element of the theory, equation \refeq{eq:hhe}. 
 We also note that there is a discontinuity near $\lzed = 0$
 -- this is consistent with the matching condition \refeq{eq:leading-order-matching-lowbeta}. 
 We must choose an appropriate
 normalisation for the eigenmodes of $\ptl$ and $\apar$:
 we select $\jintplus(\lzedmax)$, the value of $\jintplus(\lzed)$
 at the maximum value of $|\jintplus(\lzed)|$. 
 In figure \ref{fig:eigenmodes-electromagnetic-phi}, we plot the real and imaginary parts of $\ptl$. 
 The eigenmodes show oscillatory behaviour due to the 
 $2\pi$ periodic variation of the magnetic geometry. The envelope
 of the eigenmodes overlay well, 
 confirming that the electron nonadiabatic
 response sources the leading component of $\ptl$, 
 consistent with ordering \refeq{eq:large-tail-ordering} and equation \refeq{eq:qninner}. 
 Finally, in figure \ref{fig:eigenmodes-electromagnetic-apar-Psi},
 we plot $\apar$ in the region $\lpar \sim 1$, 
 and the auxiliary field \beqn \aptl(\lpar) = 
 \int^{\lpar}_{-\infty} \frac{\apar(\lparprim) }{\kparprim}d\lparprim
 - \frac{1}{2}\int^{\infty}_{-\infty}\frac{\apar(\lparprim) }{\kparprim}d\lparprim.
 \label{eq:aptl-definition}\eeqn
 Note that
 $\Delta \aptl = \aptl(\infty) - \aptl(-\infty)$.
 We see that $\apar$ is well localised to $\lpar \sim 1$, the ordering \refeq{eq:apar-ordering-outer}
 is satisfied, and that the variation in $\aptl$ is increasingly localised
 to $\lzed \approx 0$ as $\massrt \rightarrow 0$, consistent with a constant
 $\Delta \aptl$ observed by the inner region in the matching.

\begin{figure}
\begin{minipage}{0.49\textwidth}
\begin{center}
\includegraphics[clip, trim=0cm 0cm 0cm 0cm, page=1, width=1.0\textwidth]{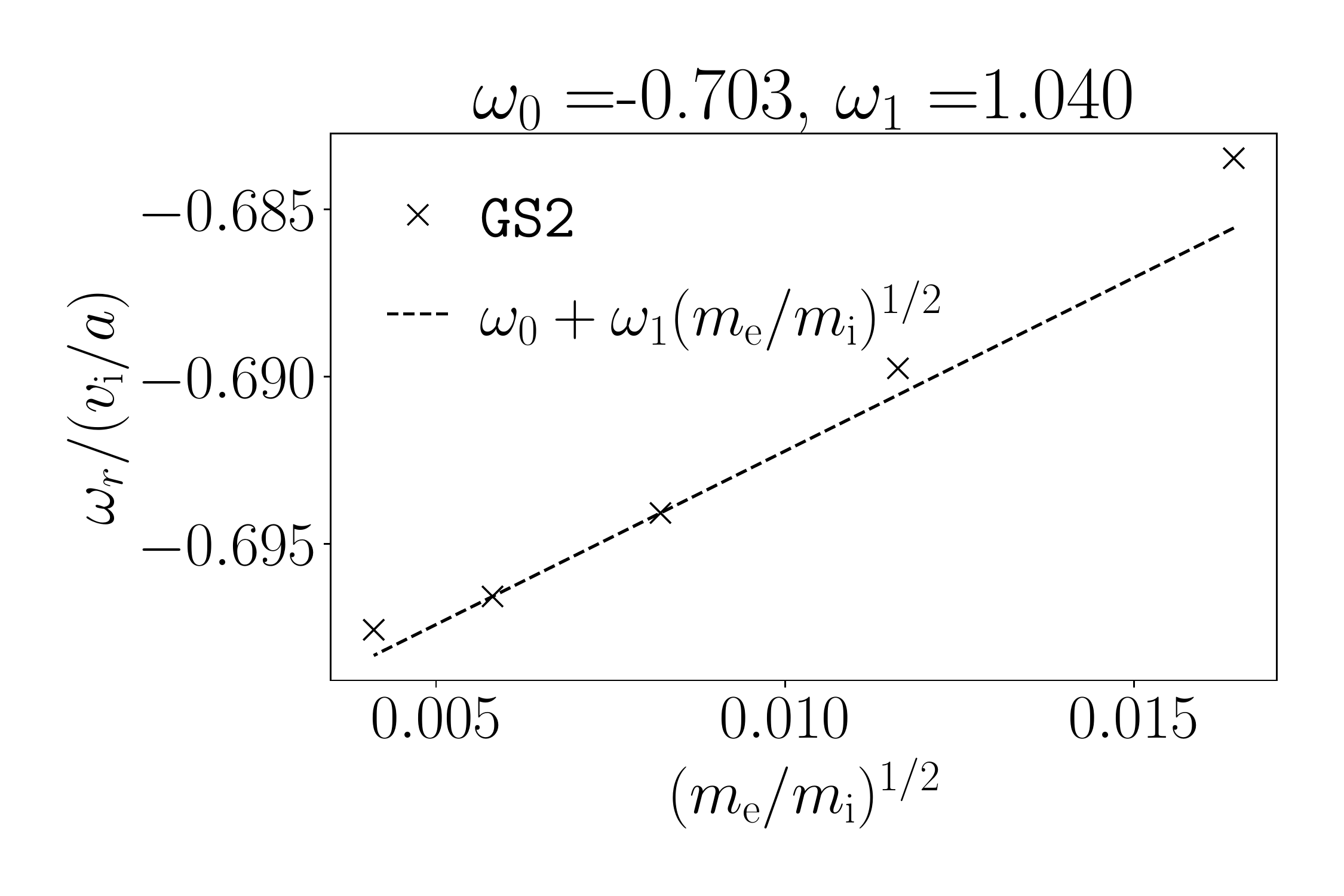}
\end{center}
\end{minipage}
\begin{minipage}{0.49\textwidth}
\begin{center}
\includegraphics[clip, trim=0cm 0cm 0cm 0cm,  page=1, width=1.0\textwidth]{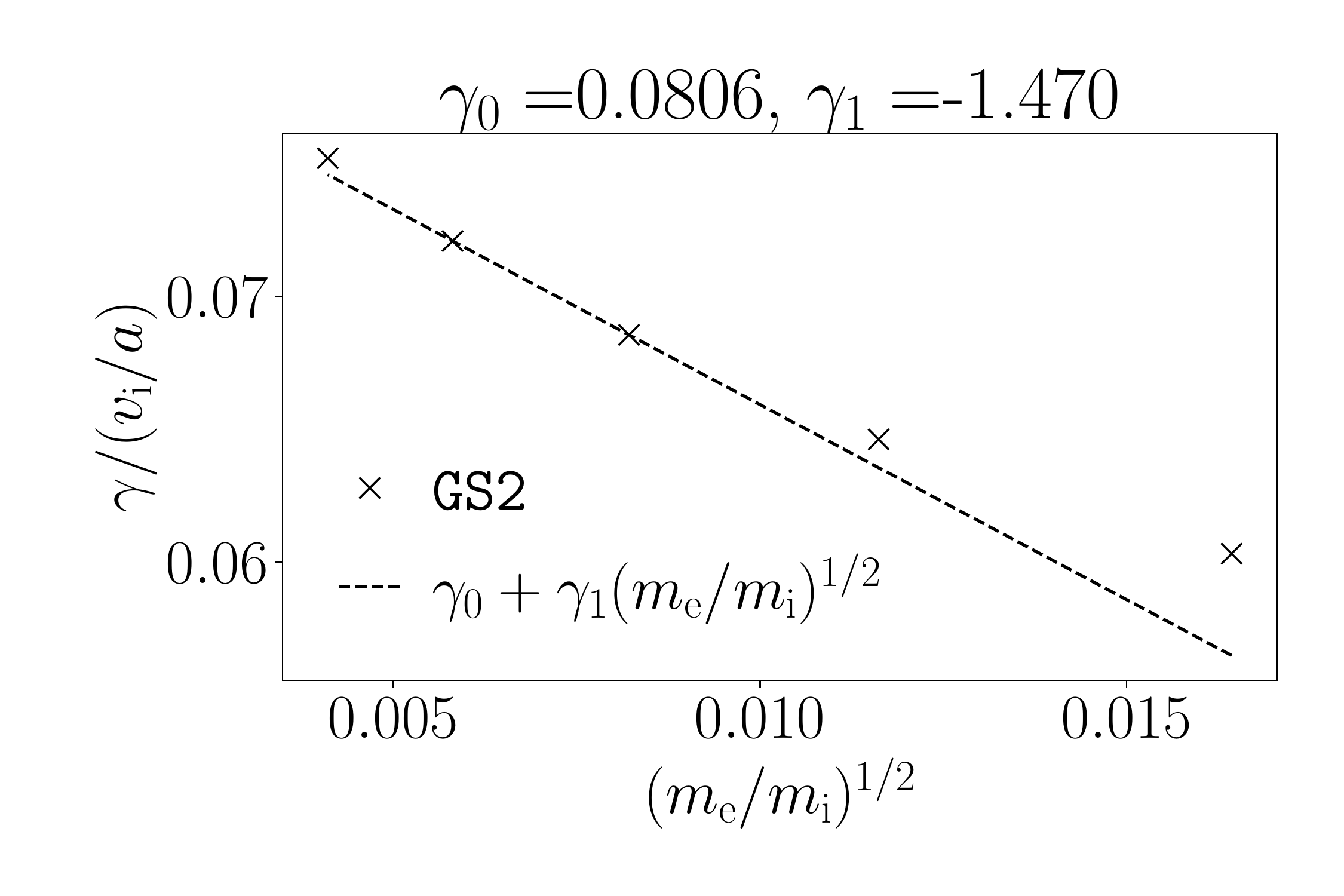}
\end{center}
\end{minipage}
\caption{ The real frequency $\wfreqr$ and growth rate $\growth$ of the MTM
 at $(\kky\gyrdi,\thetaz)= (0.8,0.0)$ as a function of the expansion parameter $\massrt$. 
 Linear fits are provided to show that the variation in $\wfreqr$ 
 and $\growth$ are consistent with the $\massrt$ expansion: the fit coefficients
 are of order unity. The growth rate increases for reducing $\massrt$, possibly 
 consistent with calculations of the impact of higher-order corrections
 from ions on tearing modes \cite{Cowley_1986PoFtearing}.}
 \label{fig:frequencies-electromagnetic}
\end{figure}

\begin{figure}
\begin{minipage}{0.49\textwidth}
\begin{center}
\includegraphics[clip, trim=0cm 0cm 0cm 0cm, page=1, width=1.0\textwidth]{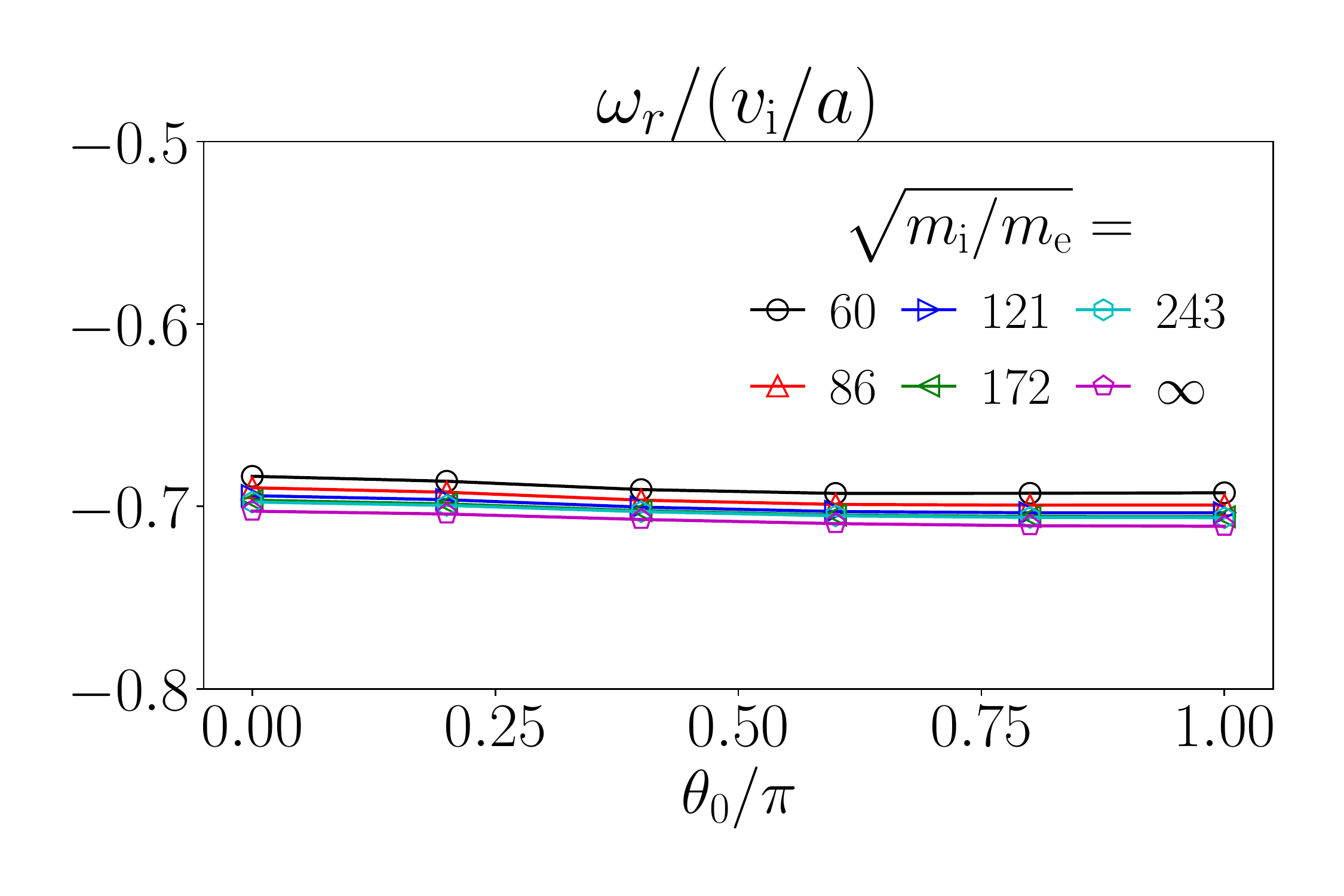}
\end{center}
\end{minipage}
\begin{minipage}{0.49\textwidth}
\begin{center}
\includegraphics[clip, trim=0cm 0cm 0cm 0cm,  page=1, width=1.0\textwidth]{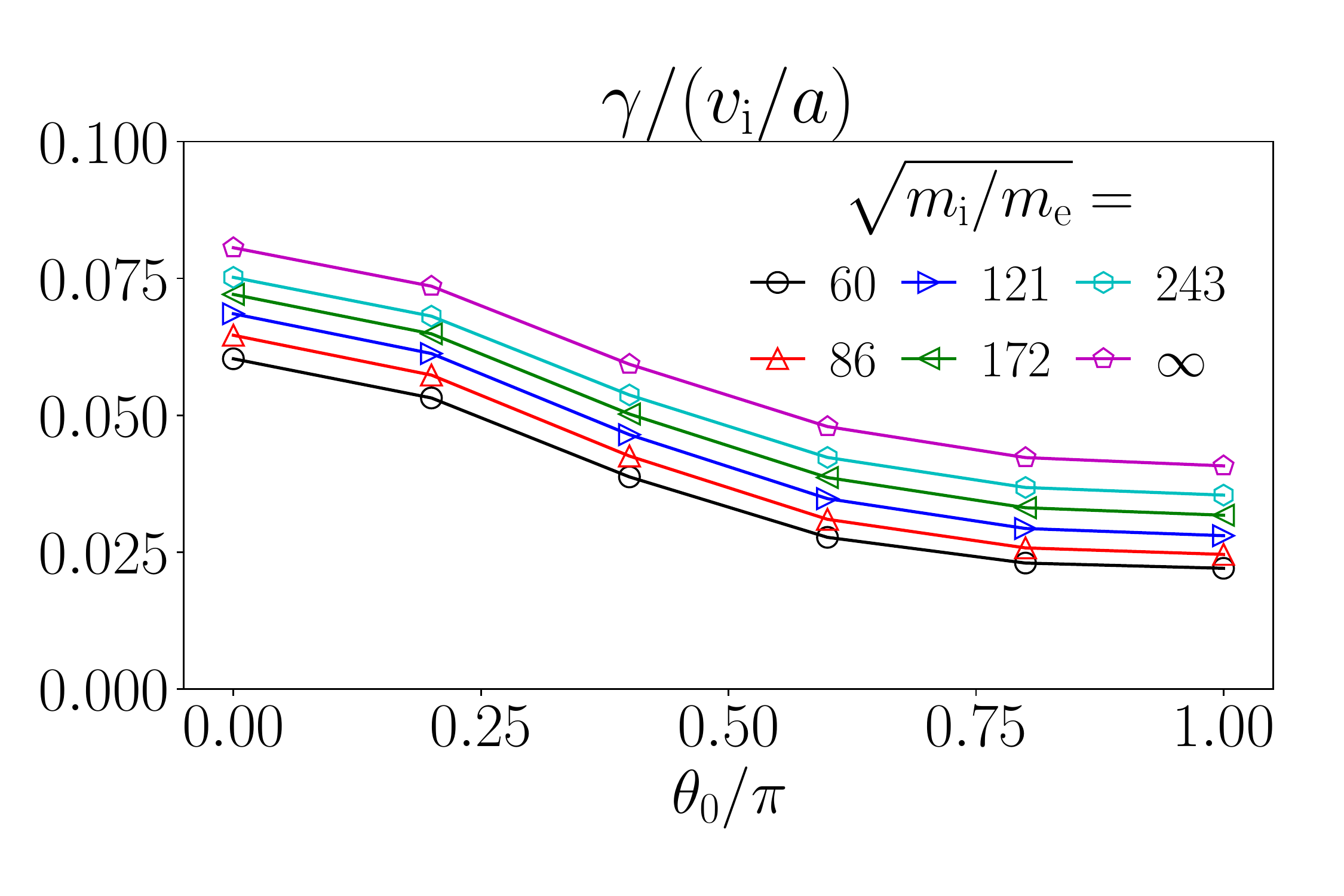}
\end{center}
\end{minipage}
\caption{ The real frequency $\wfreqr$ and growth rate $\growth$ of the MTMs
 at $\kky\gyrdi = 0.8$ as a function of $\thetaz$ and the expansion parameter $\massrut$.
 The points in the curve labelled by 
 $\massrut = \infty$ are computed from linear fits,
 see figure \ref{fig:frequencies-electromagnetic}. In section \ref{section:effective-beta-test},
 we show that 
 the variation of $\growth$ with $\thetaz$ is
 explained by the $\thetaz$ dependence of $\pbetaeff$. }
 \label{fig:frequencies-theta0-electromagnetic}
\end{figure}

 In addition to considering the eigenmodes, it is important to observe the impact
 of varying $\massrt$ on the complex frequency $\wfreq = \wfreqr + \imag \growth$.
 We should anticipate a leading order component of $\wfreq$ that is independent of $\massrt$,
  described by the leading-order asymptotic theory, and a correction that is small in $\massrt$. 
 In figure \ref{fig:frequencies-electromagnetic}, we plot the real frequency $\wfreqr$ 
 and the growth rate $\growth$ as a function of $\massrt$.
 We provide linear fits to indicate size of the variation with $\massrt$.
 Although the leading-order growth rate $\growth_0$ is numerically small 
 compared to $\wfreq_0$, the fit coefficients are of order unity,
 consistent with the theory.  We note that $\growth$ increases as $\massrt \rightarrow 0$, which
 would indicate that the $\massrt$ small corrections to 
 $\growth_0$ are stabilising. Although we have not computed these higher-order corrections here,
 we note that this result is consistent with the
 calculation for the higher-order impact of ions on tearing modes in sheared-slab geometry,
 see \cite{Cowley_1986PoFtearing}.
 Finally, in figure \ref{fig:frequencies-theta0-electromagnetic}, we plot $\wfreqr$ and $\growth$
 as a function of $\thetaz$ at fixed $\kky\gyrdi = 0.8$.
 We observe that $\growth$ has variation with $\thetaz$
 even as $\massrt \rightarrow 0$. In section \ref{section:effective-beta-test}, we prove that
 this variation in $\thetaz$ is explained by the $\thetaz$
 dependence of $\pbetaeff$ through the geometrical factor $\geofunc(\thetaz)$. 

\subsection{Testing the $\massrt$ scalings: the electrostatic mode}\label{section:ETG-mass-scan}

\begin{figure}
\begin{minipage}{0.49\textwidth}
\begin{center}
\includegraphics[clip, trim=0cm 0cm 0cm 0cm, page=1, width=1.0\textwidth]{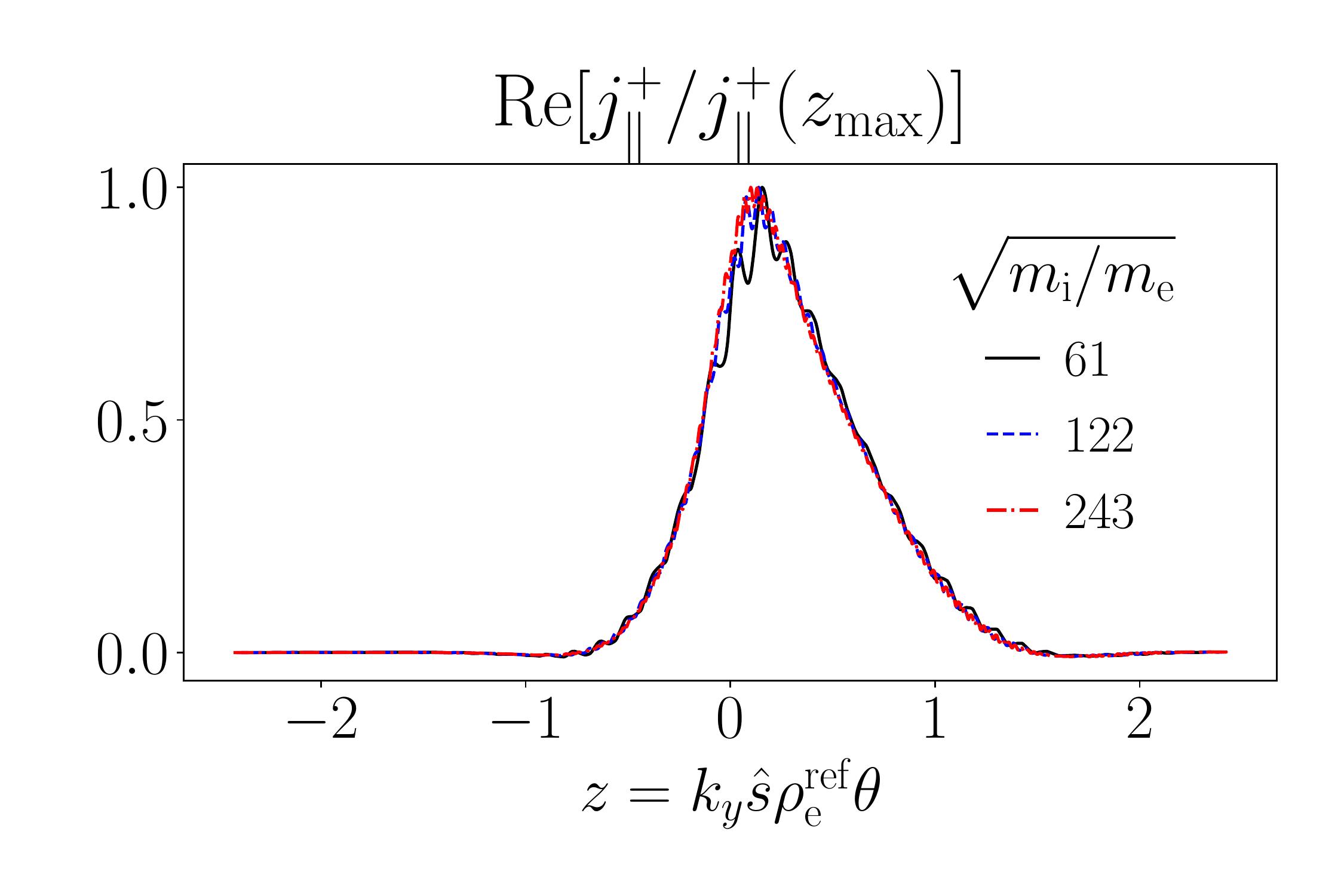}
\end{center}
\end{minipage}
\begin{minipage}{0.49\textwidth}
\begin{center}
\includegraphics[clip, trim=0cm 0cm 0cm 0cm,  page=1, width=1.0\textwidth]{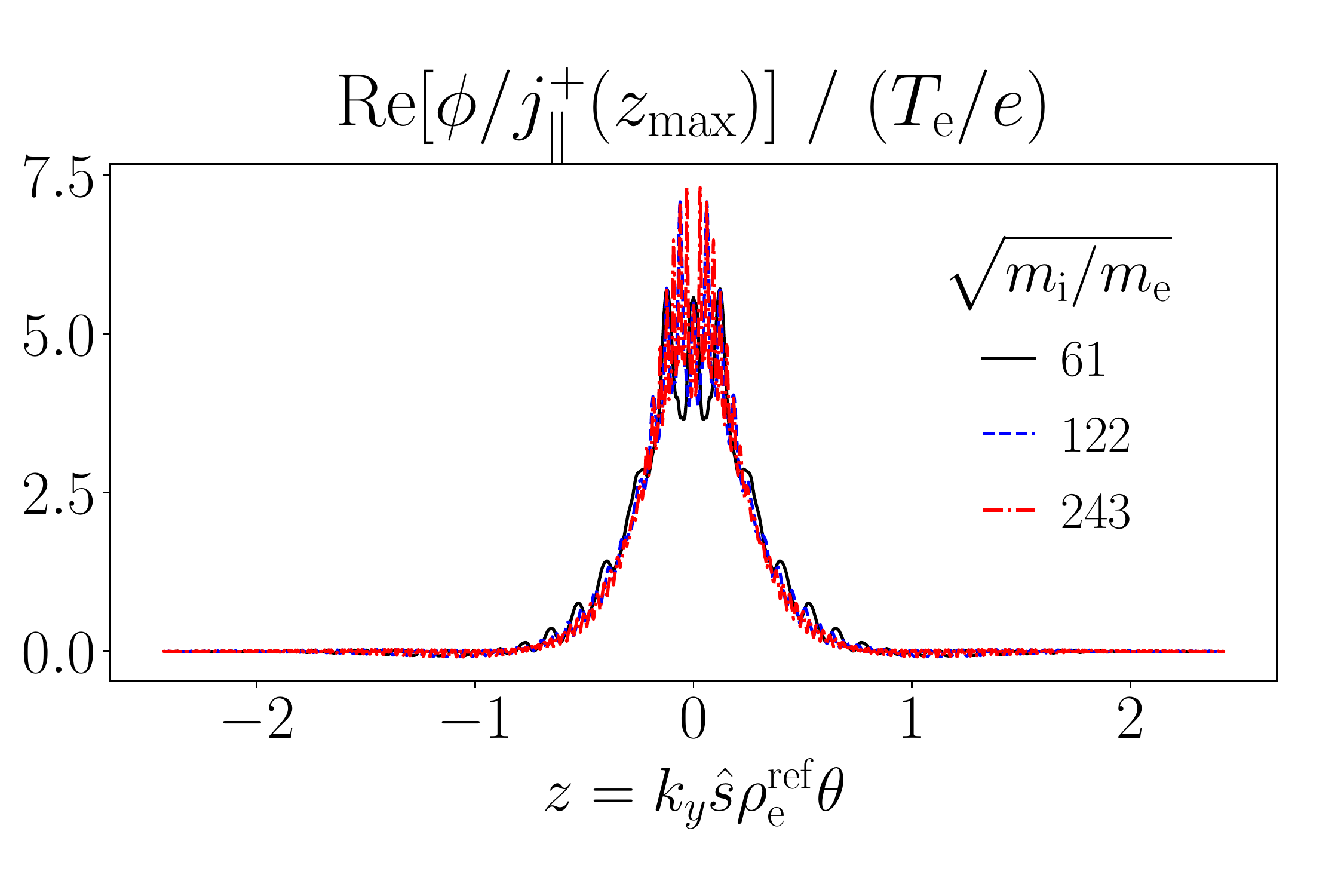}
\end{center}
\end{minipage}
\caption{ The real parts of the current-like function
 $\jintplus$ and the electrostatic potential $\ptl$ for the electrostatic mode at
 $(\kky\gyrdi,\thetaz) = (2.2,0.0)$. We normalise the eigenmode 
 by $\jintplus$ where $|\jintplus(\lzed)|$ is maximum. The field $\jintplus$
 is a measure of the forward-going electron distribution function. 
 The fact that the curves overlay indicates that the ordering \refeq{eq:large-tail-ordering}
 is satisfied, and that the width of the mode satisfies $\kky\shat\gyrderef\lpar \sim 1$.  }
 \label{fig:eigenmodes-electrostatic}
\end{figure}

We perform the same $\massrt \rightarrow 0$ analysis for the predominantly
 electrostatic mode at $(\kky\gyrdi,\thetaz) = (2.2,0.0)$. In figure \ref{fig:eigenmodes-electrostatic},
 we plot the real parts of the current-like field $\jintplus$, 
 and the potential $\ptl$. The imaginary parts have a similar order of magnitude and structure.
 Figure \ref{fig:eigenmodes-electrostatic} reveals that the mode is an
 electron-driven, electrostatic mode with large tails, like those observed in
 \cite{hardman_extended_tails}. Because the envelopes of the eigenmodes overlay well for different $\massrut$, 
 we conclude that the width of the mode satisfies 
 $\lzed = \kky \shat \gyrderef \lpar \sim 1$, and the ordering \refeq{eq:large-tail-ordering}
 is satisfied. 

\begin{figure}
\begin{minipage}{0.49\textwidth}
\begin{center}
\includegraphics[clip, trim=0cm 0cm 0cm 0cm, page=1, width=1.0\textwidth]{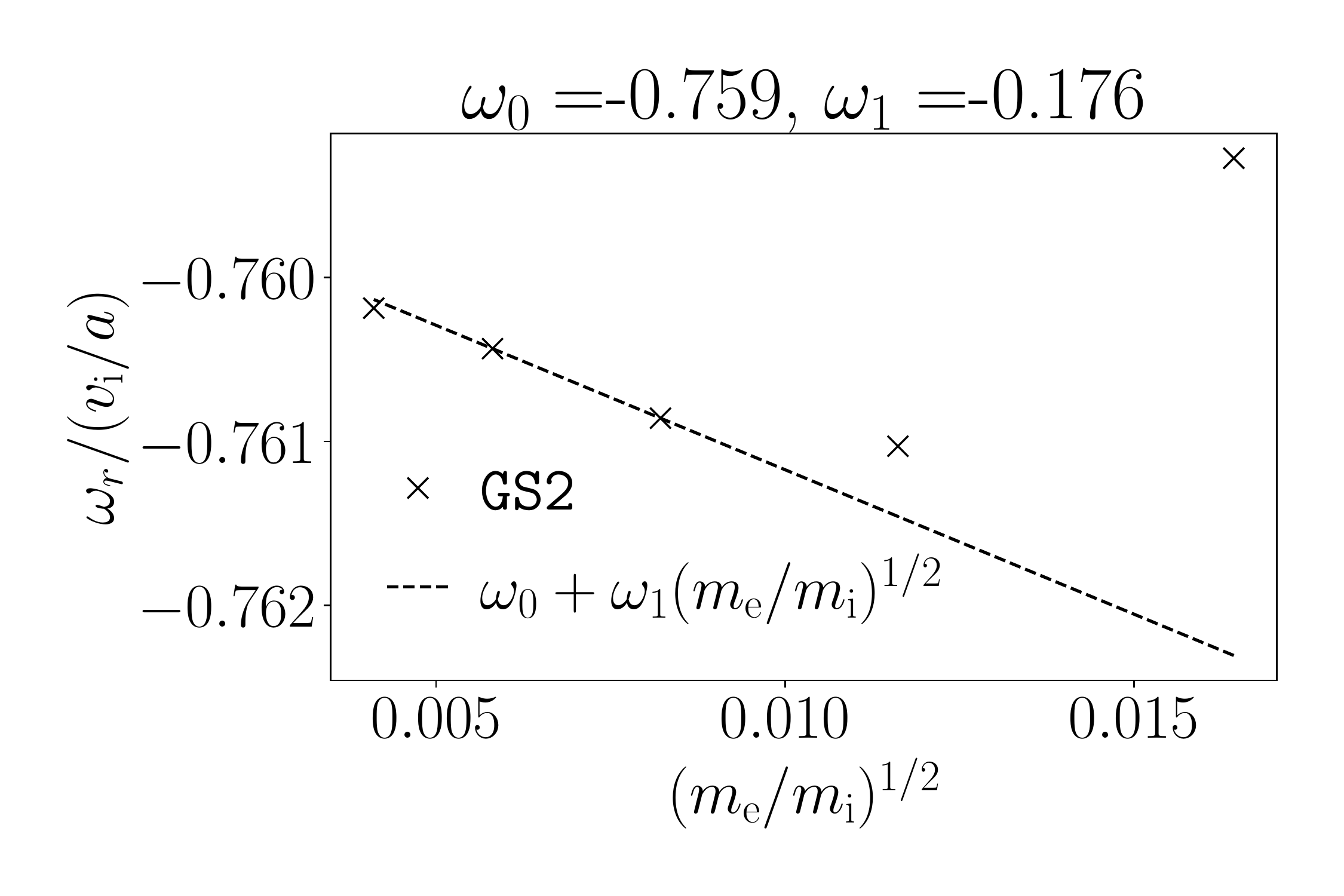}
\end{center}
\end{minipage}
\begin{minipage}{0.49\textwidth}
\begin{center}
\includegraphics[clip, trim=0cm 0cm 0cm 0cm,  page=1, width=1.0\textwidth]{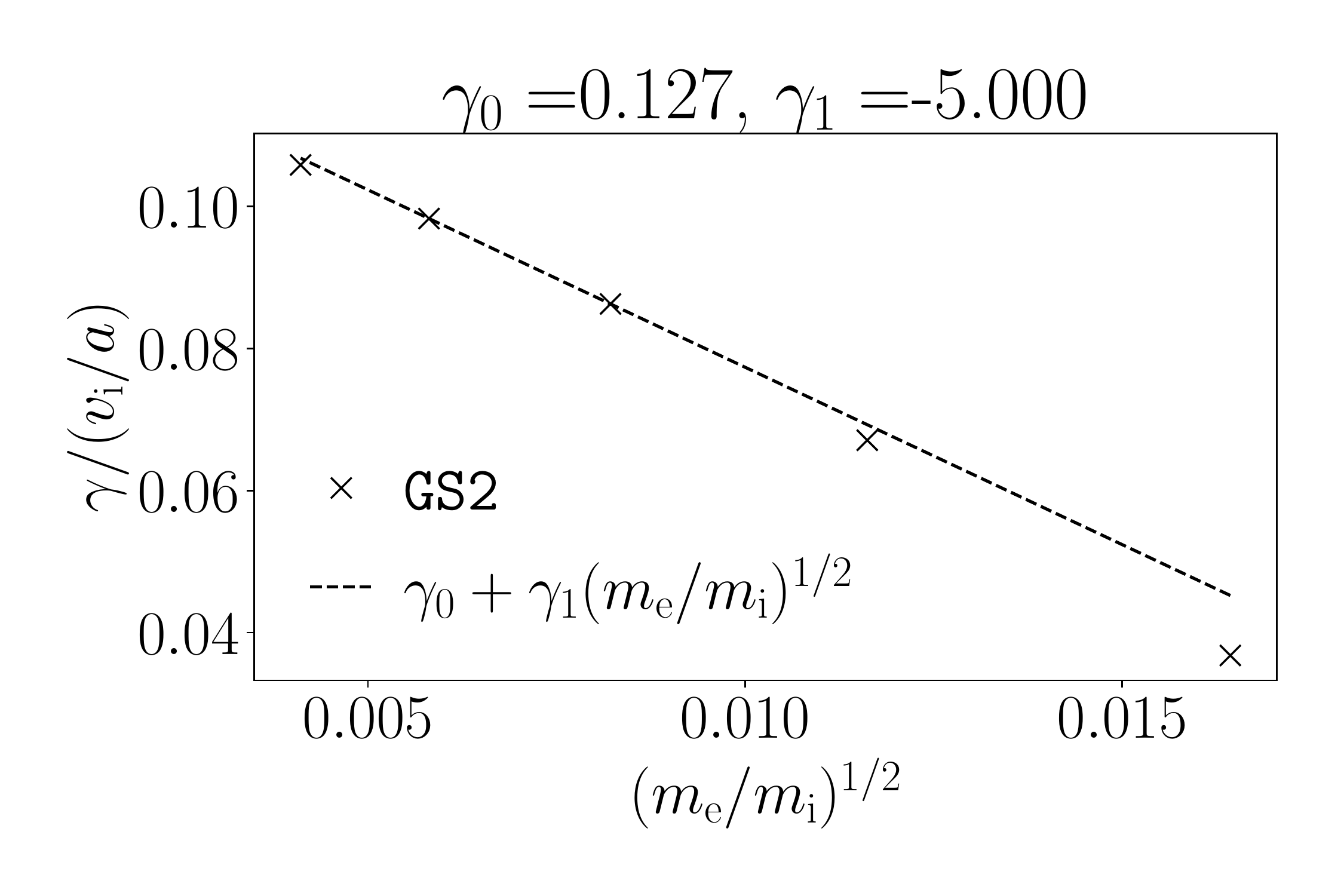}
\end{center}
\end{minipage}
\caption{ The growth rate $\growth$ and real frequency $\wfreqr$ of the electrostatic mode at
$(\kky\gyrdi,\thetaz) = (2.2,0.0)$, plotted for different values of the expansion parameter
 $\massrt$. We would expect that for small $\massrt$, $\growth$ and $\wfreqr$ 
 should be well represented by a constant plus a linear function of $\massrt$: 
 linear fits are provided to indicate the order of the variation with $\massrt$.
 The values of $\wfreq_0$, $\wfreq_1$, $\growth_0$, and $\growth_1$ 
 can be considered to be of order unity, although $\growth_0$ is surprisingly small,
 and $\growth_1$ is surprisingly large. Even though a nonasymptotic trend is observed in $\wfreqr$
 for the larger values of $\massrt$, the overall variation in $\wfreqr$ is small.
 }
 \label{fig:frequencies-electrostatic}
\end{figure}
\begin{figure}
\begin{minipage}{0.49\textwidth}
\begin{center}
\includegraphics[clip, trim=0cm 0cm 0cm 0cm, page=1, width=1.0\textwidth]{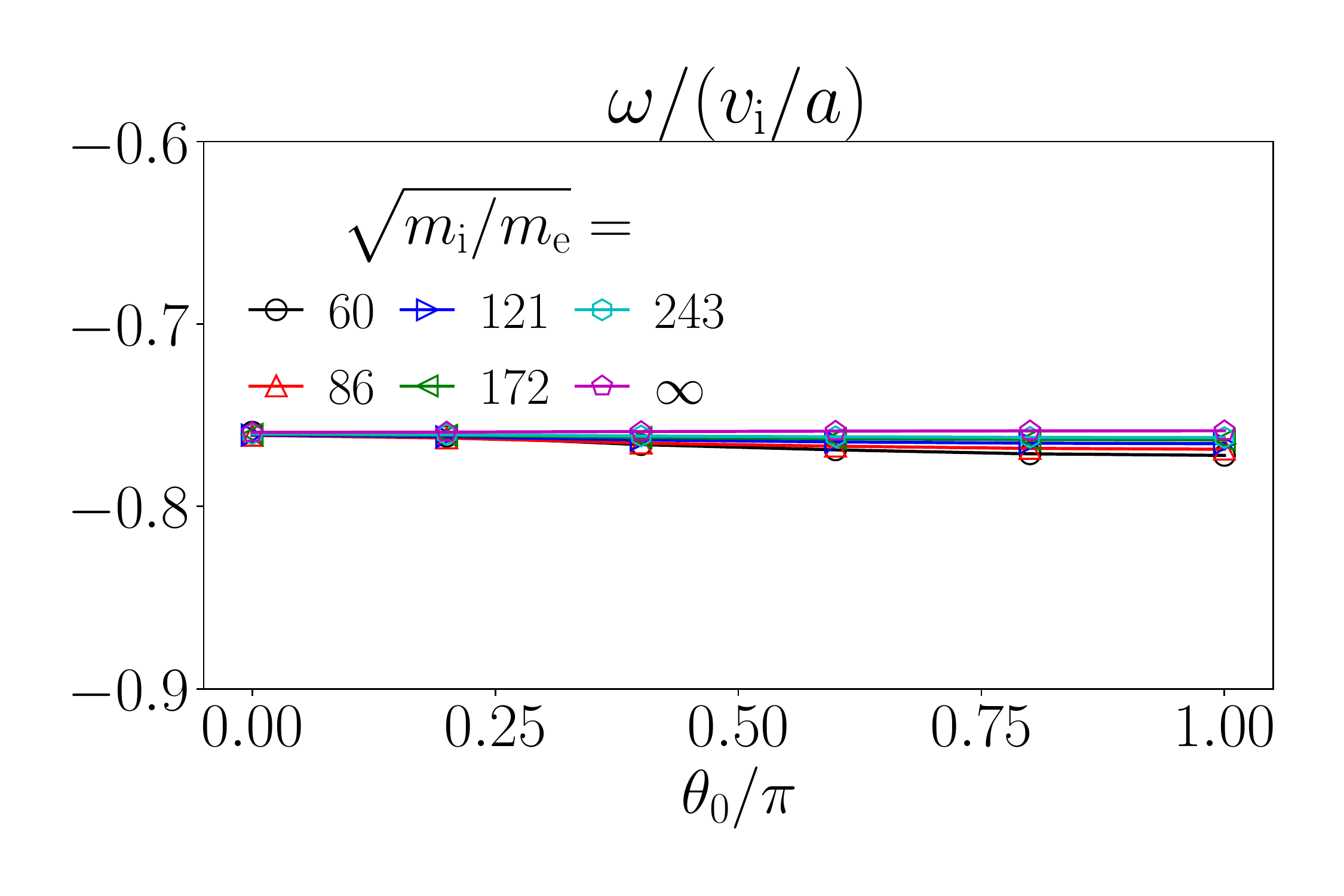}
\end{center}
\end{minipage}
\begin{minipage}{0.49\textwidth}
\begin{center}
\includegraphics[clip, trim=0cm 0cm 0cm 0cm,  page=1, width=1.0\textwidth]{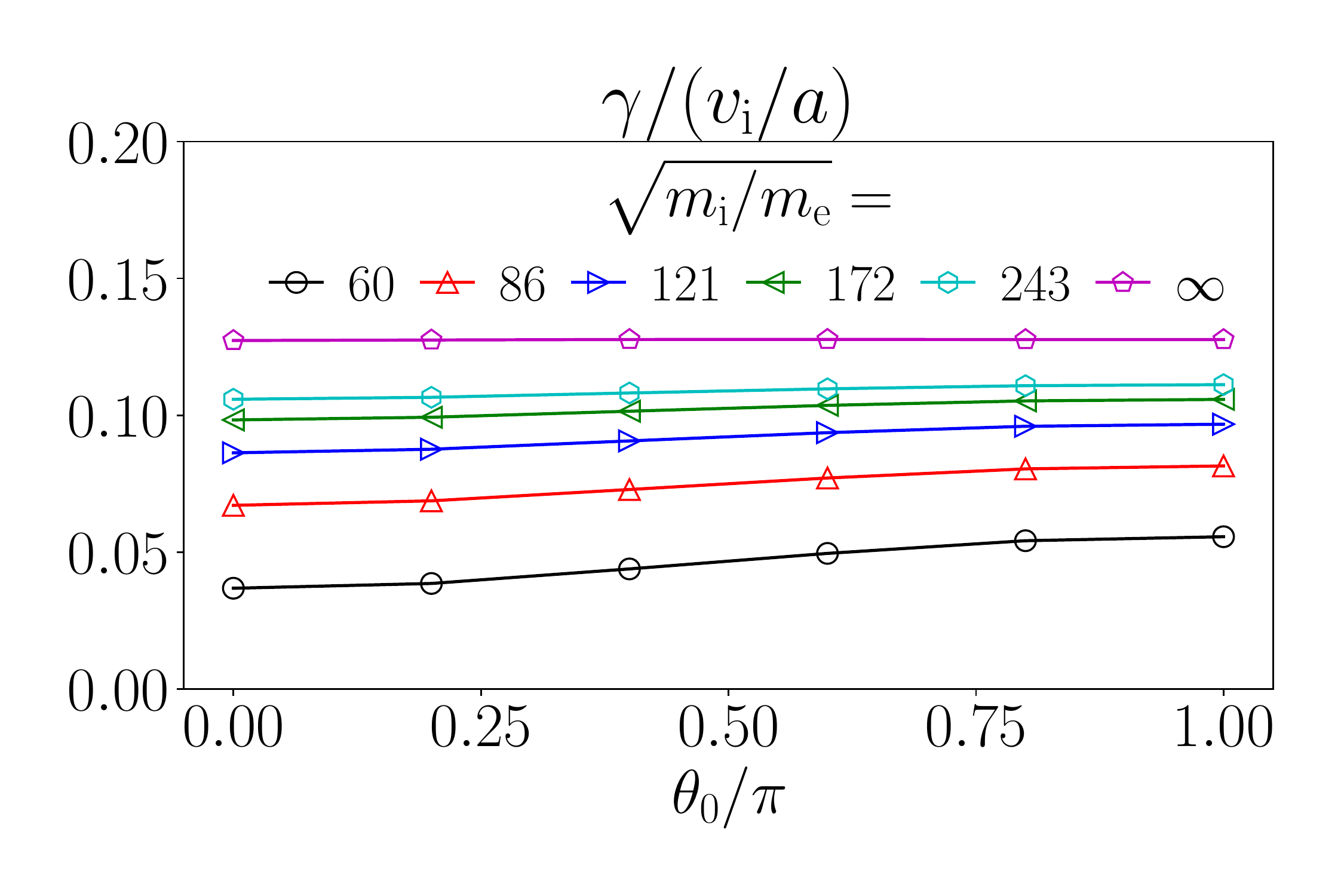}
\end{center}
\end{minipage}
\caption{The  real frequency $\wfreqr$ and growth rate $\growth$ of the electrostatic modes at
$\kky\gyrdi = 2.2$, as a function of $\thetaz$, plotted for different values of the expansion parameter
 $\massrut$. 
The points in the curve labelled by 
 $\massrut = \infty$ are computed from linear fits,
 see figure \ref{fig:frequencies-electrostatic}.
 Note that as $\massrt \rightarrow 0$, $\growth$ and $\wfreqr$ tend to constants that are independent
 of $\thetaz$, consistent with the leading-order electrostatic theory \cite{hardman_extended_tails}. }
 \label{fig:frequencies-theta0-electrostatic}
\end{figure}
 
In figure \ref{fig:frequencies-electrostatic}, we plot the real frequency and
 growth rate of the mode as a function of $\massrt$. This reveals that $\wfreqr$ 
 is well converged for the physical $\massrt = 1/61$: $\wfreqr$ changes only in the third decimal place. However, $\growth$ 
 has a surprisingly large linear variation with $\massrt$, meaning that 
 $\growth_0$ differs noticeably from $\growth$ for $\massrt = 1/61$. 
 This result may be caused by the fact that the mode is sufficiently
 close to marginal stability for first-order corrections to matter.
 We note that the fact that the mode is more unstable for smaller 
 $\massrt$ (as in the MTM case, cf. figure \ref{fig:frequencies-electromagnetic})
 suggests that the leading-order model of section \ref{section:reduced-model-equations}
 might provide a conservative upper bound on any linear growth rate.
 We examine the $\thetaz$ dependence of $\wfreq$ in figure \ref{fig:frequencies-theta0-electrostatic}.
 The real frequency $\wfreqr$ is a constant in $\thetaz$,
 consistent with the electrostatic limit described 
 in section \ref{section:subsidiary-limits} \cite{hardman_extended_tails}.
 The growth rate varies with $\thetaz$ for the physical 
mass ratio $\massrut = 61$, but increasing $\massrut$
 causes $\growth$ to tend to a constant that is independent 
of $\thetaz$, consistent with the leading-order theory.

\section{Testing the effective $\pbetae$}\label{section:effective-beta-test}
\begin{figure}
\begin{minipage}{0.49\textwidth}
\begin{center}
\includegraphics[clip, trim=0cm 0cm 0cm 0cm, page=1, width=1.0\textwidth]{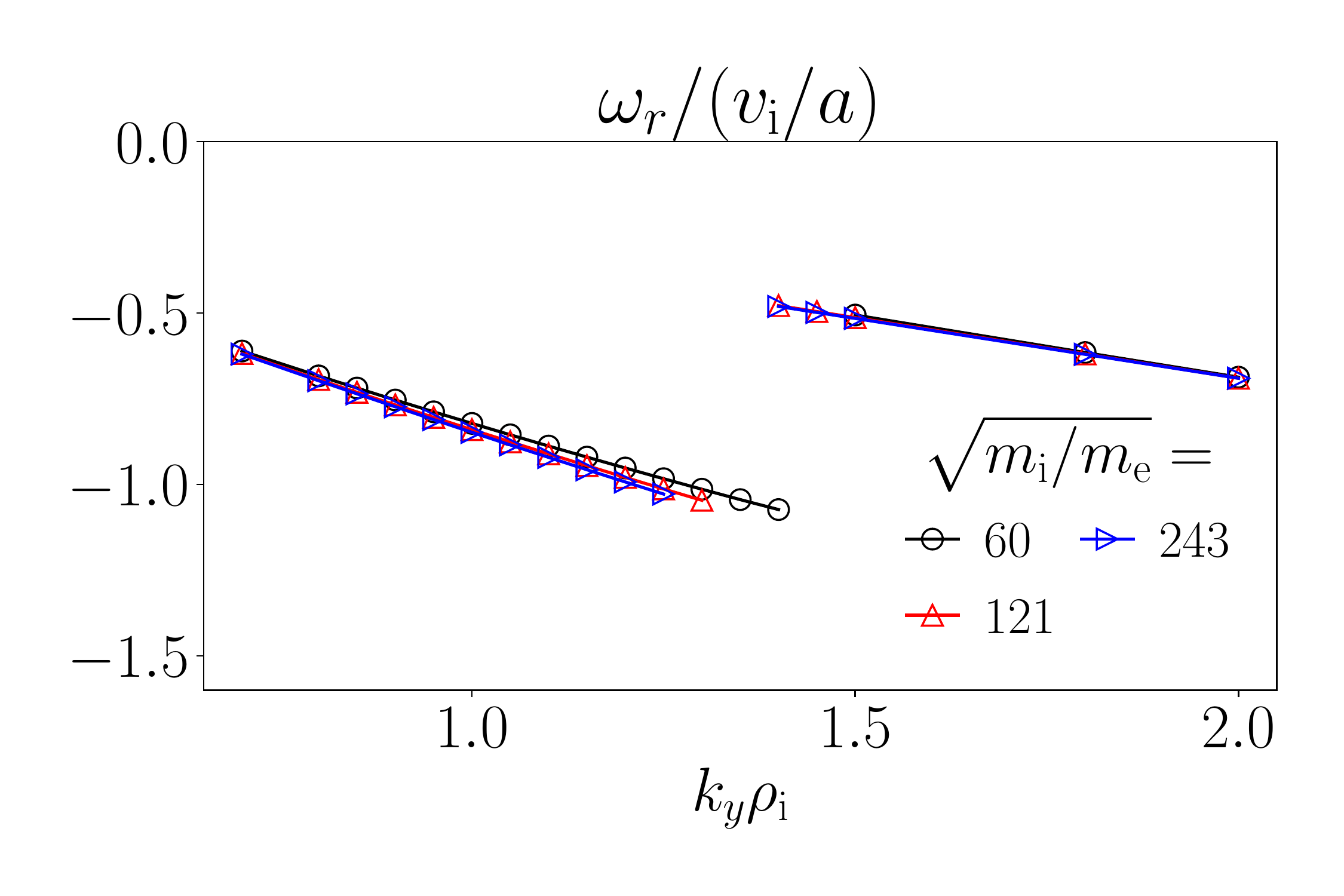}
\end{center}
\end{minipage}
\begin{minipage}{0.49\textwidth}
\begin{center}
\includegraphics[clip, trim=0cm 0cm 0cm 0cm,  page=1, width=1.0\textwidth]{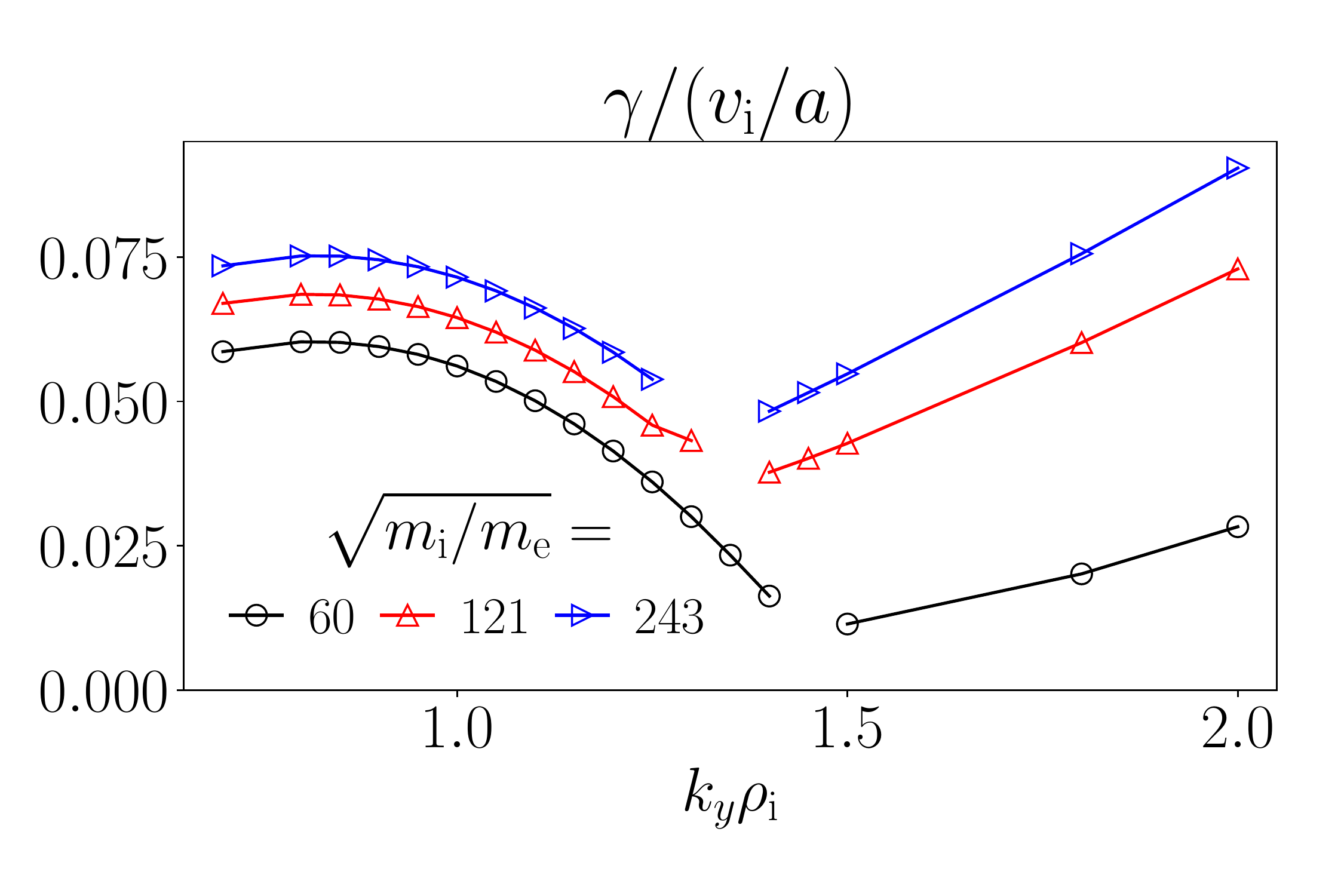}
\end{center}
\end{minipage}
\caption{The  real frequency $\wfreqr$ and growth rate $\growth$ spectra as a function 
of $\kky\gyrdi$ and the expansion parameter $\massrut\propto \pbetae^{-1}$, near the transition 
between the MTM mode branch (left of the discontinuity in $\wfreqr$) 
and the electrostatic mode branch (right of the discontinuity).
 Compare to figure \ref{fig:ky-spectrum-MAST-6252}.
 The boundary between the two types of modes remains almost
 constant in $\kky\gyrdi$ as the expansion parameter $\massrut $
 is varied by a factor of 4. Since each value of $\kky\gyrdi$ corresponds to a unique value of $\pbetaeff$, 
 a critical $\pbetaeff$ marks the boundary between the electromagnetic and the electrostatic modes. }
 \label{fig:frequencies-mass-ky-scan}
\end{figure}

 The asymptotic theory predicts that $\pbetae$ enters
 the dispersion relation only through $\pbetaeff$, an effective $\pbetae$ 
 that is defined by equation \refeq{eq:pbetaeff}.
 This is a strong insight that we now proceed to test.

 First, we examine the transition between the extended electrostatic modes
 and the MTM observed in figure \ref{fig:ky-spectrum-MAST-6252}.
 In figure \ref{fig:frequencies-mass-ky-scan}, we plot the result of simulations
 around the transition in $\kky\gyrdi$, for different $\massrut$
 and fixed $\thetaz = 0$. 
 As with the other $\massrut$ scans presented in this paper,
 we scale $\pbetae \propto \massrt$ in the scan so that $\pbetaeff$
 is held constant at each $\kky\gyrdi$.
 We note that the transition in $\kky\gyrdi$ remains close
 to $\kky\gyrdi \approx 1.4$ even as $\massrut$ is varied by a factor of 4.
 Since there is a unique correspondance between $\pbetaeff$ and $\kky\gyrdi$ for $\pbetae \propto \massrt$,
 this proves that the transition in $\kky\gyrdi$ occurs at a fixed critical $\pbetaeff \sim 1$. In other words, the critical
 $\pbetae$ for the onset of MTM instability scales with $\massrt$.  

\begin{figure}
\begin{minipage}{0.49\textwidth}
\begin{center}
\includegraphics[clip, trim=0cm 0cm 0cm 0cm, page=1, width=1.0\textwidth]{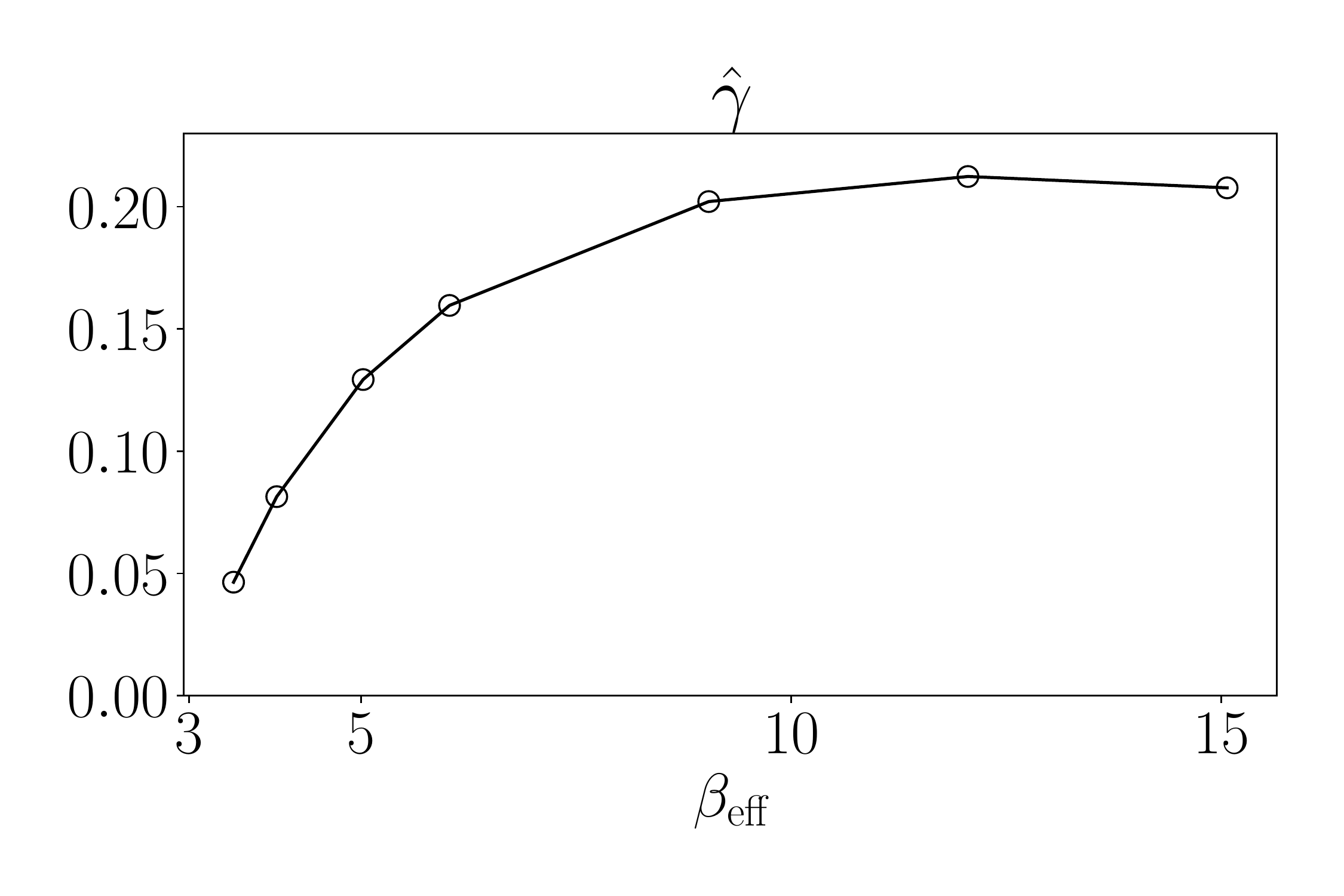}
\end{center}
\end{minipage}
\begin{minipage}{0.49\textwidth}
\begin{center}
\includegraphics[clip, trim=0cm 0cm 0cm 0cm,  page=1, width=1.0\textwidth]{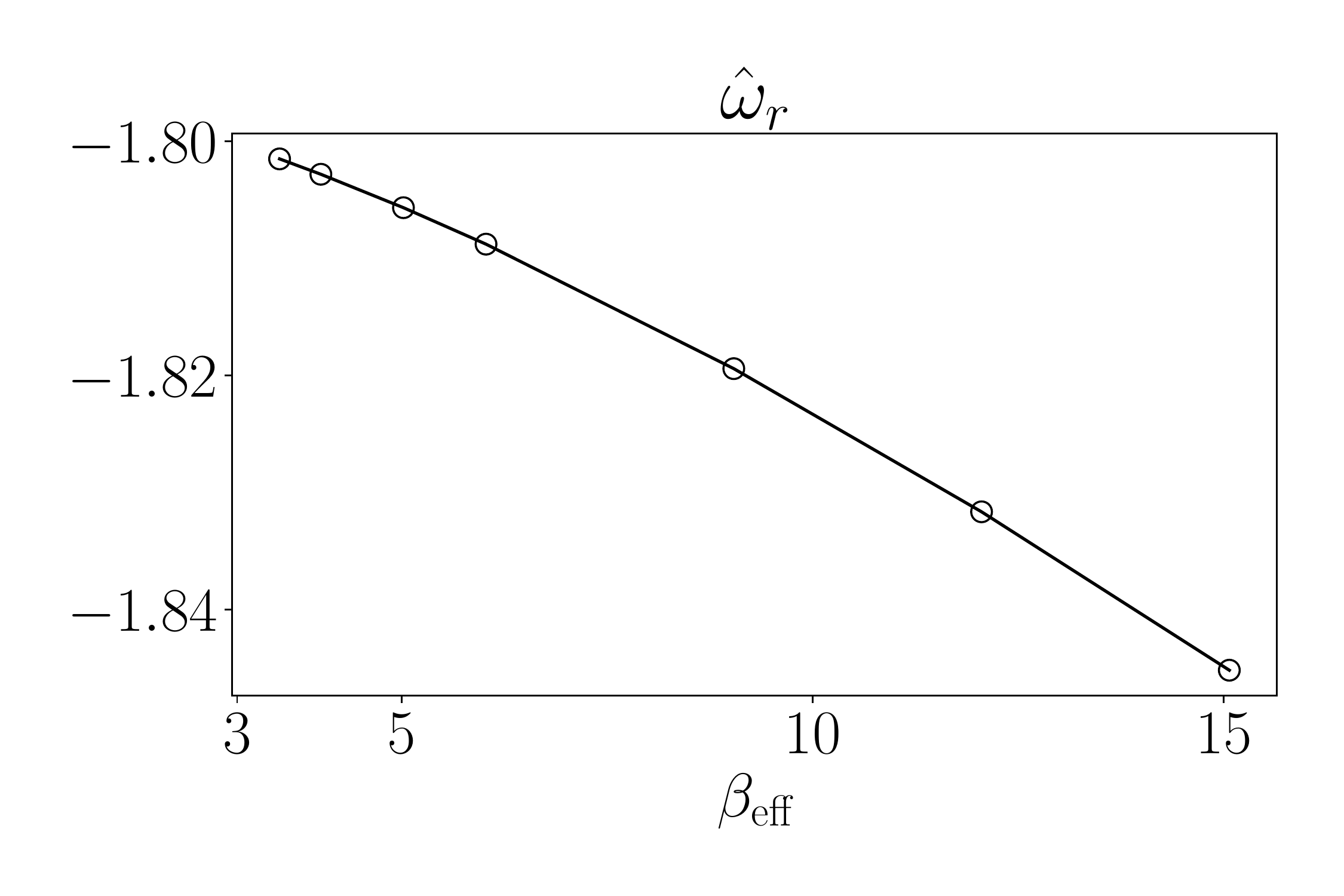}
\end{center}
\end{minipage}
\caption{ The dimensionless real frequency $\wfreqrhat$ 
and the dimensionless growth rate $\growthhat$, defined through $\wfreqhat = \wfreqrhat + \imag \growthhat$ 
and equation \refeq{eq:dispersion-relation}, plotted as a function of $\pbetaeff$,
 at fixed geometry, profiles, 
 and $\cfreqhat = 0.797$ ($\lscal\cfreqee / \vtheri = 0.303$). The data is generated through 
 a scan in $\pbetae = [0.0289,0.124]$ at fixed $(\kky\gyrdi,\thetaz)=(0.8,0.0)$ and $\massrt = 1/61$.
 Whilst the frquency $\wfreqrhat$ is almost independent of 
 $\pbetaeff$, the dependence of $\growthhat$ on $\pbetaeff$
 matches the predictions of the asymptotic theory: the critical $\pbetaeff$
 is of order unity, and $\growthhat$ appears to approach a constant for a small range 
 of $\pbetaeff$ as $\pbetaeff$ becomes large.
 }
 \label{fig:frequencies-betaeff-scan}
\end{figure}

\begin{figure}
\begin{minipage}{0.49\textwidth}
\begin{center}
\includegraphics[clip, trim=0cm 0cm 0cm 0cm, page=1, width=1.0\textwidth]{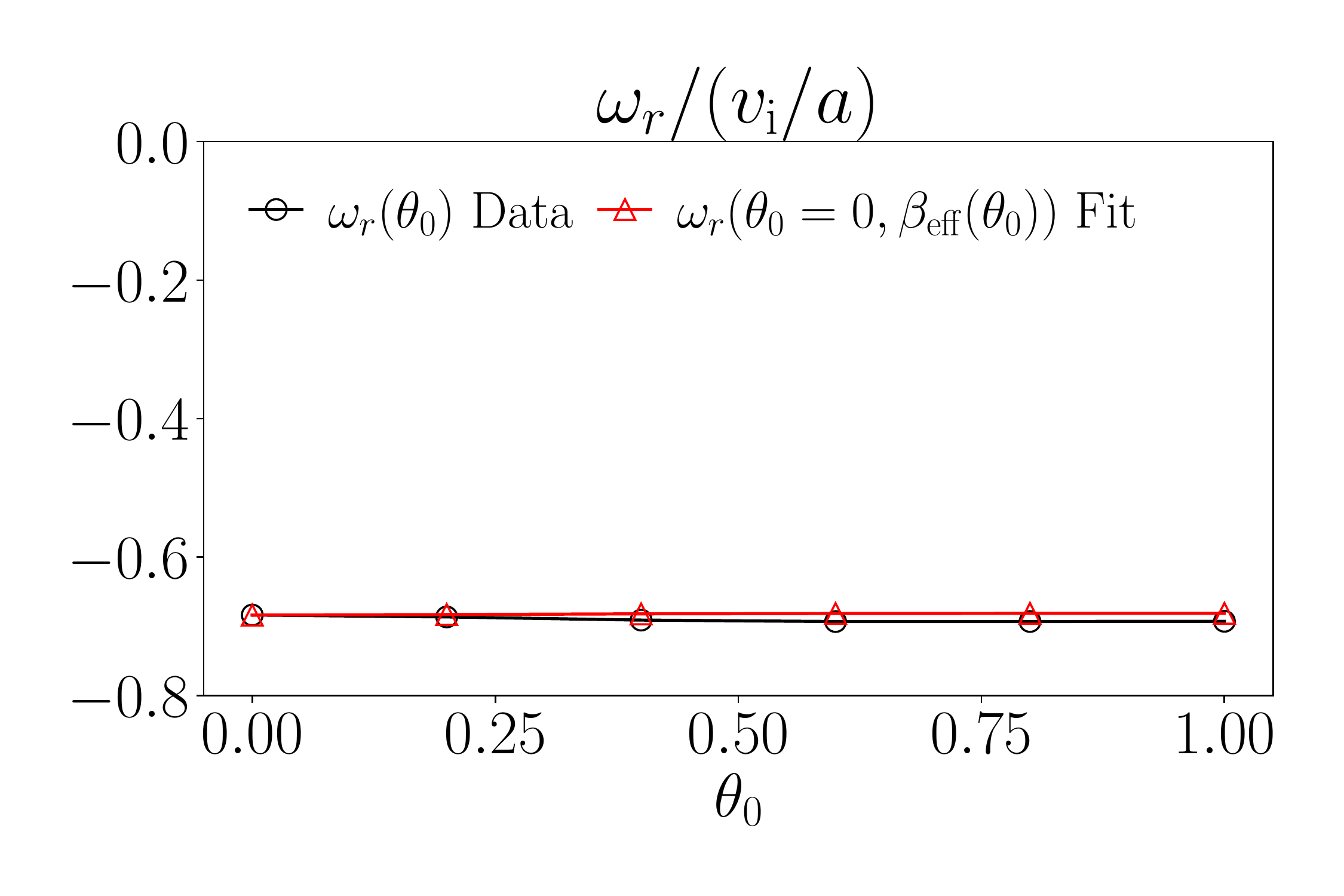}
\end{center}
\end{minipage}
\begin{minipage}{0.49\textwidth}
\begin{center}
\includegraphics[clip, trim=0cm 0cm 0cm 0cm,  page=1, width=1.0\textwidth]{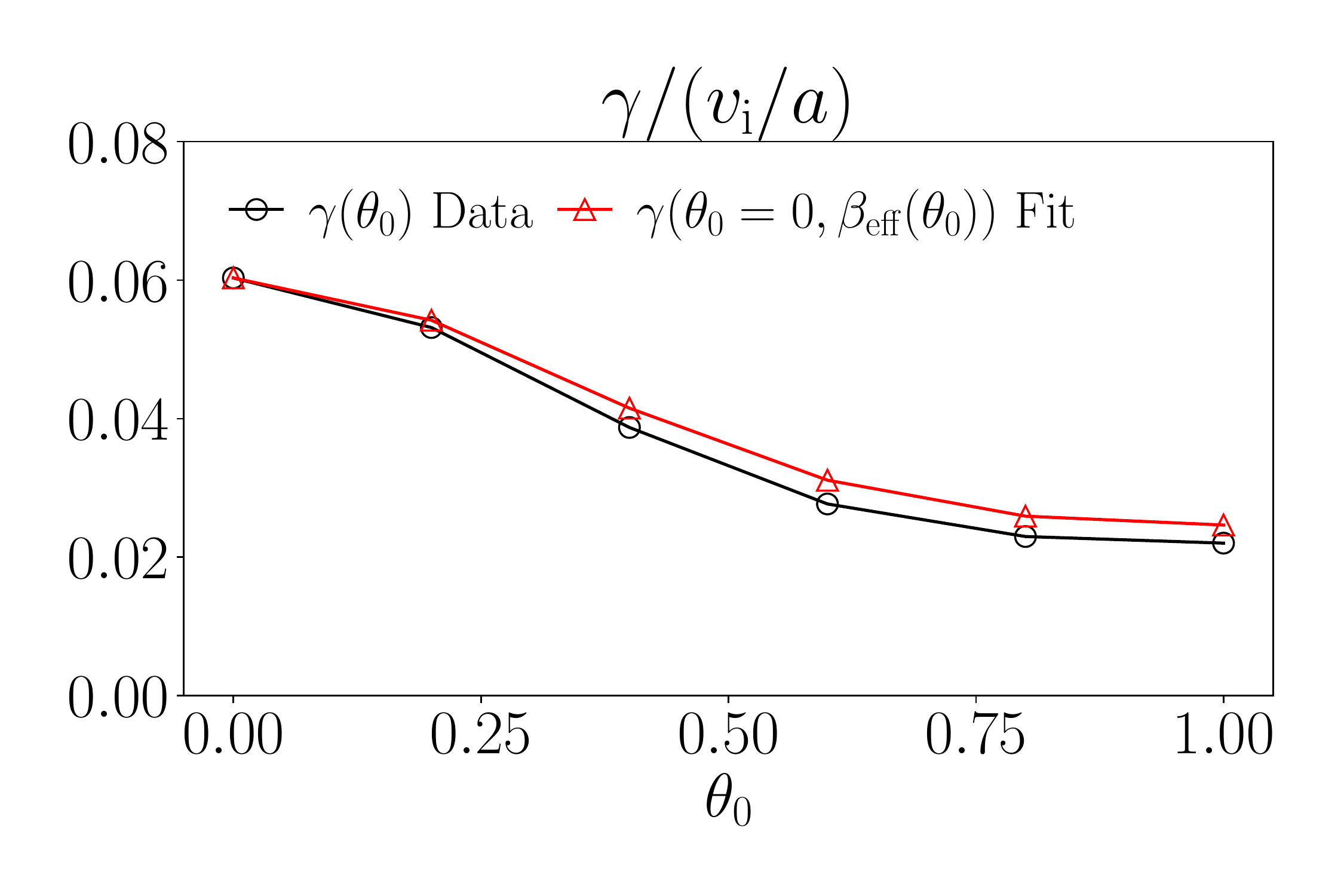}
\end{center}
\end{minipage}
\caption{ The real frequency $\wfreqr$ and growth rate $\growth$ of the MTM at $\kky\gyrdi = 0.8$,
 as a function of $\thetaz$. We compare two methods of computing the growth rate. 
 First, the result of calculating $\wfreqr(\thetaz)$ and $\growth(\thetaz)$
 directly with \gstwo~ using the nominal values of $\pbetae$ in the MAST equilibrium.
 Second, noting that $\thetaz$ only enters the dispersion relation \refeq{eq:dispersion-relation} through
 $\pbetaeff$, the result of using predictions for $\wfreqr$ and $\growth$ based 
 on the values of $\wfreqr(\thetaz=0,\pbetaeff)$ and $\growth(\thetaz=0,\pbetaeff)$
 and the value of $\pbetaeff(\thetaz)$.
 The excellent agreement of the two curves suggests that
 the matching condition \refeq{eq:leading-order-matching-lowbeta}
 is the correct one to describe the microinstabilities in MAST discharge \#6252. 
  }
 \label{fig:frequencies-theta0-scan-test}
\end{figure}

 Second, we test the detailed geometrical dependence of $\pbetaeff$
 on the function $\geofunc(\thetaz)$, defined by equation \refeq{eq:geofunc}.
 Since the only leading-order dependence of the modes on $\thetaz$ enters
 through $\pbetaeff$, we would expect that varying $\thetaz$ would have the
 same effect as varying $\pbetae$ by an appropriate amount. We test this
 hypothesis by performing a scan in $\pbetae$ at fixed $(\kky\gyrdi,\thetaz)=(0.8,0.0)$
 for the physical mass ratio $\massrt = 1/61$. In figure \ref{fig:frequencies-betaeff-scan},
 we show the result of calculating the dimensionless
 frequency $\wfreqhat = \wfreqrhat + \imag \growthhat$ 
 as a function of $\pbetaeff$, with all other parameters in
 equation \refeq{eq:dispersion-relation} held fixed. 
 The range of the $\pbetaeff$ scan is limited by the appearance of 
 competing instabilities. Below $\pbetaeff \lesssim 3$ the fastest growing instability is the
 electrostatic electron-driven mode, whereas for $\pbetaeff \gtrsim 15$ 
 a highly localised mode in $\lpar$ appears with a $\wfreqr$ 
 in the ion diamagnetic direction. 
 We use the data in figure \ref{fig:frequencies-betaeff-scan}
 to obtain $\growth(\pbetaeff)$ and $\wfreqr(\pbetaeff)$ for the mode at $\thetaz=0$.
 These fits can then be used to estimate the growth rate and frequency dependence on $\thetaz$ simply by computing $\pbetaeff(\thetaz)$ for all $\thetaz$, using the definition \refeq{eq:pbetaeff}.
 In figure \ref{fig:frequencies-theta0-scan-test} we compare these model predictions against gyrokinetic simulation results from \gstwo: 
 we see excellent agreement for all $\thetaz$.
 This procedure starts to break down for $\kky\gyrdi$ too small,
 but works well for any $\kky\gyrdi$ that is sufficiently close 
 to the electromagnetic-electrostatic mode transition
 highlighted in figures \ref{fig:ky-spectrum-MAST-6252}
 and \ref{fig:frequencies-mass-ky-scan}. This is likely 
 a result of the fact that the asymptotic theory is valid only
 when $\pbetae \sim \massrt$ -- when $\pbetae$ approaches values
 of order unity ($\pbetaeff \sim \massrut \gg 1$) then a different
 high-$\pbeta$ asymptotic theory is required.
 
    \section{Discussion}\label{section:discussion}

 In this paper, we have proposed a model for electron-driven electromagnetic linear instabilities
 that are localised to mode-rational surfaces,
 in the limit of $\pbetae \sim \kky\gyrde \sim \massrt \ll 1$. The model consists of
 the orbit-averaged equations \refeq{eq:electron_firstorder_passing_normalised},
 \refeq{eq:electron_firstorder_trapped_normalised}, and \refeq{eq:qninner} for the electron current layer, 
 and a matching condition, equation \refeq{eq:leading-order-matching-lowbeta},
 that connects the current layer to the large scale electromagnetic perturbation. 
 Physically, the matching condition represents the streaming 
 of electrons along perturbed magnetic field lines at large spatial scales:
 the magnetic field perturbations reconnect the equilibrium field lines
 and are driven by current carried by the electrons near the mode rational surface.
 The binormal scale of the mode $\kky\gyrdi\sim 1$ means that the radial width of the mode
 is of order $\gyrdi$, whilst the current 
 layer is taken to be of order $\gyrde$. As such, nonlocal physics due to the radial variation of equilibrium quantities
 of the scale of $\lscal \gg \gyrdi\gg \gyrde$ is neglected.
 
 Besides the orderings \refeq{eq:collisionless-ordering} and \refeq{eq:beta-ordering},
 no further assumptions are made  in deriving the model. 
 Hence, the model is valid in arbitrary axisymmetric
 toroidal geometry and can treat the strong shaping observed in spherical tokamaks, and include a trapped particle response.
 Passing electrons are critical to the mode, since only passing electrons carry the current that can drive reconnection.
 Trapped electrons can contribute to the drive of the instability by dragging on
 passing particles through interparticle collisions \cite{cattoPoF1981trapped}. 
 However, numerical investigations into the role of trapped particles in MTMs 
 show that trapped particles may be stabilising or destabilising \cite{Applegate_2007}. 
 
 The model makes the prediction that the dispersion
 relation has the form given by equation \refeq{eq:dispersion-relation}.
 As a consequence, we observe that $\pbetae$ only appears through the parameter $\pbetaeff$,
 defined by equation \refeq{eq:pbetaeff}. 
 Physically, the parameter $\pbetaeff$  
 determines how the binormal wavenumber and the shaping in the magnetic geometry affects the amount by which electrons
 are forced to cross equilibrium magnetic field lines, and hence drive reconnection.
 The amount of field-line-crossing and the form of $\pbetaeff$ is determined by the size of the surface-averaged 
 radial magnetic field induced by a given electron current from the rational-surface layer, see equations \refeq{eq:hhe-jump} and \refeq{eq:radial-B}. 
 We revisit MAST discharge \#6252 \cite{Applegate_2007},
 and we find that the MTMs there well satisfy the predictions made by the asymptotic theory.
 In particular, we are able to verify that $\pbetaeff$ accounts for the variation
 of the growth rate of the fastest-growing MTM mode in $\thetaz$. We also observe an electrostatic
 electron-driven mode with extended ballooning tails, similar to those observed
 in other magnetic geometries \cite{HallatschekgiantelPRL2005,hardman_extended_tails}.

 The fact that we are able to identify a $\pbetaeff$ that takes into account
 the shaping of the local flux surface means that the model has the potential
 to make an immediate impact on transport modelling. In principle, nonlinear simulations
 of MTM-driven turbulence are required to evaluate heat
 fluxes in discharges that have MTMs appearing 
 in the linear $\kky$ spectra. However, there have
 been persistent problems saturating this kind of turbulence --
 very few saturated MTM-driven turbulence simulations have been reported in the literature
 \cite{MTMWGuttenfelder2011PRL,Doerk2012mtm,maeyama2017supression,Ajay2022mtm}. One possible
 route to turbulence saturation is via equilibrium flow shear
 \cite{Newton_2010,HighcockPhysRevLett2010,BarnesPhysRevLett2011}, which is effective provided
 that unstable microinstabilities are localised to $\thetaz \approx 0$
 (the outboard midplane). Unfortunately, MTMs often have a healthy
 growth rate for all $-\pi <\thetaz \leq \pi $: the $\pbetaeff$ that we propose in this
 paper could act as a simple diagnostic to determine by how much the MTM growth
 rate $\growth$ varies in $\thetaz$. The stronger the variation in $\pbetaeff$
 between $\thetaz = 0$ and $\pm \pi$, the more likely it is that the MTMs
 will be stable at the inboard side of the device, 
 provided that $\pbetaeff \sim 1 \ll \massrut$. This insight could be
 used to target flux surfaces in design studies which are particularly amenable
 to MTM turbulence saturation through equilibrium flow shear.

 Further impact from the model could be derived from a numerical implementation
 of the equations, or an analytical solution in tractable asymptotic limits.
 A numerical implementation or analytical solution would be desirable because
 of the resolution requirements and computational costs associated with
 simulating long-wavelength MTMs with conventional gyrokinetic codes such as \gstwo.
 The large $\lpar \sim (\kky\shat\gyrde)^{-1} \gg 1$ extent in ballooning angle must be
 resolved at the same time as the fine geometrical structure in each $2\pi$
 segment in ballooning angle. Short timescales due to electron parallel streaming
 must be resolved at the same time as the slower timescales associated with the
 drives of instability at wavelengths long compared to the electron gyroradius.
 The model that we propose could potentially achieve computational savings through
 two routes: first, by eliminating the fast timescales due to electron parallel
 streaming; and second, by allowing us to represent the fine $2\pi$ structure
 of the eigenmodes with an expansion of just a few poloidal harmonics.

 \par \textit{ The authors are grateful for productive discussions with
 T. Adkins, A. A. Schekochihin, A. Zocco, P. Helander, 
 J. Larakers, D. Kennedy, R. Gaur, M. Anastopoulos-Tzanis,
 J. Maurino-Alperovich, S. Trinczek and G. Acton. 
 This work has received funding from EPSRC [Grant Number EP/R034737/1].
 This work was supported by the U.S. Department of Energy under contract number DE-AC02-09CH11466.
 The United States Government retains a non-exclusive, paid-up, irrevocable, world-wide license to publish
 or reproduce the published form of this manuscript, or allow others to do so, for United States Government purposes.
 The author acknowledges the use of the EUROfusion High Performance Computer
 (Marconi-Fusion) under projects OXGK and MULTISCA.
 This work made use of computational support by CoSeC, the Computational Science Centre 
 for Research Communities, through CCP Plasma (EP/M022463/1) and HEC Plasma (EP/R029148/1).
 The simulations were performed
 using the \gstwo~branch} \verb+https://bitbucket.org/gyrokinetics/gs2/branch/ms_pgelres+,\textit{ with commit ade5780.
The data that support the findings of this study are available upon reasonable request from the authors.
 The \gstwo~input files used to perform the gyrokinetic simulations in this study
 are publicly available \cite{Hardman_MTM_supplementary}.}

\appendix

\section {The $\kkr\gyrde \sim 1$ scale}
\label{section:kr-gyrde-sim-1-low-beta}

    The expansion in the inner region is carried out in powers
    of $\massrt \sim \kky\shat\gyrde \ll 1$. At the leading order
    in the expansion of equation \refeq{eq:lineargyrokinetic},
    we find a balance between parallel streaming and the radial magnetic drift:
 \beqn \vpar \kpar \drv{\hhez}{\lpar} 
  - \imag \kkfldl \lchi \saffacprim  \vme\cdot\nbl\flxl \; \hhez
  = 0.\label{eq:zeroth-order}\eeqn 
  In equation \refeq{eq:zeroth-order},
  we have introduced $\lchi =\lpar$ where the ballooning angle appears secularly
  in the gyrokinetic equation.  We treat $\lpar$ and $\lchi$ as
  independent variables in a multi-scale expansion, and we preserve $\lpar$
  as the argument of periodic functions. 
 
  Using the identity
  for the radial magnetic drift \cite{hintonRMP76,helander},
  \beqn \vme\cdot\nbl\flxl = 
  \vpar\kpar\drv{}{\lpar}\left(\frac{\bcur \vpar}{\cycfe}\right),
  \label{eq:radial-magnetic-drift} \eeqn
  allows us to integrate equation \refeq{eq:zeroth-order} using an integrating factor $\expo{-\imag\kkfldl\saffacprim \lchi \bcur \vpar /\cycfe}$,
  and  obtain the relationship \refeq{eq:hhe}, in terms of the coordinate $\lzed = \kky\shat|\gyrderef|\lchi$.
  At first order in the expansion the leading-order 
  source due to the fields $\sfunce = - \imag (\wstare - \wfreq)\besen{0} \charge \ptl \eqlbe /\tempe$ enters into the right hand side 
  of the electron gyrokinetic equation. We have that 
\beqn \fl \vpar \kpar \left(\drv{\hhez}{\lchi} + \drv{\hheo}{\lpar}\right)
  - \imag \kkfldl \lchi \saffacprim \vme\cdot\nbl\flxl \; \hheo \nonumber\eeqn \beqn 
  + \imag\left(\kkfldl\vme\cdot(\nbl \fldl + \lpar \nbl\saffac )
  - \wfreq\right)\hhez 
  - \cope\left[\hhez\right]= \sfunce. \label{eq:first-order}\eeqn 
  To close equation \refeq{eq:first-order}, in the passing part of phase space,
  we must impose 
  $2\pi$-periodicity in $\lpar$ on $\hheo$, i.e.,
  $\hheo(\lchi,\lpar)=\hheo(\lchi,\lpar+2\pi)$. In the trapped part of phase space
  we must impose that $\hheo$ satisfies the bounce condition for trapped 
  particles that $\hhe(\lparpm,\sign=1)=\hhe(\lparpm,\sign=-1)$ at the upper and lower bounce points $\lparpm$.
  This is achieved by multiplying equation \refeq{eq:first-order} by the factor $\expo{-\imag\kkfldl\saffacprim \lchi \bcur \vpar /\cycfe}$, applying the transit and
  bounce averages, defined by equations
  \refeq{eq:transitav} and \refeq{eq:bounceav}, respectively,
  to find the solvability conditions on $\hhez$. 
  After normalising the results, using the coordinate $\lzed = \kky\shat|\gyrderef|\lchi$,
  we have the equations
  \refeq{eq:electron_firstorder_passing_normalised}
  and \refeq{eq:electron_firstorder_trapped_normalised}.

\section{The electron drift-kinetic collision operator}
\label{section:drift-kinetic-collisions}
 In this appendix, we explicitly define the drift-kinetic 
 collision operator appearing in equation
 \refeq{eq:collision_operator}. The drift-kinetic operator 
 \beqn \copbothe\left[\fulldistf\right] = \coplandaue\left[\fulldistf\right]
  + \coplorentze\left[\fulldistf\right],\label{eq:DK-collisions}\eeqn 
 where $\fulldistf = \fulldistf(\pvel)$, 
 $\coplandaue[\cdot]$ is the linearised electron landau self-collision operator,
 and   $\coplorentze\left[\cdot\right]$ is
 the collision operator resulting from electron-ion collisions.
 
    The landau self-collision operator $\coplandaue\left[\fulldistf\right]$
    is defined by
    \beqn \fl \coplandaue\left[ \fulldistf \right] =
    \frac{\cfreqee \vthere^3}{2} 
   \drv{}{\pvel} \cdot \intvprim{\frac{\eqlbe\eqlbeprim}{\dense}
    \Ufunc(\pvel - \pvelprim) \cdot \left(\drv{}{\pvel}\left(\frac{\fulldistf}{\eqlbe}\right)-
    \drv{}{\pvelprim}\left(\frac{\fulldistfprim}{\eqlbeprim}\right)\right)} , \label{eq:coplandaus} \eeqn
    where the electron self-collision frequency  
    $ \cfreqee$ is defined through equation \refeq{eq:cfreqss}, 
    with the shorthand notation
    $\fulldistf = \fulldistf(\pvel)$,
    $\fulldistfprim = \fulldistf(\pvelprim)$,
    $\eqlbe = \eqlbe(\pvel)$, $\eqlbeprim = \eqlbe(\pvelprim)$, 
    \beqn\Ufunc(\pvel - \pvelprim) =
    \frac{\eye |\pvel - \pvelprim|^2 - (\pvel - \pvelprim)(\pvel - \pvelprim)}{|\pvel - \pvelprim|^3},
    \label{eq:ufuncdef}
    \eeqn
    and $\eye$ the identity matrix.
    Similarly, the electron-ion collision operator is defined by 
    \beqn {\coplorentze\left[\fulldistf \right] = \frac{\cfreqei \vthere^3}{2} 
    \drv{}{\pvel} \cdot \left(\frac{\vmag^2 \eye - \pvel \pvel}{\vmag^3}
    \cdot \drv{\fulldistf}{\pvel}\right),} \label{eq:coplorentze}\eeqn 
    with the electron-ion collision frequency 
    $\cfreqei$ defined by equation \refeq{eq:cfreqss}.
    Note that the ion mean velocity does not appear in equation \refeq{eq:coplorentze}
    because the nonadiabatic response of ions is small for $\kkr\gyrdi \gg 1$. 
          
 \section {The $\kkr\dske \sim 1$ scale for $\pbetae \ll 1$}\label{section:kr-dske-sim-1-low-beta}

 In the $\pbetae$ ordering \refeq{eq:beta-ordering},
 the electron inertial scale $\dske = \gyrde/\sqrt{\pbeta}$ is intermediate to the ion and electron gyroradius scales,
 i.e., $\gyrde \ll \dske \ll \gyrdi$. Since we assume that $\kky \gyrdi \sim 1$ and $\shat \sim 1$,
 to treat the scale of $\kkr \dske \sim 1$ we must consider large ballooning angles in the range
 \beqn \lpar \sim (\kky\shat\dske)^{-1} \sim \massrup{1/4} \gg 1. \label{eq:low-beta-theta-scale} \eeqn
 We use a similar multi-scale expansion as used
 in \ref{section:kr-gyrde-sim-1-low-beta} to derive the drift-orbit-averaged
 equations in the $\kkr\dske \sim 1$ region. Here,
 we expand in powers of $\massrtp{1/4}$,
 and we take $\lpar \sim  \massrutp{1/4}$.
 We introduce the independent variable $\lchi = \lpar$ where ballooning angle appears
 securlarly in the gyrokinetic equation, and we reserve $\lpar$
 for argument of periodic functions.
    We estimate the sizes of terms in the electron gyrokinetic equation
    using the non-dimensionalised Amp\`{e}re's law, 
    equation \refeq{eq:ampere-nondim}, and the
    basic ordering \refeq{eq:large-tail-ordering}.

 At leading-order in the expansion we find that  
 \beqn \vpar \kpar \drv{\hhez}{\lpar} = 0,\eeqn 
 i.e., $\hhez = \hhez(\lchi,\energy,\pitch,\sign)$ is 
 independent of $2\pi$-geometric variation through $\lpar$.
 At first-order in the expansion, we find that $\hhez$ is also independent of $\lchi$:
 \beqn \vpar \kpar \left(\drv{\hhez}{\lchi} + \drv{\hheh}{\lpar} \right)
 - \imag \kkfldl \lchi \saffacprim \vme\cdot\nbl\flxl \; \hhez= 0.\label{eq:first-order-dske-low-beta}\eeqn
 Applying the transit average  $\transav{\cdot}$,
 defined by equation  \refeq{eq:transitav}, and using the identity
 \refeq{eq:radial-magnetic-drift},
 we find that  \beqn \drv{\hhez}{\lchi}= 0,\eeqn
 i.e., in fact, $\hhez = \hhez(\energy,\pitch,\sign)$.
 
 This discussion shows that the leading-order passing electron distribution function
 is constant across the intermediate region where $\kkr\dske \sim 1$. 
 The leading-order passing electron distribution function at $\kkr\dske \sim 1$ is determined by the incoming
 boundary conditions that are imposed on the $\kkr\dske \sim 1$ region from matching to
 the $\kkr\gyrdi \sim 1$ and $\kkr\gyrde \sim 1$ regions.
 This comes about because in the ordering \refeq{eq:collisionless-ordering},
 electron parallel streaming dominates over the sources in the gyrokinetic equation 
 at this scale. 
\section{Numerical resolutions}
\label{section:resolutions}
 
 The simulations presented in sections \ref{section:numerical-evidence} and \ref{section:effective-beta-test}
 of this paper use the following common set of numerical resolutions:
 the timestep size is taken to be $\delt = 0.05/\kky\gyrdi$; we take 
$\ntheta = 33$ points per $2\pi$ element in the ballooning angle grid;
 $\npitch = 27$ points in the pitch angle grid, which is formed with a
 Radau-Gauss grid for passing particles and an unevenly spaced grid for trapped particles;
 and $\negrid =24$ points in the spectral energy grid \cite{barnesPoP10a}.
 The number of $2\pi$ elements in the ballooning grid $\ntwopi$
 is chosen to be large enough to resolve the eigenmode and frequency.
 The ballooning-space extent of electron-driven modes varies with $\kky\gyrde$,
 leading to different requirements on $\ntwopi$ for modes of different 
 $\massrt$ and $\kky\gyrdi$. For the mode at $\kky\gyrdi = 0.1$ and
 $\massrt = 1/61$ in figure \ref{fig:eigenmodes-MAST-6252},
 $\ntwopi = 399$ is required, whereas for the modes at $\kky\gyrdi = 0.8$ ($2.2$) $\ntwopi = 79$ ($39$)
 is found to be adequate. When $\massrt$ is varied at fixed  $\kky\gyrdi$,
 we increase $\ntwopi \propto \massrut$.

\bibliography{hardman_lowbeta_electromagnetic_modes.bbl} 

\end{document}